\documentclass[longauth]{aa}

\usepackage{graphicx}
\usepackage{natbib}
\usepackage{scalerel}
\usepackage{comment}

\usepackage[table]{xcolor}
\usepackage{lscape}
\usepackage{lastpage}
\usepackage{booktabs}

\usepackage{txfonts}
\usepackage[pdfencoding=auto,psdextra]{hyperref}
\hypersetup{
    colorlinks=true,
    linkcolor=blue,
    filecolor=magenta,      
    urlcolor=blue,
    citecolor=blue
}
\makeatletter
\renewcommand*\aa@pageof{, page \thepage{} of \pageref*{LastPage}}
\makeatother

\usepackage[utf8]{inputenc}

\usepackage[switch, modulo]{lineno}
\usepackage[version=4]{mhchem}
\usepackage{xspace}
              
\renewcommand{\linenumbers}[0]{}

\usepackage{euclid}

\newcommand{\NNEW}{202\xspace}% Number in new 
\newcommand{\NNEWSPECTRA}{33\xspace}%Number in new with confident Q1 spt, ie those in table 2 if we only count >M7
\newcommand{\NKNWSPECTRA}{167\xspace}%Number in known with confident Q1 spt
\newcommand{\NSPECTRA}{200\xspace}%sum of nnewspectra and nkwnspectra
\newcommand{\NNEWINTSPECTRA}{46\xspace}%Number in new with interesting spectra
\newcommand{\NDES}{326\xspace}% Number in DES and Q1
\newcommand{\NKNW}{12\xspace}% Number in master file and Q1
\newcommand{\EDFN}{EDF-N\xspace}
\newcommand{\EDFF}{EDF-F\xspace}
\newcommand{\EDFS}{EDF-S\xspace}

\providecommand{\gmag}{\ensuremath{G}}
\providecommand{\gbp}{\ensuremath{G_{\mathrm{BP}}}}
\providecommand{\grp}{\ensuremath{G_{\mathrm{RP}}}}
\date{Received 1 April 2026 / Accepted dd June 2026}
\begin{document}
% \AtBeginDocument{%
%   \nolinenumbers
%   \renewcommand{\linenumbers}{}
% }

\title{Euclid: Quick Data Release (Q1) -- Ultracool dwarfs in the Euclid Deep Fields 
\thanks{This paper is published on behalf of the Euclid Consortium}
}
% \subtitle{Ultracool dwarfs in the Euclid Deep Fields}

%=============================================
%%%% Version Friday 27th of March 2026 10:23:55 AM UT												
%%%% Please do not edit the author list -- contact ECEB Bureau for changes
%\newcommand{\orcid}[1]{} %% define as link to https://orcid.org/#1 if needed
\newcommand{\orcid}[1]{} %% if already defined in aa.cls: comment, or use renewcommand			   
% \author{A.~Mohandasan\orcid{0000-0001-5182-0330}\thanks{\email{anjana.mohandasan@univ-fcomte.fr}}\inst{\ref{aff1}}
\author{A.~Mohandasan\orcid{0000-0001-5182-0330}\thanks{\email{anjana.mohandasan@univ-fcomte.fr}}
\inst{\ref{aff1}}
\and R.~L.~Smart\orcid{0000-0002-4424-4766}\inst{\ref{aff2},\ref{aff3}}
\and C.~Reyl\'e\orcid{0000-0003-2258-2403}\inst{\ref{aff1}}
\and V.~Le~Brun\orcid{0000-0002-5027-1939}\inst{\ref{aff4}}
\and A.~P\'erez-Garrido\orcid{0000-0002-5139-1975}\inst{\ref{aff5}}
\and E.~Ba\~nados\orcid{0000-0002-2931-7824}\inst{\ref{aff6}}
\and H.~R.~A.~Jones\inst{\ref{aff3}}
\and T.~Dupuy\orcid{0000-0001-9823-1445}\inst{\ref{aff7}}
\and B.~Goldman\orcid{0000-0002-2729-7276}\inst{\ref{aff8},\ref{aff9}}
\and S.~L.~Casewell\orcid{0000-0003-2478-0120}\inst{\ref{aff10}}
\and M.~R.~Zapatero~Osorio\orcid{0000-0001-5664-2852}\inst{\ref{aff11}}
\and M.~Rejkuba\orcid{0000-0002-6577-2787}\inst{\ref{aff12}}
\and E.~L.~Mart\'in\orcid{0000-0002-1208-4833}\inst{\ref{aff13},\ref{aff14}}
\and S.~Mu\~{n}oz~Torres\orcid{0000-0003-4269-4779}\inst{\ref{aff13},\ref{aff14}}
\and C.~Dom\'{\i}nguez-Tagle\orcid{0000-0002-0463-6212}\inst{\ref{aff13},\ref{aff14}}
\and M.~{\v Z}erjal\orcid{0000-0001-6023-4974}\inst{\ref{aff13},\ref{aff14}}
\and N.~Hu\'elamo\inst{\ref{aff10}}
\and N.~Lodieu\orcid{0000-0002-3612-8968}\inst{\ref{aff15},\ref{aff14}}
\and P.~Cruz\orcid{0000-0003-1793-200X}\inst{\ref{aff10}}
\and F.~Z.~Majidi\orcid{0000-0002-8407-5282}\inst{\ref{aff16}}
\and R.~Rebolo\orcid{0000-0003-3767-7085}\inst{\ref{aff13},\ref{aff17},\ref{aff14}}
\and M.~W.~Phillips\orcid{0000-0001-6041-7092}\inst{\ref{aff11}}
\and J.-Y.~Zhang\orcid{0000-0001-5392-2701}\inst{\ref{aff13},\ref{aff14}}
\and N.~Aghanim\orcid{0000-0002-6688-8992}\inst{\ref{aff18}}
\and B.~Altieri\orcid{0000-0003-3936-0284}\inst{\ref{aff19}}
\and A.~Amara\inst{\ref{aff20}}
\and S.~Andreon\orcid{0000-0002-2041-8784}\inst{\ref{aff21}}
\and N.~Auricchio\orcid{0000-0003-4444-8651}\inst{\ref{aff22}}
\and C.~Baccigalupi\orcid{0000-0002-8211-1630}\inst{\ref{aff23},\ref{aff24},\ref{aff25},\ref{aff26}}
\and M.~Baldi\orcid{0000-0003-4145-1943}\inst{\ref{aff27},\ref{aff22},\ref{aff28}}
\and A.~Balestra\orcid{0000-0002-6967-261X}\inst{\ref{aff29}}
\and S.~Bardelli\orcid{0000-0002-8900-0298}\inst{\ref{aff22}}
\and P.~Battaglia\orcid{0000-0002-7337-5909}\inst{\ref{aff22}}
\and A.~Biviano\orcid{0000-0002-0857-0732}\inst{\ref{aff24},\ref{aff23}}
\and A.~Bonchi\orcid{0000-0002-2667-5482}\inst{\ref{aff30}}
\and E.~Branchini\orcid{0000-0002-0808-6908}\inst{\ref{aff31},\ref{aff32},\ref{aff21}}
\and M.~Brescia\orcid{0000-0001-9506-5680}\inst{\ref{aff33},\ref{aff16}}
\and J.~Brinchmann\orcid{0000-0003-4359-8797}\inst{\ref{aff34},\ref{aff35}}
\and S.~Camera\orcid{0000-0003-3399-3574}\inst{\ref{aff36},\ref{aff37},\ref{aff2}}
\and G.~Ca\~nas-Herrera\orcid{0000-0003-2796-2149}\inst{\ref{aff38},\ref{aff39},\ref{aff40}}
\and V.~Capobianco\orcid{0000-0002-3309-7692}\inst{\ref{aff2}}
\and C.~Carbone\orcid{0000-0003-0125-3563}\inst{\ref{aff41}}
\and J.~Carretero\orcid{0000-0002-3130-0204}\inst{\ref{aff42},\ref{aff43}}
\and S.~Casas\orcid{0000-0002-4751-5138}\inst{\ref{aff44}}
\and M.~Castellano\orcid{0000-0001-9875-8263}\inst{\ref{aff45}}
\and G.~Castignani\orcid{0000-0001-6831-0687}\inst{\ref{aff22}}
\and S.~Cavuoti\orcid{0000-0002-3787-4196}\inst{\ref{aff16},\ref{aff46}}
\and K.~C.~Chambers\orcid{0000-0001-6965-7789}\inst{\ref{aff47}}
\and A.~Cimatti\inst{\ref{aff48}}
\and C.~Colodro-Conde\inst{\ref{aff13}}
\and G.~Congedo\orcid{0000-0003-2508-0046}\inst{\ref{aff11}}
\and C.~J.~Conselice\orcid{0000-0003-1949-7638}\inst{\ref{aff49}}
\and L.~Conversi\orcid{0000-0002-6710-8476}\inst{\ref{aff50},\ref{aff19}}
\and Y.~Copin\orcid{0000-0002-5317-7518}\inst{\ref{aff51}}
\and A.~Costille\inst{\ref{aff4}}
\and F.~Courbin\orcid{0000-0003-0758-6510}\inst{\ref{aff52},\ref{aff53}}
\and H.~M.~Courtois\orcid{0000-0003-0509-1776}\inst{\ref{aff54}}
\and M.~Cropper\orcid{0000-0003-4571-9468}\inst{\ref{aff55}}
\and A.~Da~Silva\orcid{0000-0002-6385-1609}\inst{\ref{aff56},\ref{aff57}}
\and H.~Degaudenzi\orcid{0000-0002-5887-6799}\inst{\ref{aff58}}
\and G.~De~Lucia\orcid{0000-0002-6220-9104}\inst{\ref{aff24}}
\and H.~Dole\orcid{0000-0002-9767-3839}\inst{\ref{aff18}}
\and M.~Douspis\orcid{0000-0003-4203-3954}\inst{\ref{aff18}}
\and F.~Dubath\orcid{0000-0002-6533-2810}\inst{\ref{aff58}}
\and X.~Dupac\inst{\ref{aff19}}
\and S.~Dusini\orcid{0000-0002-1128-0664}\inst{\ref{aff59}}
\and S.~Escoffier\orcid{0000-0002-2847-7498}\inst{\ref{aff60}}
\and M.~Farina\orcid{0000-0002-3089-7846}\inst{\ref{aff61}}
\and F.~Faustini\orcid{0000-0001-6274-5145}\inst{\ref{aff45},\ref{aff30}}
\and S.~Ferriol\inst{\ref{aff51}}
\and S.~Fotopoulou\orcid{0000-0002-9686-254X}\inst{\ref{aff62}}
\and M.~Frailis\orcid{0000-0002-7400-2135}\inst{\ref{aff24}}
\and E.~Franceschi\orcid{0000-0002-0585-6591}\inst{\ref{aff22}}
\and S.~Galeotta\orcid{0000-0002-3748-5115}\inst{\ref{aff24}}
\and K.~George\orcid{0000-0002-1734-8455}\inst{\ref{aff63}}
\and W.~Gillard\orcid{0000-0003-4744-9748}\inst{\ref{aff60}}
\and B.~Gillis\orcid{0000-0002-4478-1270}\inst{\ref{aff11}}
\and C.~Giocoli\orcid{0000-0002-9590-7961}\inst{\ref{aff22},\ref{aff28}}
\and J.~Gracia-Carpio\inst{\ref{aff64}}
\and B.~R.~Granett\orcid{0000-0003-2694-9284}\inst{\ref{aff21}}
\and A.~Grazian\orcid{0000-0002-5688-0663}\inst{\ref{aff29}}
\and F.~Grupp\inst{\ref{aff64},\ref{aff63}}
\and S.~V.~H.~Haugan\orcid{0000-0001-9648-7260}\inst{\ref{aff65}}
\and J.~Hoar\inst{\ref{aff19}}
\and W.~Holmes\inst{\ref{aff66}}
\and I.~M.~Hook\orcid{0000-0002-2960-978X}\inst{\ref{aff67}}
\and F.~Hormuth\inst{\ref{aff68}}
\and A.~Hornstrup\orcid{0000-0002-3363-0936}\inst{\ref{aff69},\ref{aff70}}
\and K.~Jahnke\orcid{0000-0003-3804-2137}\inst{\ref{aff6}}
\and M.~Jhabvala\inst{\ref{aff71}}
\and E.~Keih\"anen\orcid{0000-0003-1804-7715}\inst{\ref{aff72}}
\and S.~Kermiche\orcid{0000-0002-0302-5735}\inst{\ref{aff60}}
\and A.~Kiessling\orcid{0000-0002-2590-1273}\inst{\ref{aff66}}
\and B.~Kubik\orcid{0009-0006-5823-4880}\inst{\ref{aff51}}
\and K.~Kuijken\orcid{0000-0002-3827-0175}\inst{\ref{aff40}}
\and M.~K\"ummel\orcid{0000-0003-2791-2117}\inst{\ref{aff63}}
\and M.~Kunz\orcid{0000-0002-3052-7394}\inst{\ref{aff73}}
\and H.~Kurki-Suonio\orcid{0000-0002-4618-3063}\inst{\ref{aff74},\ref{aff75}}
\and Q.~Le~Boulc'h\inst{\ref{aff76}}
\and A.~M.~C.~Le~Brun\orcid{0000-0002-0936-4594}\inst{\ref{aff77}}
\and D.~Le~Mignant\orcid{0000-0002-5339-5515}\inst{\ref{aff4}}
\and S.~Ligori\orcid{0000-0003-4172-4606}\inst{\ref{aff2}}
\and P.~B.~Lilje\orcid{0000-0003-4324-7794}\inst{\ref{aff65}}
\and V.~Lindholm\orcid{0000-0003-2317-5471}\inst{\ref{aff74},\ref{aff75}}
\and I.~Lloro\orcid{0000-0001-5966-1434}\inst{\ref{aff78}}
\and G.~Mainetti\orcid{0000-0003-2384-2377}\inst{\ref{aff76}}
\and D.~Maino\inst{\ref{aff79},\ref{aff41},\ref{aff80}}
\and E.~Maiorano\orcid{0000-0003-2593-4355}\inst{\ref{aff22}}
\and O.~Mansutti\orcid{0000-0001-5758-4658}\inst{\ref{aff24}}
\and S.~Marcin\inst{\ref{aff81}}
\and O.~Marggraf\orcid{0000-0001-7242-3852}\inst{\ref{aff82}}
\and M.~Martinelli\orcid{0000-0002-6943-7732}\inst{\ref{aff45},\ref{aff83}}
\and N.~Martinet\orcid{0000-0003-2786-7790}\inst{\ref{aff4}}
\and F.~Marulli\orcid{0000-0002-8850-0303}\inst{\ref{aff84},\ref{aff22},\ref{aff28}}
\and R.~Massey\orcid{0000-0002-6085-3780}\inst{\ref{aff85}}
\and E.~Medinaceli\orcid{0000-0002-4040-7783}\inst{\ref{aff22}}
\and S.~Mei\orcid{0000-0002-2849-559X}\inst{\ref{aff86},\ref{aff87}}
\and Y.~Mellier\thanks{Deceased}\inst{\ref{aff88},\ref{aff89}}
\and M.~Meneghetti\orcid{0000-0003-1225-7084}\inst{\ref{aff22},\ref{aff28}}
\and E.~Merlin\orcid{0000-0001-6870-8900}\inst{\ref{aff45}}
\and G.~Meylan\inst{\ref{aff90}}
\and A.~Mora\orcid{0000-0002-1922-8529}\inst{\ref{aff91}}
\and M.~Moresco\orcid{0000-0002-7616-7136}\inst{\ref{aff84},\ref{aff22}}
\and L.~Moscardini\orcid{0000-0002-3473-6716}\inst{\ref{aff84},\ref{aff22},\ref{aff28}}
\and R.~Nakajima\orcid{0009-0009-1213-7040}\inst{\ref{aff82}}
\and C.~Neissner\orcid{0000-0001-8524-4968}\inst{\ref{aff92},\ref{aff43}}
\and S.-M.~Niemi\orcid{0009-0005-0247-0086}\inst{\ref{aff38}}
\and C.~Padilla\orcid{0000-0001-7951-0166}\inst{\ref{aff92}}
\and S.~Paltani\orcid{0000-0002-8108-9179}\inst{\ref{aff58}}
\and F.~Pasian\orcid{0000-0002-4869-3227}\inst{\ref{aff24}}
\and K.~Pedersen\inst{\ref{aff93}}
\and W.~J.~Percival\orcid{0000-0002-0644-5727}\inst{\ref{aff94},\ref{aff95},\ref{aff96}}
\and V.~Pettorino\inst{\ref{aff38}}
\and S.~Pires\orcid{0000-0002-0249-2104}\inst{\ref{aff97}}
\and G.~Polenta\orcid{0000-0003-4067-9196}\inst{\ref{aff30}}
\and M.~Poncet\inst{\ref{aff98}}
\and L.~A.~Popa\inst{\ref{aff99}}
\and L.~Pozzetti\orcid{0000-0001-7085-0412}\inst{\ref{aff22}}
\and F.~Raison\orcid{0000-0002-7819-6918}\inst{\ref{aff64}}
\and A.~Renzi\orcid{0000-0001-9856-1970}\inst{\ref{aff100},\ref{aff59}}
\and J.~Rhodes\orcid{0000-0002-4485-8549}\inst{\ref{aff66}}
\and G.~Riccio\inst{\ref{aff16}}
\and E.~Romelli\orcid{0000-0003-3069-9222}\inst{\ref{aff24}}
\and M.~Roncarelli\orcid{0000-0001-9587-7822}\inst{\ref{aff22}}
\and R.~Saglia\orcid{0000-0003-0378-7032}\inst{\ref{aff63},\ref{aff64}}
\and Z.~Sakr\orcid{0000-0002-4823-3757}\inst{\ref{aff101},\ref{aff102},\ref{aff103}}
\and D.~Sapone\orcid{0000-0001-7089-4503}\inst{\ref{aff104}}
\and B.~Sartoris\orcid{0000-0003-1337-5269}\inst{\ref{aff63},\ref{aff24}}
\and J.~A.~Schewtschenko\orcid{0000-0002-4913-6393}\inst{\ref{aff11}}
\and M.~Schirmer\orcid{0000-0003-2568-9994}\inst{\ref{aff6}}
\and P.~Schneider\orcid{0000-0001-8561-2679}\inst{\ref{aff82}}
\and T.~Schrabback\orcid{0000-0002-6987-7834}\inst{\ref{aff105}}
\and A.~Secroun\orcid{0000-0003-0505-3710}\inst{\ref{aff60}}
\and G.~Seidel\orcid{0000-0003-2907-353X}\inst{\ref{aff6}}
\and S.~Serrano\orcid{0000-0002-0211-2861}\inst{\ref{aff106},\ref{aff107},\ref{aff108}}
\and P.~Simon\inst{\ref{aff82}}
\and C.~Sirignano\orcid{0000-0002-0995-7146}\inst{\ref{aff100},\ref{aff59}}
\and G.~Sirri\orcid{0000-0003-2626-2853}\inst{\ref{aff28}}
\and L.~Stanco\orcid{0000-0002-9706-5104}\inst{\ref{aff59}}
\and J.~Steinwagner\orcid{0000-0001-7443-1047}\inst{\ref{aff64}}
\and C.~Surace\orcid{0000-0003-2592-0113}\inst{\ref{aff4}}
\and P.~Tallada-Cresp\'{i}\orcid{0000-0002-1336-8328}\inst{\ref{aff42},\ref{aff43}}
\and A.~N.~Taylor\inst{\ref{aff11}}
\and I.~Tereno\orcid{0000-0002-4537-6218}\inst{\ref{aff56},\ref{aff109}}
\and R.~Toledo-Moreo\orcid{0000-0002-2997-4859}\inst{\ref{aff110}}
\and F.~Torradeflot\orcid{0000-0003-1160-1517}\inst{\ref{aff43},\ref{aff42}}
\and A.~Tsyganov\inst{\ref{aff111}}
\and I.~Tutusaus\orcid{0000-0002-3199-0399}\inst{\ref{aff102}}
\and L.~Valenziano\orcid{0000-0002-1170-0104}\inst{\ref{aff22},\ref{aff112}}
\and J.~Valiviita\orcid{0000-0001-6225-3693}\inst{\ref{aff74},\ref{aff75}}
\and T.~Vassallo\orcid{0000-0001-6512-6358}\inst{\ref{aff63},\ref{aff24}}
\and G.~Verdoes~Kleijn\orcid{0000-0001-5803-2580}\inst{\ref{aff113}}
\and A.~Veropalumbo\orcid{0000-0003-2387-1194}\inst{\ref{aff21},\ref{aff32},\ref{aff31}}
\and D.~Vibert\orcid{0009-0008-0607-631X}\inst{\ref{aff4}}
\and Y.~Wang\orcid{0000-0002-4749-2984}\inst{\ref{aff114}}
\and J.~Weller\orcid{0000-0002-8282-2010}\inst{\ref{aff63},\ref{aff64}}
\and A.~Zacchei\orcid{0000-0003-0396-1192}\inst{\ref{aff24},\ref{aff23}}
\and G.~Zamorani\orcid{0000-0002-2318-301X}\inst{\ref{aff22}}
\and F.~M.~Zerbi\inst{\ref{aff21}}
\and E.~Zucca\orcid{0000-0002-5845-8132}\inst{\ref{aff22}}
\and J.~Mart\'{i}n-Fleitas\orcid{0000-0002-8594-569X}\inst{\ref{aff91}}
\and V.~Scottez\inst{\ref{aff88},\ref{aff115}}}
										   
%%%% please do not edit the affiliation list -- contact ECEB Bureau for changes
\institute{Universit\'e Marie et Louis Pasteur, CNRS, Observatoire des Sciences de l’Univers THETA Franche-Comt\'e Bourgogne, Institut UTINAM, Observatoire de Besan\c{c}on, BP~1615, 25010 Besan\c{c}on Cedex, France\label{aff1}
\and
INAF-Osservatorio Astrofisico di Torino, Via Osservatorio 20, 10025 Pino Torinese, Italy\label{aff2}
\and
Department of Physics, Astronomy and Mathematics, University of Hertfordshire, College Lane, Hatfield AL10 9AB, United Kingdom\label{aff3}
\and
Aix-Marseille Universit\'e, CNRS, CNES, LAM, Marseille, France\label{aff4}
\and
Departamento F\'isica Aplicada, Universidad Polit\'ecnica de Cartagena, Campus Muralla del Mar, 30202 Cartagena, Murcia, Spain\label{aff5}
\and
Max-Planck-Institut f\"ur Astronomie, K\"onigstuhl 17, 69117 Heidelberg, Germany\label{aff6}
\and
International Space University, 1 rue Jean-Dominique Cassini, 67400 Illkirch-Graffenstaden, France\label{aff7}
\and
Universit\'e de Strasbourg, CNRS, Observatoire astronomique de Strasbourg, UMR 7550, 67000 Strasbourg, France\label{aff8}
\and
School of Physics and Astronomy, University of Leicester, University Road, Leicester, LE1 7RH, United Kingdom\label{aff9}
\and
Centro de Astrobiolog\'ia (CAB), CSIC-INTA, ESAC Campus, Camino Bajo del Castillo s/n, 28692 Villanueva de la Ca\~nada, Madrid, Spain\label{aff10}
\and
Institute for Astronomy, University of Edinburgh, Royal Observatory, Blackford Hill, Edinburgh EH9 3HJ, United Kingdom\label{aff11}
\and
European Southern Observatory, Karl-Schwarzschild-Str.~2, 85748 Garching, Germany\label{aff12}
\and
Instituto de Astrof\'{\i}sica de Canarias, 38205 La Laguna, Tenerife, Spain\label{aff13}
\and
Universidad de La Laguna, Dpto. Astrof\'\i sica, 38206 La Laguna, Tenerife, Spain\label{aff14}
\and
 Instituto de Astrof\'{\i}sica de Canarias, 38205 La Laguna; Universidad de La Laguna, Dpto. Astrof\'\i sica, 38206 La Laguna, Tenerife, Spain\label{aff15}
\and
INAF-Osservatorio Astronomico di Capodimonte, Via Moiariello 16, 80131 Napoli, Italy\label{aff16}
\and
Consejo Superior de Investigaciones Cientificas, Calle Serrano 117, 28006 Madrid, Spain\label{aff17}
\and
Universit\'e Paris-Saclay, CNRS, Institut d'astrophysique spatiale, 91405, Orsay, France\label{aff18}
\and
European Space Agency/ESAC, Camino Bajo del Castillo, s/n., Urb. Villafranca del Castillo, 28692 Villanueva de la Ca\~nada, Madrid, Spain\label{aff19}
\and
School of Mathematics and Physics, University of Surrey, Guildford, Surrey, GU2 7XH, United Kingdom\label{aff20}
\and
INAF-Osservatorio Astronomico di Brera, Via Brera 28, 20122 Milano, Italy\label{aff21}
\and
INAF-Osservatorio di Astrofisica e Scienza dello Spazio di Bologna, Via Piero Gobetti 93/3, 40129 Bologna, Italy\label{aff22}
\and
Institute for Fundamental Physics of the Universe (IFPU), via Beirut 2, 34151 Trieste, Italy\label{aff23}
\and
INAF-Osservatorio Astronomico di Trieste, Via G. B. Tiepolo 11, 34143 Trieste, Italy\label{aff24}
\and
INFN, Sezione di Trieste, Via Valerio 2, 34127 Trieste, Italy\label{aff25}
\and
SISSA, International School for Advanced Studies, Via Bonomea 265, 34136 Trieste, Italy\label{aff26}
\and
Dipartimento di Fisica e Astronomia, Universit\`a di Bologna, Via Gobetti 93/2, 40129 Bologna, Italy\label{aff27}
\and
INFN-Sezione di Bologna, Viale Berti Pichat 6/2, 40127 Bologna, Italy\label{aff28}
\and
INAF-Osservatorio Astronomico di Padova, Via dell'Osservatorio 5, 35122 Padova, Italy\label{aff29}
\and
Space Science Data Center, Italian Space Agency, via del Politecnico snc, 00133 Roma, Italy\label{aff30}
\and
Dipartimento di Fisica, Universit\`a di Genova, Via Dodecaneso 33, 16146 Genova, Italy\label{aff31}
\and
INFN-Sezione di Genova, Via Dodecaneso 33, 16146 Genova, Italy\label{aff32}
\and
Department of Physics ``E. Pancini'', University Federico II, Via Cinthia 6, 80126 Napoli, Italy\label{aff33}
\and
Instituto de Astrof\'isica e Ci\^encias do Espa\c{c}o, Universidade do Porto, CAUP, Rua das Estrelas, PT4150-762 Porto, Portugal\label{aff34}
\and
Faculdade de Ci\^encias da Universidade do Porto, Rua do Campo de Alegre, 4150-007 Porto, Portugal\label{aff35}
\and
Dipartimento di Fisica, Universit\`a degli Studi di Torino, Via P. Giuria 1, 10125 Torino, Italy\label{aff36}
\and
INFN-Sezione di Torino, Via P. Giuria 1, 10125 Torino, Italy\label{aff37}
\and
European Space Agency/ESTEC, Keplerlaan 1, 2201 AZ Noordwijk, The Netherlands\label{aff38}
\and
Institute Lorentz, Leiden University, Niels Bohrweg 2, 2333 CA Leiden, The Netherlands\label{aff39}
\and
Leiden Observatory, Leiden University, Einsteinweg 55, 2333 CC Leiden, The Netherlands\label{aff40}
\and
INAF-IASF Milano, Via Alfonso Corti 12, 20133 Milano, Italy\label{aff41}
\and
Centro de Investigaciones Energ\'eticas, Medioambientales y Tecnol\'ogicas (CIEMAT), Avenida Complutense 40, 28040 Madrid, Spain\label{aff42}
\and
Port d'Informaci\'{o} Cient\'{i}fica, Campus UAB, c. Albareda s/n, 08193 Bellaterra (Barcelona), Spain\label{aff43}
\and
Institut für Theoretische Teilchenphysik und Kosmologie (TTK), RWTH Aachen University, 52056 Aachen, Germany\label{aff44}
\and
INAF-Osservatorio Astronomico di Roma, Via Frascati 33, 00078 Monteporzio Catone, Italy\label{aff45}
\and
INFN section of Naples, Via Cinthia 6, 80126 Napoli, Italy\label{aff46}
\and
Institute for Astronomy, University of Hawaii, 2680 Woodlawn Drive, Honolulu, HI 96822, USA\label{aff47}
\and
Dipartimento di Fisica e Astronomia ``Augusto Righi'', Alma Mater Studiorum Universit\`a di Bologna, Viale Berti Pichat 6/2, 40127 Bologna, Italy\label{aff48}
\and
Jodrell Bank Centre for Astrophysics, Department of Physics and Astronomy, University of Manchester, Oxford Road, Manchester M13 9PL, United Kingdom\label{aff49}
\and
European Space Agency/ESRIN, Largo Galileo Galilei 1, 00044 Frascati, Roma, Italy\label{aff50}
\and
Universit\'e Claude Bernard Lyon 1, CNRS/IN2P3, IP2I Lyon, UMR 5822, Villeurbanne, 69100, France\label{aff51}
\and
Institut de Ci\`{e}ncies del Cosmos (ICCUB), Universitat de Barcelona (IEEC-UB), Mart\'{i} i Franqu\`{e}s 1, 08028 Barcelona, Spain\label{aff52}
\and
Instituci\'o Catalana de Recerca i Estudis Avan\c{c}ats (ICREA), Passeig de Llu\'{\i}s Companys 23, 08010 Barcelona, Spain\label{aff53}
\and
Université Claude Bernard Lyon 1 (UCBL), CNRS/IN2P3, IUF, IP2I Lyon, 4 rue Enrico Fermi, 69622 Villeurbanne, France\label{aff54}
\and
Mullard Space Science Laboratory, University College London, Holmbury St Mary, Dorking, Surrey RH5 6NT, United Kingdom\label{aff55}
\and
Departamento de F\'isica, Faculdade de Ci\^encias, Universidade de Lisboa, Edif\'icio C8, Campo Grande, PT1749-016 Lisboa, Portugal\label{aff56}
\and
Instituto de Astrof\'isica e Ci\^encias do Espa\c{c}o, Faculdade de Ci\^encias, Universidade de Lisboa, Campo Grande, 1749-016 Lisboa, Portugal\label{aff57}
\and
Département d’astronomie, Université de Genève,, ch. d'Ecogia 16, 1290 Versoix, Switzerland\label{aff58}
\and
INFN-Padova, Via Marzolo 8, 35131 Padova, Italy\label{aff59}
\and
Aix-Marseille Universit\'e, CNRS/IN2P3, CPPM, Marseille, France\label{aff60}
\and
INAF-Istituto di Astrofisica e Planetologia Spaziali, via del Fosso del Cavaliere, 100, 00100 Roma, Italy\label{aff61}
\and
School of Physics, HH Wills Physics Laboratory, University of Bristol, Tyndall Avenue, Bristol, BS8 1TL, United Kingdom\label{aff62}
\and
Universit\"ats-Sternwarte M\"unchen, Fakult\"at f\"ur Physik, Ludwig-Maximilians-Universit\"at M\"unchen, Scheinerstr.~1, 81679 M\"unchen, Germany\label{aff63}
\and
Max-Planck-Institut für extraterrestrische Physik, Giessenbachstr. 1, 85748 Garching, Germany\label{aff64}
\and
Institutt for teoretisk astrofysikk, University of Oslo, P.O. Box 1029 Blindern, 0315 Oslo, Norway\label{aff65}
\and
Jet Propulsion Laboratory, California Institute of Technology, 4800 Oak Grove Drive, Pasadena, CA, 91109, USA\label{aff66}
\and
Department of Physics, Lancaster University, Lancaster, LA1 4YB, United Kingdom\label{aff67}
\and
Felix Hormuth Engineering, Goethestraße 17, 69181 Leimen, Germany\label{aff68}
\and
Danmarks Tekniske Universitet, Elektrovej 327, 2800 Kgs. Lyngby, Denmark\label{aff69}
\and
Cosmic Dawn Center (DAWN), Denmark\label{aff70}
\and
NASA Goddard Space Flight Center, Greenbelt, MD 20771, USA\label{aff71}
\and
Department of Physics and Helsinki Institute of Physics, Gustaf H\"allstr\"omin katu 2, University of Helsinki, 00014 Helsinki, Finland\label{aff72}
\and
Universit\'e de Gen\`eve, D\'epartement de Physique Th\'eorique and Centre for Astroparticle Physics, 24 quai Ernest-Ansermet, CH-1211 Gen\`eve 4, Switzerland\label{aff73}
\and
Department of Physics, P.O. Box 64, University of Helsinki, 00014 Helsinki, Finland\label{aff74}
\and
Helsinki Institute of Physics, Gustaf H{\"a}llstr{\"o}min katu 2, University of Helsinki, 00014 Helsinki, Finland\label{aff75}
\and
Centre de Calcul de l'IN2P3/CNRS, 21 avenue Pierre de Coubertin 69627 Villeurbanne Cedex, France\label{aff76}
\and
Laboratoire d'etude de l'Univers et des phenomenes eXtremes, Observatoire de Paris, Universit\'e PSL, Sorbonne Universit\'e, CNRS, 92190 Meudon, France\label{aff77}
\and
SKAO, Jodrell Bank, Lower Withington, Macclesfield SK11 9FT, United Kingdom\label{aff78}
\and
Dipartimento di Fisica ``Aldo Pontremoli'', Universit\`a degli Studi di Milano, Via Celoria 16, 20133 Milano, Italy\label{aff79}
\and
INFN-Sezione di Milano, Via Celoria 16, 20133 Milano, Italy\label{aff80}
\and
Fachhochschule Nordwestschweiz, Hochschule f\"ur Informatik, 5210 Windisch, Switzerland\label{aff81}
\and
Universit\"at Bonn, Argelander-Institut f\"ur Astronomie, Auf dem H\"ugel 71, 53121 Bonn, Germany\label{aff82}
\and
INFN-Sezione di Roma, Piazzale Aldo Moro, 2 - c/o Dipartimento di Fisica, Edificio G. Marconi, 00185 Roma, Italy\label{aff83}
\and
Dipartimento di Fisica e Astronomia ``Augusto Righi'', Alma Mater Studiorum Universit\`a di Bologna, via Piero Gobetti 93/2, 40129 Bologna, Italy\label{aff84}
\and
Department of Physics, Institute for Computational Cosmology, Durham University, South Road, Durham, DH1 3LE, United Kingdom\label{aff85}
\and
Universit\'e Paris Cit\'e, CNRS, Astroparticule et Cosmologie, 75013 Paris, France\label{aff86}
\and
CNRS-UCB International Research Laboratory, Centre Pierre Bin\'etruy, IRL2007, CPB-IN2P3, Berkeley, USA\label{aff87}
\and
Institut d'Astrophysique de Paris, 98bis Boulevard Arago, 75014, Paris, France\label{aff88}
\and
Institut d'Astrophysique de Paris, UMR 7095, CNRS, and Sorbonne Universit\'e, 98 bis boulevard Arago, 75014 Paris, France\label{aff89}
\and
Institut de Physique, Laboratoire d’Astrophysique, Ecole Polytechnique F\'ed\'erale de Lausanne (EPFL), Observatoire de Sauverny, 1290 Versoix, Switzerland\label{aff90}
\and
Aurora Technology for European Space Agency (ESA), Camino bajo del Castillo, s/n, Urbanización Villafranca del Castillo, Villanueva de la Ca\~nada, 28692 Madrid, Spain\label{aff91}
\and
Institut de F\'{i}sica d'Altes Energies (IFAE), Barcelona Institute of Science and Technology, Campus UAB, 08193 Bellaterra, Barcelona, Spain\label{aff92}
\and
DARK, Niels Bohr Institutet, K\o benhavns Universitet, Jagtvej 155, 2200 Copenhagen, Denmark\label{aff93}
\and
Waterloo Centre for Astrophysics, University of Waterloo, Waterloo, Ontario N2L 3G1, Canada\label{aff94}
\and
Department of Physics and Astronomy, University of Waterloo, Waterloo, Ontario N2L 3G1, Canada\label{aff95}
\and
Perimeter Institute for Theoretical Physics, Waterloo, Ontario N2L 2Y5, Canada\label{aff96}
\and
Universit\'e Paris-Saclay, Universit\'e Paris Cit\'e, CEA, CNRS, AIM, 91191 Gif-sur-Yvette, France\label{aff97}
\and
Centre National d'Etudes Spatiales, Centre spatial de Toulouse, 18 avenue Edouard Belin, 31401 Toulouse Cedex 9, France\label{aff98}
\and
Institutul de \c{S}tiin\c{t}e Spa\c{t}iale, Str. Atomi\c{s}tilor nr.~409, M\u{a}gurele, Ilfov, 077125, Romania\label{aff99}
\and
Dipartimento di Fisica e Astronomia ``G. Galilei'', Universit\`a di Padova, Via Marzolo 8, 35131 Padova, Italy\label{aff100}
\and
Institut f\"ur Theoretische Physik, Universität Heidelberg, Philosophenweg 16, 69120 Heidelberg, Germany\label{aff101}
\and
Institut de Recherche en Astrophysique et Plan\'etologie (IRAP), Universit\'e de Toulouse, CNRS, UPS, CNES, 14 Av. Edouard Belin, 31400 Toulouse, France\label{aff102}
\and
Universit\'e Saint-Joseph, Facult\'e des Sciences, Beirut, Lebanon\label{aff103}
\and
Departamento de F\'isica, Facultad de Ciencias F\'{\i}sicas y Matem\'aticas (FCFM), Universidad de Chile, Blanco Encalada 2008, Santiago, Chile\label{aff104}
\and
Universit\"at Innsbruck, Institut f\"ur Astro- und Teilchenphysik, Technikerstr. 25/8, 6020 Innsbruck, Austria\label{aff105}
\and
Institut d'Estudis Espacials de Catalunya (IEEC),  Edifici RDIT, Campus UPC, 08860 Castelldefels, Barcelona, Spain\label{aff106}
\and
Satlantis, University Science Park, Sede Bld, 48940 Leioa, Bilbao, Spain\label{aff107}
\and
Institute of Space Sciences (ICE, CSIC), Campus UAB, Carrer de Can Magrans s/n, 08193 Barcelona, Spain\label{aff108}
\and
Instituto de Astrof\'isica e Ci\^encias do Espa\c{c}o, Faculdade de Ci\^encias, Universidade de Lisboa, Tapada da Ajuda, 1349-018 Lisboa, Portugal\label{aff109}
\and
Universidad Polit\'ecnica de Cartagena, Departamento de Electr\'onica y Tecnolog\'ia de Computadoras, Plaza del Hospital 1, 30202 Cartagena, Spain\label{aff110}
\and
Centre for Information Technology, University of Groningen, P.O. Box 11044, 9700 CA Groningen, The Netherlands\label{aff111}
\and
INFN-Bologna, Via Irnerio 46, 40126 Bologna, Italy\label{aff112}
\and
Kapteyn Astronomical Institute, University of Groningen, PO Box 800, 9700 AV Groningen, The Netherlands\label{aff113}
\and
Infrared Processing and Analysis Center, California Institute of Technology, Pasadena, CA 91125, USA\label{aff114}
\and
Institut Catholique de Lille (ICL), Junia, Universit\'e Catholique de Lille, LITL, 59000 Lille, France\label{aff115}}    
%=============================================
%=============================================

%=============================================
%=============================================
% ABSTRACT
\abstract
{ Ultracool dwarfs (UCDs) encompass the lowest-mass stars and brown dwarfs at the stellar--substellar boundary. These intrinsically faint objects are challenging to detect. However, the wide coverage and deep sensitivity of the \Euclid survey will increase the number of confirmed and well-characterised UCDs by several orders of magnitude.
We carried out a systematic study using \Euclid{} Quick Data Release (Q1) low-resolution near-infrared (NIR) spectra to identify and characterise UCDs. A sample of previously known UCDs in Q1 was compiled to validate the analysis, which was then applied to newly identified UCD candidates in Euclid Deep Field North (\EDFN; $22.9\deg^2$, down to $\JE{} \approx 24.5$\,mag). 
We confirm \NNEWSPECTRA{} new UCDs in \EDFN, spanning spectral types (SpTs) M7--T1 ($\JE=17$--21\,mag) and distances of 50--600\,pc. Six of them were also independently identified in other Q1 studies. We followed up on two of these objects with ground-based optical and NIR observations using the Gran Telescopio Canarias, providing an independent validation of the \Euclid spectral classification and the ability of Q1 spectra to constrain the stellar parameters of UCDs.  
We determined the SpTs from \Euclid\ Q1 spectra of \NNEWSPECTRA{} newly identified UCDs and \NKNWSPECTRA{} previously known UCDs by fitting spectral standards from the SpeX Prism Library Analysis Toolkit. The known UCD sample includes 103 objects in common with the Dark Energy Survey (M7--T0, $\JE = 17$--22\,mag, and $d = 50$--600\,pc), 6 objects from literature compilations of UCD sources (M8--T7, $\JE = 18$--20\,mag, and $d = 30$--300\,pc), and 58 in common with \textit{Gaia} (M7--L1, $\JE = 17$--24\,mag, and $d = 70$--300\,pc).
We derived independent stellar parameter estimates through atmospheric model fitting; the resulting effective temperatures are in good agreement with \Euclid SpTs. Combined with complementary ground- and space-based data, these results demonstrate the potential for determining proper motions and identifying companions. Additionally, we investigated \textit{Gaia} UCDs in Q1 by combining their mean XP spectra with \Euclid NIR spectra, extending the wavelength coverage and enabling more comprehensive determinations of stellar parameters, including radii.
To support broader classification efforts, we analysed the UCDs in colour–colour space and assessed their separation from quasar loci. 
We present a publicly available, machine-readable catalogue that includes \Euclid observations of known and newly identified sources, with spectral classifications and stellar parameters.
This work demonstrates \Euclid{}’s capability to identify and characterise the faintest stellar populations, establishing a foundation for large-scale UCD studies with forthcoming survey data.
}
%=============================================

%=============================================
% Provide up to five keywords:
\keywords{stars: brown dwarfs, stars: low-mass $-$ methods: observational, data analysis $-$ techniques: photometric, spectroscopic}
%    from the list in
%     https://www.aanda.org/for-authors/latex-issues/information-files#pop}
% Add short versions of title and author list for page headings
%
   \titlerunning{New ultracool dwarfs in the Euclid Deep Fields}
   \authorrunning{Mohandasan et al.}
   \maketitle
%=============================================

%=============================================
%=============================================
%=============================================
% MAIN TEXT
\section{\label{sc:Intro}Introduction}

Ultracool dwarfs (UCDs) are the lowest-mass, coldest, and faintest products of star formation \citep{1963ApJ...137.1121K, 1963PThPh..30..460H}. \citet{1997AJ....113.1421K} defined them as objects with spectral types (SpTs) M7 and later, effective temperatures ($T_{\mathrm{eff}}$) below $\sim$2800\,K \citep[Fig.~4 in][]{2024A&A...685A...6R}, and masses $M \lesssim 0.1\,M_\odot$ \citep{2017ApJS..231...15D}. They are valuable probes in stellar astrophysics since they encompass both very low-mass stars that slowly fuse hydrogen \citep{chabrier1997, 2004ApJS..155..191B} and brown dwarfs (BDs), which have insufficient mass to sustain hydrogen fusion \citep[below $\sim$0.075\,$M_\odot$ at solar metallicity, e.g.,][]{2001RvMP...73..719B, baraffe2003}.
With lifetimes exceeding the age of the Universe \citep{1997ApJ...482..420L}, they are present in all Galactic components, from young stellar associations to the halo, making them excellent tracers for Milky Way studies \citep{1997AJ....113.2246R, 2021A&A...650A.201R, 2024ApJS..271...55K}. 
Furthermore, they provide crucial insights into stellar--substellar formation, such as the low-mass cutoff of the initial mass function and the mechanisms responsible for BD formation, a process that is poorly understood and continues to be an active research focus \citep{whiteworth2006, luhman2012, 2019A&A...624A..94M, vowell2025}.

\vspace{0.05cm}

The unique nature of UCDs arises from distinct processes in their interior and surface layers, which set them apart from typical stars \citep{2019ApJ...879...94F, 2023A&A...671A.119C}. Their cool but complex atmospheres, with dust, condensation, and cloud coverage, are challenging to model \citep{cushing2008, 2012ApJ...756..172M, 2021MNRAS.506.1944B}. Strong molecular absorption, mainly due to \ce{H2O}, as well as \ce{CO}, \ce{NH3}, and \ce{CH4}, causes UCD spectra to deviate significantly from blackbody radiation, which produces a pseudo-continuum \citep{1999ApJ...519..802K, 2019AJ....157..101M}. Atmospheric properties depend on factors such as $T_{\mathrm{eff}}$, metallicity ([Fe/H]), surface gravity ($\log{g}$), cloud characteristics, and vertical mixing \citep{2015A&A...577A..42B, 2020A&A...637A..38P}, which are essential for spectral classification and modelling energy distributions.

Ultracool dwarfs are classified into SpTs \textup{M}, \textup{L}, \textup{T}, and \textup{Y}, based on their spectral features. M dwarfs exhibit \ce{TiO} and \ce{VO} bands, which weaken in L dwarfs, where \ce{H2O} and \ce{CO} absorption dominate. T dwarfs show characteristic \ce{CH4} absorption in the near-infrared (NIR), which strengthens in Y dwarfs and \ce{NH3} absorption emerges \citep{2006ApJ...637.1067B,2011ApJ...743...50C}. Objects of SpTs \textup{M7} to $\sim$ \textup{L4} include both stars and young BDs, while objects beyond $\sim$ \textup{L4} are more likely to be BDs \citep{2017ApJS..231...15D}. As UCDs cool over time, they transition through later SpTs, resulting in objects with the same SpT having different masses and $T_{\mathrm{eff}}$ \citep{1997ApJ...491..856B, 2004ApJS..155..191B}.

Ultracool dwarfs, in particular the coolest M dwarfs, are also emerging as key targets for exoplanet detection, through both radial velocity (RV) and transit surveys \citep{bonfils2013, 2016Natur.533..221G, 2017Natur.542..456G, 2018SPIE10700E..1ID}, since their small radii enhance the detectability of orbiting planets. However, the significant stellar activity of M dwarfs remains a major limitation, inhibiting confirmation of exoplanet detection around these systems.
The atmospheres of low-mass BDs, when they are bright enough for spectroscopic observation, are much easier to study than those of exoplanets, whose atmospheric signals are often overwhelmed by the light of their host stars. Several systems with substellar companions have been identified using direct imaging \citep{2005A&A...438L..25C, 2005A&A...438L..29C, marois2008, legrange2010}.

% =====================================
\begin{table*}[htbp!]{
\label{tab:masterucds}
\caption{Selected UCDs with photometric and spectroscopic SpTs from the literature. }
\begin{center}
\begin{tabular}{lllccccc}
\toprule
\toprule
Short name  &  \Euclid name                & Simbad name            &  \JE   & $J$  & Ref. $J$ & SpT & Ref. SpT\\
  &   &    &  [mag]   & [mag]  & &  &  \\
\midrule
J0332$-$2733 &  J033234.44$-$273333.9 & WISEA J033234.35$-$273333.8 & 18.301 & 18.17 & {P1} & M9 & {S1} \\           
J0352$-$4910 &  J035231.98$-$491058.8 & WISEA J035231.80$-$491059.4 & 19.520 & 18.77 & P1  & T7 & S1\\           
J0359$-$4740 &  J035909.93$-$474057.4 & WISEA J035909.75$-$474056.8 & 19.763 & 19.02 & P1 & T7.5 & S1\\         
J0402$-$4704 &  J040254.89$-$470440.2 & WISEA J040254.85$-$470440.9 & 20.050 & 19.88 & P1 & L4.5 & S1\\         
J0403$-$4916 &  J040351.20$-$491603.5 & CWISE J040351.12$-$491605.4 & 21.907 & 21.12 & P2 & (T7) & S2 \\         
J0413$-$4750 &  J041358.31$-$475035.0 & WISE J041358.14$-$475039.3  & 21.447 & 20.55 & P2  & T9 & S3\\           
J0416$-$4721 &  J041657.68$-$472115.9 & WISE J041657.60$-$472115.0  & 19.964 & 18.42 & P1 & (T2$-$T5) & S4 \\      
J1745$+$6459 &  J174556.40$+$645937.1 & WISE J174556.65$+$645933.8  & 19.967 & 19.93 & P3 & T7 & S3 \\           
J1754$+$6712 &  J175410.42$+$671212.4 & WISEA J175410.34$+$671212.0 & 17.487 & 17.04 & P4  & L1 & S1 \\           
J1757$+$6741 &  J175730.71$+$674136.5 & 2MASS J17573073$+$6741401   & $...^{a}$ & 14.61 & P4  & M9 & S1 \\           
J1811$+$6658 &  J181125.15$+$665801.8 & CWISE J181125.34$+$665806.4 & 22.073 & 22.53 & P5  & (T9) & S5 \\         
J1819$+$6600 &  J181949.30$+$660059.1 & 2MASS J18194942$+$6600548   & $...^{a}$ & 15.43 & P4 & (L2) & S6 \\         
\bottomrule
\end{tabular}
\end{center}
Photometric references:
P1\,\cite{2021A&A...651A..69K}; P2\,
\cite{2021ApJ...918...11L}; 
P3\,estimate based on \cite{2013APJS..205....6M}, P4\,\cite{2003yCat.2246....0C}, P5\,\cite{2024ApJS..271...55K}. 
 SpT references: 
S1\,\cite{2024A&A...686A.171Z}; S2\,\cite{2024AJ....167..253R}; S3\,\cite{2013APJS..205....6M}; S4\,\cite{2014MNRAS.437.1009P}; S5\,\cite{2024ApJS..271...55K}; S6\,\cite{2021A&A...645A.100S}.
$^{a}$ Objects with saturation, and their \Euclid magnitudes are thus not included.}
\end{table*}
% ===================================

Ultracool dwarfs constitute a significant fraction of the Milky Way's stellar population, with estimates suggesting that they represent at least 15\% of star-like objects in the solar neighbourhood \citep{2017AJ....154..151B}. However, detection difficulties lead to statistical incompleteness; \cite{2019ApJ...883..205B} estimated, using a spectroscopically verified volume-limited sample within 25\,pc, that 69--80\% of \textup{M7}--\textup{L5} objects have been detected. This is consistent with later assessments based on \textit{Gaia} and combined photometric–astrometric samples \citep[][]{2021A&A...650A.201R, golovin2023, 2024ApJS..271...55K}, which generally find completeness in the range 60--90\% depending on the SpTs and selection method. Their faint luminosities and primary emission in the NIR make UCDs challenging to observe, with most being too faint for \textit{Gaia} and ground-based telescopes \citep{2013APJS..205....6M, 2017MNRAS.469..401S, 2019MNRAS.485.4423S}.

\Euclid's deep and wide surveys \citep{EuclidSkyOverview,EuclidSkyVIS,EuclidSkyNISP} are transforming UCD studies, enabling the discovery of unprecedented numbers of objects. Simulations predict \Euclid will detect nearly 1 million L dwarfs, 500\,000 T dwarfs, and a few Y dwarfs \citep{2021MNRAS.501..281S}. The resulting homogeneous sample will provide valuable insights into UCD demographics, enabling the identification of rare, faint, and nearby objects.

In searching for UCDs, one of the primary methods involves photometric searches across optical \citep[e.g.,][]{2000ApJ...536L..35L, 2006AJ....131.2722C}, NIR \citep[e.g.,][]{1999ApJ...519..802K, 2008MNRAS.390..304P},  and mid- to far-infrared wavelengths \citep{2011ApJ...743...50C, 2011ApJS..197...19K}. The most numerous contaminants in these searches are distant galaxies and quasars (QSOs), while UCDs emerge as a major contaminant in QSO searches \citep[e.g.,][]{2002ApJ...568L.107L, caballero2008, 2009A&A...505...97M, 2016ApJS..227...11B, 2023ApJS..265...29B, 2021MNRAS.508..737T, belladitta2025}. The effective separation of these objects often requires additional spectroscopic observations, which \Euclid can provide.

The first \Euclid Quick Data Release \citep[Q1;][]{Q1-TP001} has motivated several UCD studies. \citet[][]{zerjal2026} focused on discovering UCDs using photometric selection in Q1, while \cite{dominguez2025} used spectroscopic analysis of \Euclid spectra to classify UCDs and searched for new ones using spectral indices. 
\cite{zhang2025} searched for new UCDs and \cite{frank2026} analysed T-type dwarfs using Q1 data.
In this study, we established methods for characterising UCDs based on \Euclid data and applied them to a newly discovered sample of 202 UCD candidates in Euclid Deep Field North \citep[\EDFN;][]{Scaramella-EP1}.

The paper is structured as follows. Section~\ref{sec:sources} presents the UCD sample, including a comparison sample of spectroscopically and photometrically identified known UCDs in Q1, and the selection of newly identified UCD candidates in the \EDFN. In Sect.~\ref{sec:spt}, we determine the \Euclid SpTs of the UCDs by performing spectral standard fitting on their Q1 spectra, and discuss some of the difficulties in using the Q1 spectra with illustrative examples. Section~\ref{sec:combinedtable} examines \Euclid colours, proper motions (PMs), and distances of the UCDs, as well as the derivation and analysis of stellar parameters from Q1 spectra. Section~\ref{sec:gaia_euclid_spec_comb} describes the combined \Euclid--\textit{Gaia} spectra with a larger wavelength range and the comprehensive analysis we carried out to obtain various stellar parameters, including stellar radii, from them. Section~\ref{sec:GTC_spectra} presents the data reduction and analysis of the two newly discovered UCDs in \EDFN, for which ground-based follow-up observations were obtained using the Gran Telescopio Canarias (GTC). We summarise and present our conclusions in Sect.~\ref{sec:conclusions}. 
Appendix~\ref{app:table} provides the details of the electronic table containing the UCDs analysed in this study, including the newly discovered ones. The evaluation of various \Euclid fluxes is presented in Appendix~\ref{app:photometry}, in which we identified the most suitable choices for point-like sources. It also deals with the UCD distribution and the predicted QSO locus in various \Euclid colour-colour planes. Lastly, in Appendix~\ref{app:atmo_bt_comparison}, we compare the results of the ATMO \citep{2020A&A...637A..38P} and BT-Settl \citep{2015A&A...577A..42B} models used for stellar parameter estimation.

%=============================================
% ===============================================
\section{New UCD candidates and comparison sample}
\label{sec:sources}

In this section, we describe the sample used in this study: the newly identified candidate UCDs in the \EDFN region, one of the Q1 fields, and the previously known UCDs in Q1, identified by cross-matching published UCD catalogues with the entire Q1 dataset.

% ===============================================
\subsection{New UCD candidates in the \EDFN}

As part of the \Euclid QSO search pipeline, which identifies QSOs through \ion{C}{iii}] and \ion{Mg}{ii} emission lines, spectroscopic template fitting was performed to identify other objects that occupy the same colour space as QSOs. This approach enabled a purely spectroscopic identification of UCDs and QSOs, allowing an evaluation of the pipeline’s ability to distinguish between them \citep{2025MNRAS.542.1088B}. This procedure was applied to the Q1 spectra \citep{Q1-TP007, Q1-TP006} in \EDFN (22.9 \,deg$^2$ centred on $\rm{RA}=\ra{17;58;55.9}$ and $\rm{Dec}= \ang{+66;01;04.7}$) and resulted in \NNEW UCD candidates, which were provided to the \Euclid UCD work package for further investigation (Vincent Le Brun and Eduardo Ba\~nados, priv. comm.). In this paper, we present the analysis of this sample of new UCD candidates and the confirmation of a subset.

% ===============================================
\subsection{Previously identified UCDs found in Q1}

To understand the \Euclid observations of UCDs, we searched for previously detected UCDs appearing in Q1. 
We cross-matched the Q1 merged (MER) catalogue \citep{Q1-TP004}, generated by the \Euclid MER processing function that merges multi-wavelength data products, with a compilation of already known UCDs described below. 
The catalogue is available through the \Euclid{} science archive\footnote{\url{https://eas.esac.esa.int/sas/}}.

% ===============================================
\subsubsection{UCDs from DES+VHS+WISE} 
\label{sec:des_sources}

\cite{2019MNRAS.489.5301C} present extensive catalogues of candidate UCDs, including 11\,745 photometrically classified objects near the stellar$-$substellar boundary and BDs, with SpTs ranging from L0 to T9, and 20\,863 M dwarfs with SpTs ranging from M6 to M9. These candidates have detections in the Dark Energy Survey (DES) Year-3 data \citep{2015AJ....150..150F}, VISTA Hemisphere Survey (VHS) DR3 data \citep{2013Msngr.154...35M}, and Wide-field Infrared Survey Explorer \citep[WISE;][]{2010AJ....140.1868W}, covering 2400\,deg$^2$ in the southern sky and combining photometry in eight filters. 
Henceforth, these DES+VHS+WISE candidate UCDs will be referred to as DES objects.

We searched for counterparts of DES objects in Q1, with a search radius of $5\arcsec$, chosen to account for positional uncertainties while avoiding missed associations. Because the depth of the \Euclid data can produce multiple nearby detections within this radius, we applied a preliminary photometric consistency cut to reject obvious mismatches. We compared \Euclid{} magnitudes in the \IE and \YE bands with $i$- and $Y$-band magnitudes from DES, and likewise we compared magnitudes in the \JE and \HE bands with $J$- and $H$-band magnitudes from VHS. 
The \Euclid{} magnitudes are defined in the AB system as
\begin{equation}
m = -2.5 \log_{10}\!\left(\frac{\texttt{FLUX\_M\_2FWHM\_APER}}{3631}\right),
\end{equation}
where $\texttt{FLUX\_M\_2FWHM\_APER}$ is the flux density, in $\mathrm{Jy}$, derived from aperture photometry using an aperture of diameter twice the full width at half-maximum (FWHM), which is found to be the most suitable for point-like objects (see Sect.~\ref{sec:euclid_cl} for details). Here, $m=\IE$, \YE, \JE, or \HE.
DES~Y3 BD catalogues provide \textit{i} and \textit{Y} magnitudes in the AB system, while VHS $J$ and $H$ magnitudes are in the Vega system.
Zero points used for magnitudes to flux conversion between AB and Vega magnitude systems were taken from the Spanish Virtual Observatory (SVO) Filter Profile Services \citep{2012ivoa.rept.1015R}. Magnitude conversions between \Euclid{} and observations from other surveys were performed using the relations reported in appendix~D of \cite{EuclidSkyNISP}.
Since all surveys are calibrated in the AB system, genuine counterparts are expected to have consistent magnitudes within the photometric uncertainties. Matches with a magnitude difference of greater than $1$\,mag in any filter were rejected.

For objects passing this filtering, we defined a figure of merit (FoM) as
\begin{equation}
    {\rm FoM} = \sqrt{\sum_{{m=I, Y, J, H}}\left(\frac{\Delta m}{\sigma_m}\right)^2+\sum\left(\frac{S}{\sigma_S}\right)^2},
\end{equation}
where $S$ denotes the angular separation between the DES and Q1 positions, and $m$ represents the magnitude in the four photometric bands. The best cross-match is selected as the pair that minimises the FoM value. Using this approach, we recovered \NDES{} reliable matches in Q1 with SpTs ranging from M7 to L6 dwarfs; 220 M dwarfs and 106 L dwarfs. All of these objects also have Q1 spectra available in the archive.

% ===============================================
\subsubsection{UCDs from the literature}\label{sec.UCDsfromtheliterature}

We cross-matched the Q1 data with the catalogue of known UCDs from \cite{2019MNRAS.485.4423S}. This catalogue has been updated since publication to include all spectroscopically classified objects later than \textup{M9} published before 2024, as well as examples of earlier SpTs. 
The latest, unpublished version contains 8435 spectroscopically classified and 3821 photometrically classified UCDs, spanning SpTs \textup{M7} to \textup{Y3} and distributed across the entire sky. By including both spectroscopic and photometric classifications, this catalogue provides an expanded representation of the M7--M9 population, in addition to objects from spectroscopically confirmed catalogues such as the UltracoolSheet \citep{best_2025_15802304}.
Our cross-match identified \NKNW objects with SpTs ranging from \textup{M9} to \textup{T9} (hereafter LIT), classified spectroscopically and photometrically. These objects, listed in Table~\ref{tab:masterucds}, have  Q1 spectra available in the archive. In the table, brackets indicate photometric classifications. \JE is the \Euclid magnitude in the AB system, based on aperture photometry of aperture size $2 \times$ FWHM (see Sect.~\ref{sec:euclid_cl} and Appendix~\ref{app:photometry} for details), while $J$ is a magnitude from either the Two Micron All Sky Survey \citep[2MASS;][]{Skrutskie2006} or Mauna Kea Observatories \citep[MKO;][]{Tokunaga2002} photometric system, but converted to AB magnitudes.

The initial match was performed by selecting the nearest Q1 entry within a $5\arcsec$ radius of each UCD. When PMs were available, we updated the positions to epoch 2024.6, consistent with Q1 data. If PMs were unavailable, we used the original discovery positions, which were often based on ground-based catalogues with epochs of observations typically around the early 2000s. Multiple matches were found within a few arcseconds. The best magnitude match was not always the closest match due to PM. We resolved such cases using the FoM method described in Sect.~\ref{sec:des_sources}.

Two objects in Table~\ref{tab:masterucds}, J1757$+$6741 and J1819$+$6600, exhibit significant magnitude differences. These objects are saturated in \Euclid observations, rendering their magnitudes unreliable. However, visual inspection confirmed the matches, and their observed spectra agree with published classifications. We have removed the unrealistic magnitude estimates from the table; however, bright objects, with \JE < 18\,mag, should be visually confirmed if magnitude differences are used as a secondary match criterion.

% ===============================================
\subsubsection{\textit{Gaia} UCDs}\label{sec.gaiaUCDs}

\cite{2023A&A...669A.139S} catalogued 94\,158 \textit{Gaia} DR3 sources as UCDs, based on their $T_{\mathrm{eff}}$ obtained from the Extended stellar parametrizer-Ultracool dwarfs\footnote{\url{https://gea.esac.esa.int/archive/documentation//GDR3/Data_analysis/chap_cu8par/sec_cu8par_apsis/ssec_cu8par_apsis_espucd.html}} (ESP-UCD) pipeline. Cross-matching  \textit{Gaia} {\tt source\_id} with the {\tt gaia\_id} in the Q1 MER yielded 197 common objects (\JE $\leq$ 26\,mag). Of these 139 objects (\JE $\leq$ 24\,mag) have Q1 spectra. Four of these sources (\texttt{OBJECT\_ID} = $-$577964646489153925, $-$663296172475883538, $-$587631527497989311, and $-$596828778491172067) were also part of the DES sample.

% ===============================================
In summary, the cross-matching yielded 326 reliable matches between the DES UCD catalogues and Q1, 12 matches between the catalogue of \cite{2019MNRAS.485.4423S} and Q1, 197 matches between the \textit{Gaia} UCD sample and Q1, of which 4 overlap with the DES sample. In total, this results in a final matched sample of 531 known UCDs with \Euclid photometry. 
This known UCD sample in Q1, together with the 202 new UCD candidates in the \EDFN, are used for the spectroscopic analysis based on \Euclid red-grism (\RGE) spectra, as described in Sect. \ref{sec:spt}.

% ===============================================
% ===============================================
\section{Spectral analysis}
\label{sec:spt}

In this section, we present the analysis of \Euclid slitless \RGE spectra ($R>480$, 1.2--1.9\,\micron), processed through the \texttt{SIR} pipeline, which performs contamination mitigation, wavelength and flux calibration, and extraction of combined 1D spectra from dispersed imaging data \citep{Q1-TP006}.
The procedure for identifying reliable UCDs from Q1 spectra, as detailed in Sect. \ref{sec:spec_fit}, was first trained on the curated sample of 531 previously identified UCDs in Q1. It was then applied to the 202 new UCD candidates in \EDFN to confirm their candidacy. Spectral analysis of previously known UCDs in Q1 and the new UCD candidates in \EDFN are reported in Sects. \ref{sec:qualitycheck} and \ref{spectral_new}, respectively.

% ===============================================
\subsection{Fitting and quality control}
\label{sec:spec_fit}

The data points that make up each Q1 spectrum were filtered to refine the measurements. We excluded points shortwards of 1.22\,\micron{} and redwards of 1.88\,\micron{} from the fitting range to avoid noise at the spectral edges. In addition, only spectral points with \texttt{MASK =} 0 were retained, where the \texttt{MASK} parameter flags spectral bins affected by quality issues or unreliable calibration \citep{Q1-TP007}. These selected spectral points were then used to compute the signal-to-noise ratio (S/N). The spectra were then compared to spectral templates to determine the UCD SpTs. No explicit S/N threshold was applied, as no clear correlation was found between S/N and reliable UCD classification, as demonstrated in Sect. \ref{sec:qualitycheck}.

To determine the SpTs of the objects from Q1 spectra, we used a classify-by-standard method and spectral standards from  SpeX Prism Library Analysis Toolkit \citep[SPLAT;][]{2017ASInC..14....7B}. For each Q1 spectrum, the SPLAT routine performed a $\chi^2$ minimisation against all available standards, over the wavelength range 1.22--1.88\,\micron. \Euclid spectra were smoothed to match the spectral resolution of the SpeX templates prior to comparison. The spectral standards adopted in the fitting are taken from \cite{2006ApJ...637.1067B}, \cite{2010ApJS..190..100K}, and \cite{2011ApJ...743...50C}.

% ===============================================
\subsection{Spectral analysis: Previously identified UCDs in Q1}
\label{sec:qualitycheck}

We analysed the Q1 spectra of 326 DES objects and 12 LIT objects, performing spectral typing as described in Sect.~\ref{sec:spec_fit}. 
The resulting sample consists of 104 objects with photometrically derived SpTs (103 from the DES sample and one from the LIT sample) and five objects with spectroscopically determined SpTs (from LIT). 
The selected 103 DES objects span a \JE range of $\sim$17--22\,mag, with Q1 SpTs ranging from M7 to T0. They have, on average, a mean S/N of $\sim$ 22, with an average of 360 pixels passing the selection criterion. 
The six selected LIT objects span a \JE range of 18--20\,mag, with Q1 SpTs ranging from M8 to T7. They have a mean S/N of $\sim$ 15, with an average of 210 pixels passing the selection criterion.

% ===============================================
\begin{figure}
    \centering
    \includegraphics[width=1\linewidth]{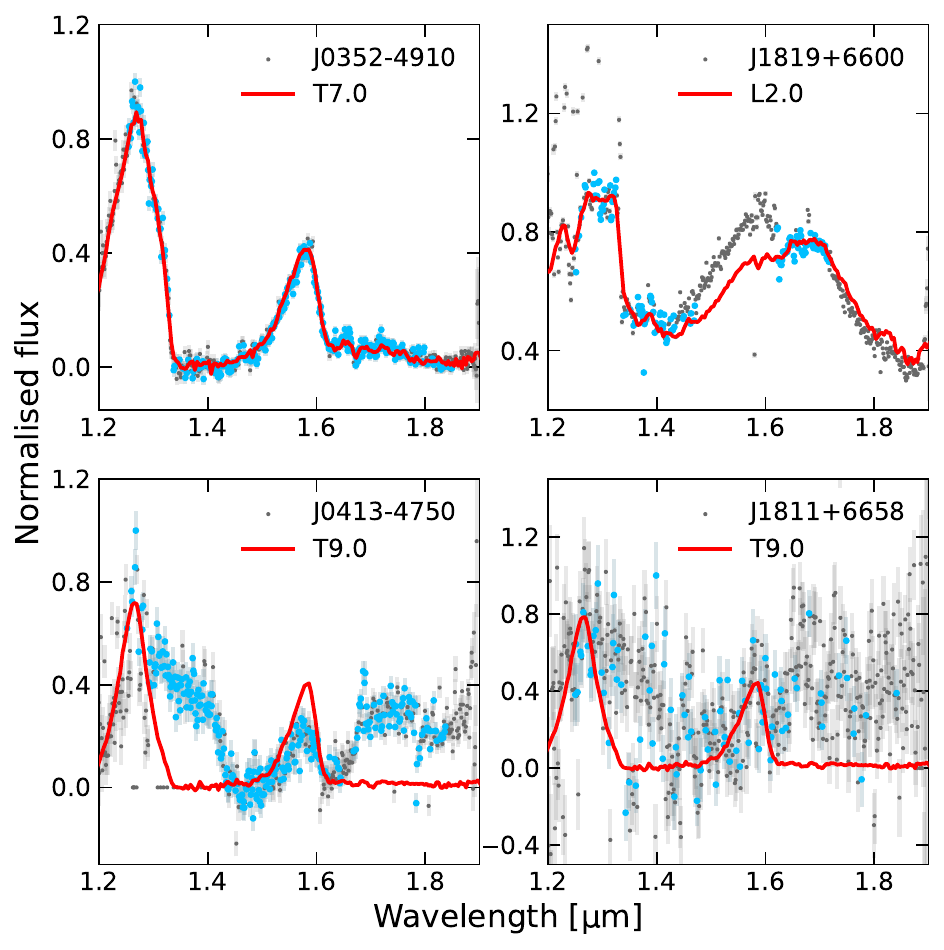}
    \caption{Q1 spectra of four UCDs with published spectral classifications. Small grey points show all measured spectral data, while big sky blue points indicate those retained after quality filtering. The solid red line represents the template spectrum corresponding to the published SpT (refer to  Sect.~\ref{sec:qualitycheck} for details).}
    \label{fig:maskeffect}
\end{figure}
% ===============================================

The SpTs derived from Q1 spectra are consistent with their previously reported photometric and spectroscopic classifications, with a standard deviation of 1.86, corresponding to an average difference of one to two spectral subclasses. Figure~\ref{fig:maskeffect} presents the Q1 spectra of four representative LIT objects. The dominant sources of uncertainty here arise from background noise, detector effects, and spectral contamination from neighbouring sources inherent to slitless spectroscopy. Only the filtered data were used for spectral standard fitting, as described in Sect.~\ref{sec:spec_fit}.

J0352$-$4910 (WISEA J035231.80$-$491059.4) is a T7 dwarf spectroscopically confirmed by \cite{2024A&A...686A.171Z}. The Q1 observations are of high quality, with only a few rejected points, and our spectral fit on \Euclid Near Infrared Spectrometer and Photometer (NISP) spectra recovers the published classification.

J1819+6600 (2MASS J18194942+6600548) has a photometrically derived SpT of L2.2 from \cite{2021A&A...645A.100S}. Our spectral fit on \Euclid NISP spectra also classifies it as an L2. This object is bright ($J=15.43$\,mag; \citealt{2003yCat.2246....0C}), leading to saturation in the \Euclid{} data and rendering the \JE{} magnitude unreliable.
Several rejected data points deviate from the SPLAT standard spectra, while selected data points match the spectral standard. Since this source is bright, it is also detected in \textit{Gaia} DR3 (\texttt{source\_id} = 2257578507499901312) with a parallax of $28.3\pm0.6$\,mas, consistent with an L2 classification. This result demonstrates that even with partial spectral data, a reliable classification can be achieved when appropriate filtering is applied.

In contrast, the T9 dwarf, J0413$-$4750 \citep[WISE J041358.14$-$475039.3;][]{2013APJS..205....6M}, exhibits a Q1 spectrum that deviates from all available spectral templates, despite few data points being flagged as bad. We examined the \Euclid{} images to determine whether the poor fit of the spectrum could be attributed to detector artefacts or contamination from a nearby source. However, no conclusive explanation has been found for the mismatch between the observed spectrum and UCD templates and leave further investigation to future data releases, which will revisit the field.

\begin{figure}
    \centering
    \includegraphics[angle=0,height=4.3cm]{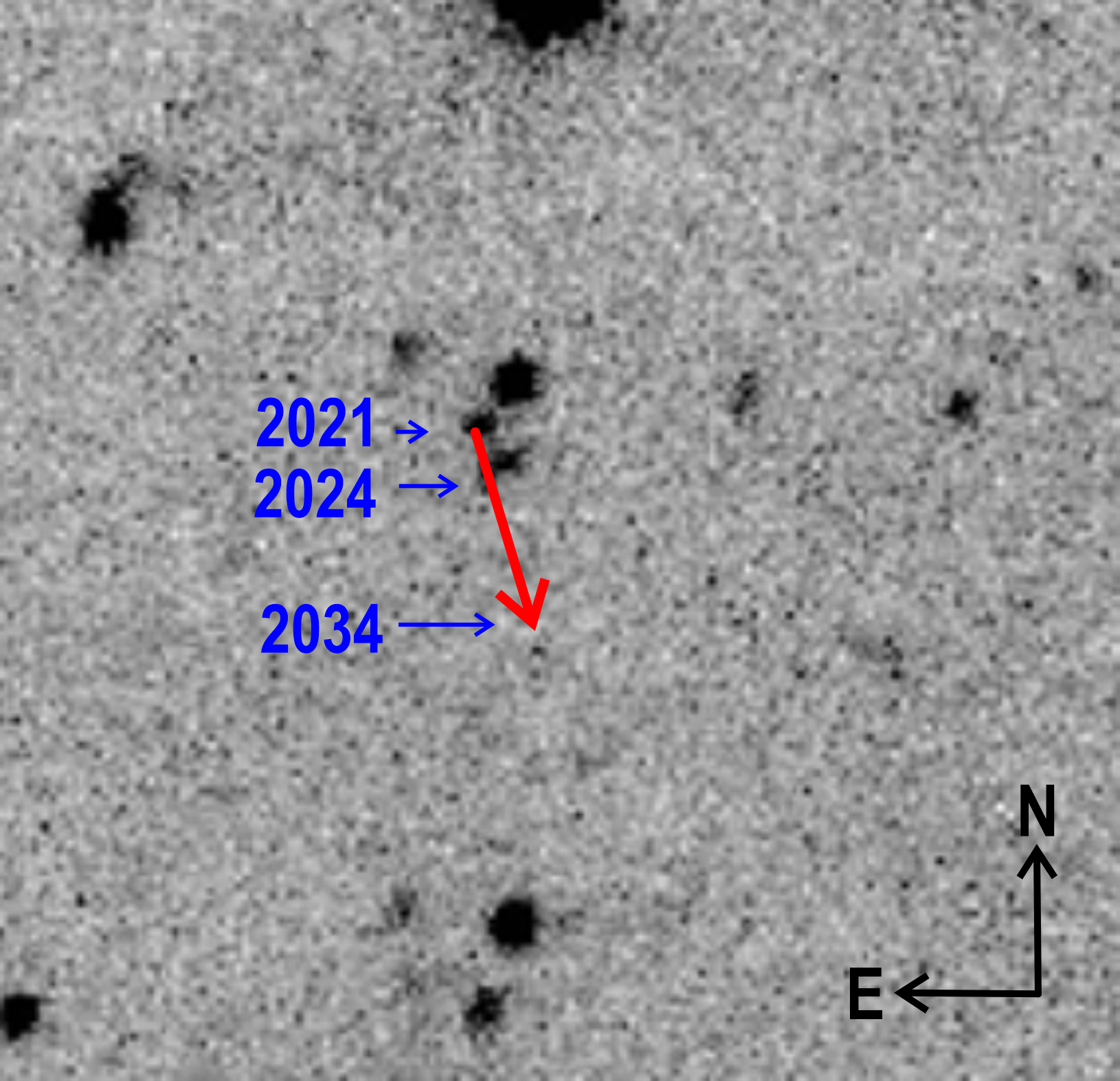}
    \includegraphics[angle=0,height=4.3cm]{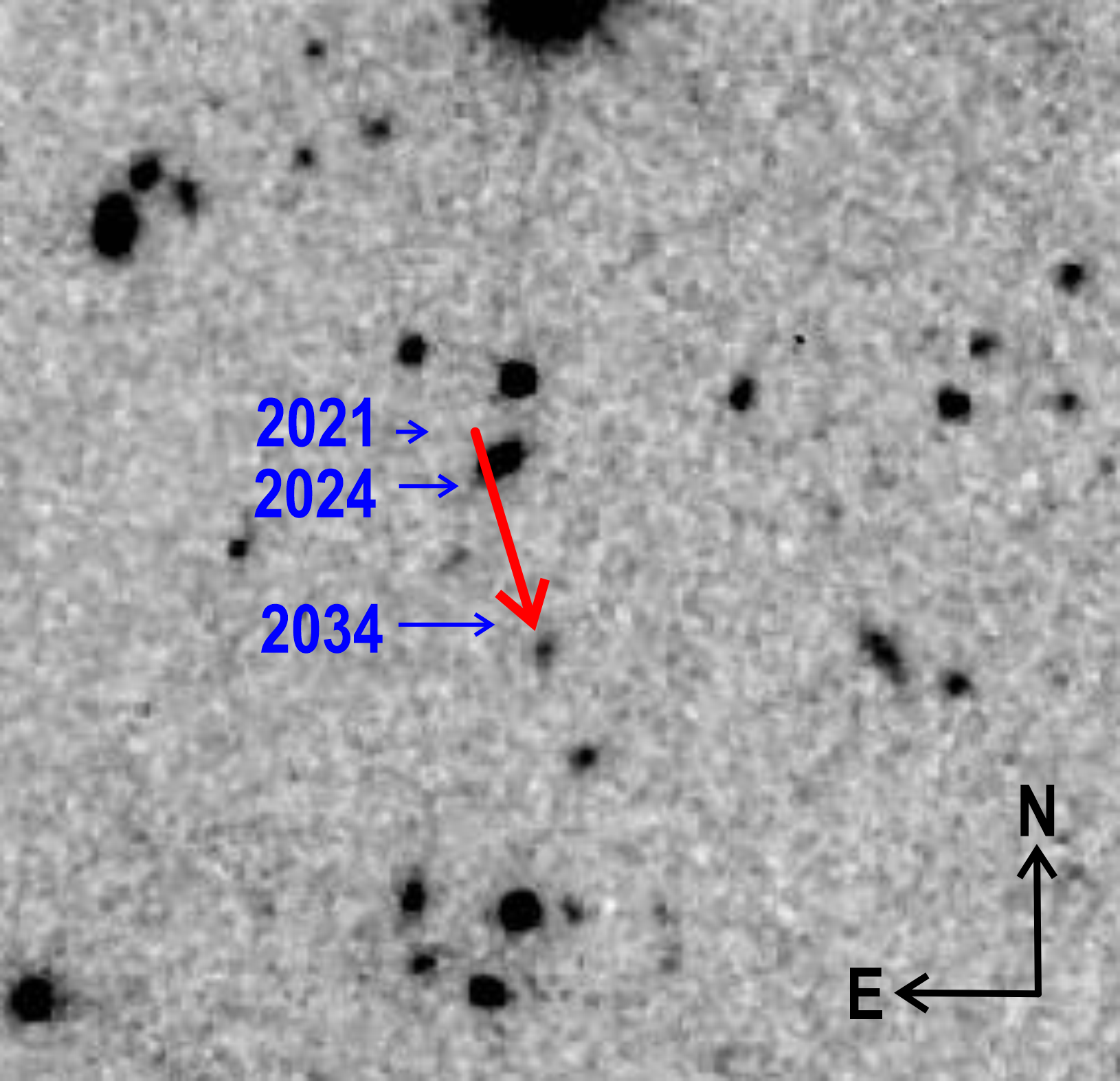}
    \caption{\textit{Left panel}: Keck MOSFIRE $J$ (MKO) band 33\,$\arcsec$ $\times$ 33\,$\arcsec$ image showing J1811$+$6658 in 2021, along with its predicted positions for 2024 and 2034. 
    \textit{Right panel}: Q1 background-subtracted \JE image from 18 July 2024 with identical position markings and sizes.}
    \label{fig:J1811+6658} 
\end{figure}

The final example shown in Fig.~\ref{fig:maskeffect} (bottom right) is J1811+6658 (CWISE J181125.34$+$665806.4), recently identified as a BD candidate within 20\,pc by \cite{2024ApJS..271...55K}.  
In that study, the authors reported a parallax of ($69.7 \pm 6.8$)\,mas and a PM exceeding $0.5\,\arcsec\,\mathrm{yr}^{-1}$. Based on this distance, apparent magnitude, and colour, they classified J1811+6658 as a T9--Y1 dwarf. However, the observed \Euclid spectra show a spread in spectral pattern, and our spectral fitting procedure assigns an unreliable SpT of L2. In Fig.~\ref{fig:J1811+6658}, we present a Keck Multi-Object Spectrometer For Infra-Red Exploration (MOSFIRE) $J$-band image of this object from 2021 alongside the \Euclid image from 2024 with its predicted trajectory through 2034 (the anticipated end of an extended \Euclid mission).  
At present, J1811$+$6658 overlaps with a background object and is assigned to the Q1 detection $\texttt{OBJECT\_ID}=2728547962669671905$. The incorrect spectral classification is therefore caused by significant contamination.  
As the field is revisited throughout the mission, the object will move away from the background source, allowing uncontaminated spectral extraction and parallax estimation.

The \textit{Gaia} UCDs, classified based on $T_{\mathrm{eff}}$, are relatively bright ($G \leq 20.5$\,mag) and predominantly exhibit low noise and minimal scatter in Q1 spectra suitable for robust classification. Using the SPLAT spectral typing procedure described in Sect.~\ref{sec:spec_fit}, a total of 58 sources were matched to UCD templates, spanning SpTs from M7 to L1 and \JE in the range  $\sim$17--24\,mag. They have a median S/N of $\sim$ 70 and, on average $\sim$ 370 pixels passing the selection criterion.

% ===============================================
Combining DES, LIT, and \textit{Gaia} samples, we obtain \NKNWSPECTRA Q1 spectra that match SPLAT UCD templates with reliable spectral typing. Having thus validated the spectral analysis method for Q1 spectra using the previously identified UCDs, we applied this method to the newly detected UCD candidates in \EDFN, as described in Sect. \ref{spectral_new}.

% ===============================================
% ===============================================

% ===============================================
\begin{table}[htbp!]
\caption{New UCDs in \EDFN discovered from Q1 data.} 
\label{tab:qsoucds}
\begin{center}
\begin{tabular}{cccl}
\toprule
\toprule
Short name  &  \Euclid name         & \JE & SpT  \\
  &    & [mag] &   \\
\midrule  
   J1821+6653  & J182133.43$+$665312.2 &  20.593 &       M7.0\\
   J1808+6621  & J180845.29$+$662118.7 &  19.971 &       M8.0\\
   J1737+6621  & J173729.32$+$662141.7 &  19.317 &       M8.0$^{1, 2}$\\
   J1802+6611  & J180242.86$+$661100.3 &  19.167 &       M8.0\\
   J1816+6540  & J181650.97$+$654019.0 &  19.693 &       M8.0\\
   J1730+6536  & J173038.36$+$653630.0 &  19.582 &       M8.0\\
   J1753+6600  & J175339.09$+$660028.0 &  20.978 &       M9.0\\
   J1801+6435  & J180119.12$+$643547.0 &  20.060 &       M9.0\\
   J1801+6832  & J180120.11$+$683231.0 &  19.334 &       M9.0\\
   J1812+6739  & J181217.66$+$673947.4 &  19.549 &       L0.0\\
  J1752+6811  & J175224.29$+$681130.8 &  20.557 &       L0.0\\
   J1752+6710  & J175224.03$+$671045.1 &  19.802 &       L0.0$^{1, 2}$\\
   J1800+6604  & J180005.51$+$660416.6 &  19.392 &       L0.0\\
   J1755+6712  & J175516.15$+$671214.6 &  18.166 &       L1.0\\
   J1755+6322  & J175557.68$+$632200.6 &  19.574 &       L1.0\\
   J1749+6326  & J174906.70$+$632637.3 &  19.273 &       L1.0\\
   J1732+6547  & J173225.69$+$654747.9 &  19.390 &       L1.0\\
   J1744+6653  & J174441.74$+$665349.4 &  19.076 &       L1.0\\
   J1800+6719  & J180046.67$+$671902.4 &  19.603 &       L1.0\\
   J1806+6623  & J180624.15$+$662345.2 &  19.902 &       L2.0$^{1, 2}$\\
   J1732+6555  & J173222.86$+$655522.6 &  20.162 &       L2.0\\
   J1745+6427  & J174526.94$+$642725.0 &  19.867 &       L2.0$^{1, 2}$\\
   J1802+6517  & J180247.62$+$651738.8 &  17.824 &       L3.0$^{1, 2}$\\
   J1801+6321  & J180120.97$+$632100.2 &  21.084 &       L4.0\\
   J1754+6551  & J175418.85$+$655123.3 &  20.339 &       L4.0\\
   J1744+6406  & J174409.24$+$640638.8 &  17.763 &       L4.0\\
   J1801+6708  & J180153.19$+$670853.1 &  19.662 &       L5.0\\
   J1801+6623  & J180142.91$+$662327.5 &  20.961 &       L7.0\\
   J1812+6616  & J181201.84$+$661609.7 &  20.212 &       L7.0\\
   J1746+6419  & J174645.56$+$641933.6 &  20.609 &       L7.0\\
   J1815+6456  & J181528.44$+$645627.1 &  20.250 &       L7.0\\
   J1807+6401  & J180754.09$+$640141.4 &  19.078 &       L7.0\\
   J1807+6642  & J180744.18$+$664252.8 &  19.906 &       T0.0$^{1}$\\
   % another two objs: M9 - L2 available
\bottomrule
\end{tabular}
\end{center}
Notes. 1: also found by \cite{dominguez2025}. 2: also found by \cite{zerjal2026}.
\end{table}
% ===============================================

% ===============================================
\begin{figure*}
%\sidecaption
    \centering
\includegraphics[angle=0,width=17cm]{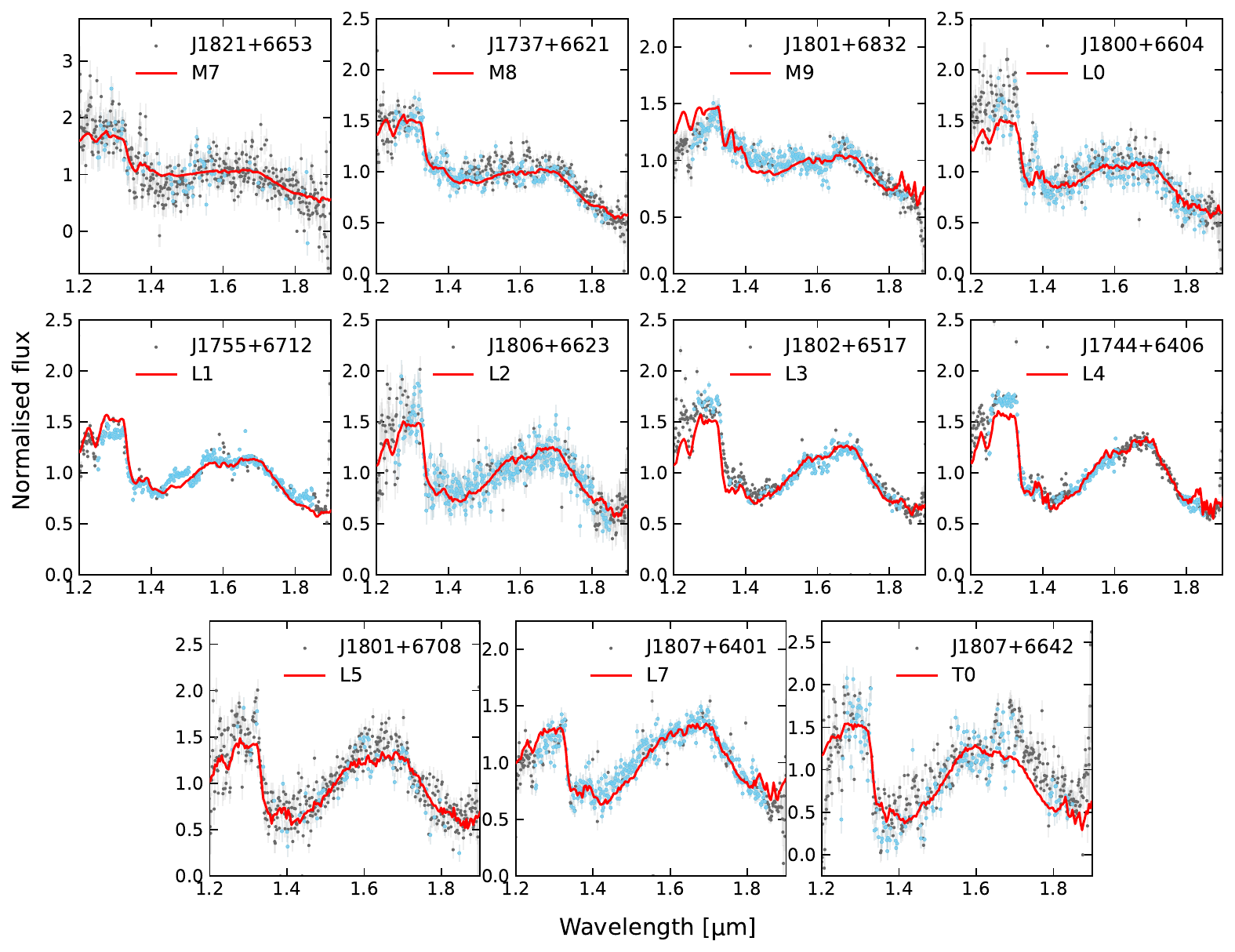}
\caption{Q1 spectra of a subset of newly discovered UCDs in the \EDFN. The colour scheme follows that used in Fig.~\ref{fig:maskeffect}. }
\label{fig:speactra_highsnr} 
\end{figure*}
% ===============================================

% ===============================================
\subsection{Spectral analysis: New UCD candidates in \EDFN}
\label{spectral_new}

Spectral analysis and spectral standard fitting are applied to the sample of \NNEW{} UCD candidates in \EDFN. Among them, \NNEWSPECTRA{} have Q1 spectra that allow us to derive confident SpTs. They have SpTs ranging from M7 to T0 and \JE\ magnitude in the range 17--21\,mag. These objects are listed in Table~\ref{tab:qsoucds}. The reported SpT is derived from Q1 NISP spectra. Spectra representative of unique SpTs from M7 to T0, along with their best-fit SPLAT spectral standards, are shown in Fig.~\ref{fig:speactra_highsnr}.

Further, \NNEWINTSPECTRA objects have interesting spectra that deviate from the UCD standard spectra while showing some similarities and notable trends. These sources warrant further investigation in the forthcoming data releases. We describe two such examples below. The rest of the sample has insufficient quality spectra to assign reliable SpTs.

\begin{figure}
\centering
\includegraphics[angle=0,width=1\linewidth]{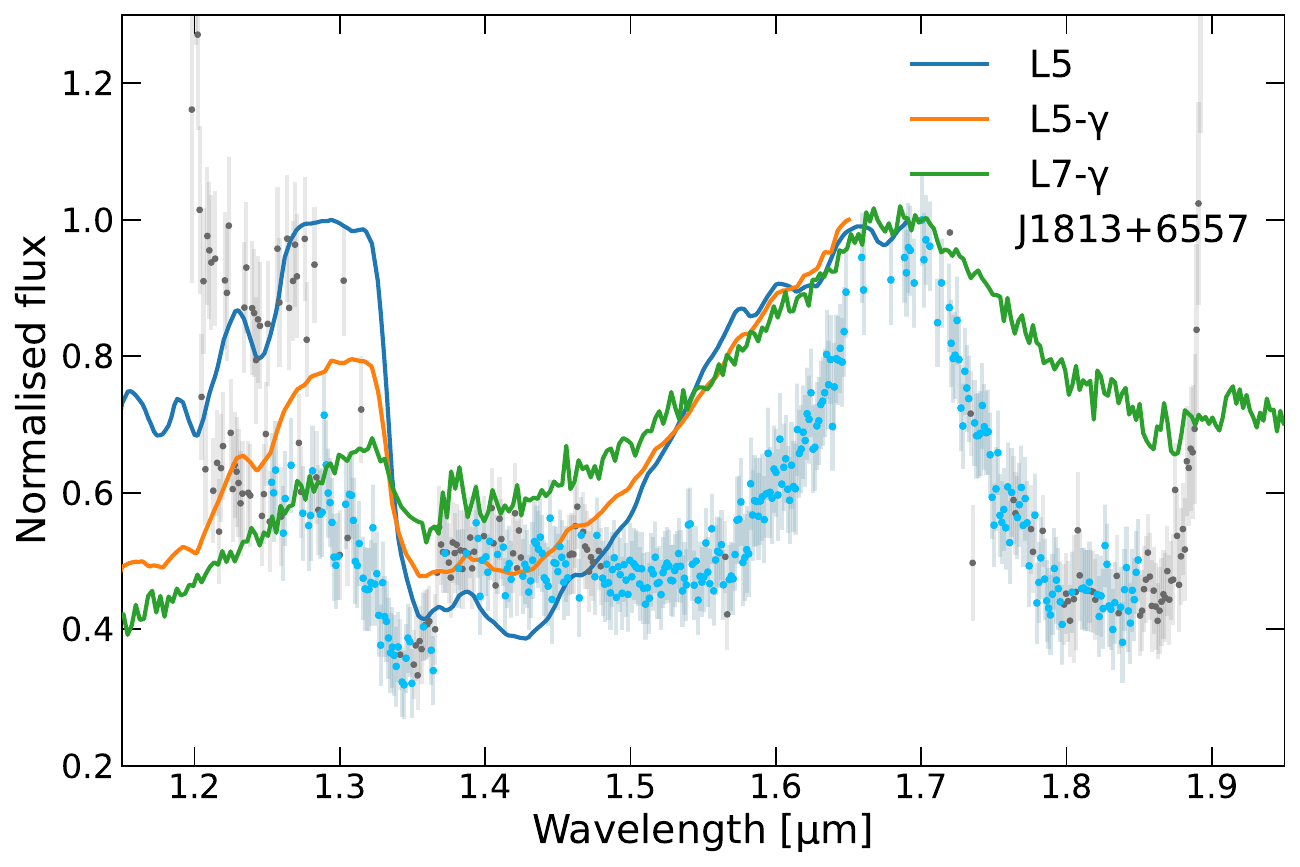}
\caption{Peculiar Q1 spectrum of J1813$+$6557 compared with low-gravity and normal L dwarf spectra. The very low-gravity L5 (L5-$\gamma$) and the normal L5 comparison spectra are from HST, while the very low-gravity L7 (L7-$\gamma$) was observed from the ground (see Sect.~\ref{spectral_new}). All spectra are normalised at 1.7\,\micron{}.}
\label{fig:speactra_new} 
\end{figure}

Figure~\ref{fig:speactra_new} shows the Q1 spectrum of J1813$+$6557 ($\texttt{OBJECT\_ID}=2734840299659521922$), which has 315 pixels meeting the quality cuts and a mean S/N of 18. The spectrum exhibits features atypical for normal UCDs, including a narrow \HE-band peak and a peculiar dip around 1.35\,\micron{}. Since the triangular \HE band is characteristic of young, low-gravity objects, we compare it to a very low-gravity L7 from ground-based observations \citep[][]{2022A&A...664A.111B} and to a very low-gravity L5, and normal L5 from the Cloud Atlas \textit{Hubble} Space Telescope (HST) Wide Field Camera 3 (WFC3) UCD spectra archive \citep{2019AJ....157..101M} in Fig.~\ref{fig:speactra_new}. Although no exact match is obtained, the comparison shows that the low-gravity objects exhibit depressed flux in the \JE band. The HST spectra, not hampered by telluric absorption, also show enhanced flux around 1.4\,\micron{} in very low-gravity objects.

% ===============================================
% ===============================================

\begin{figure}
\centering
\includegraphics[angle=0,width=0.9\linewidth]{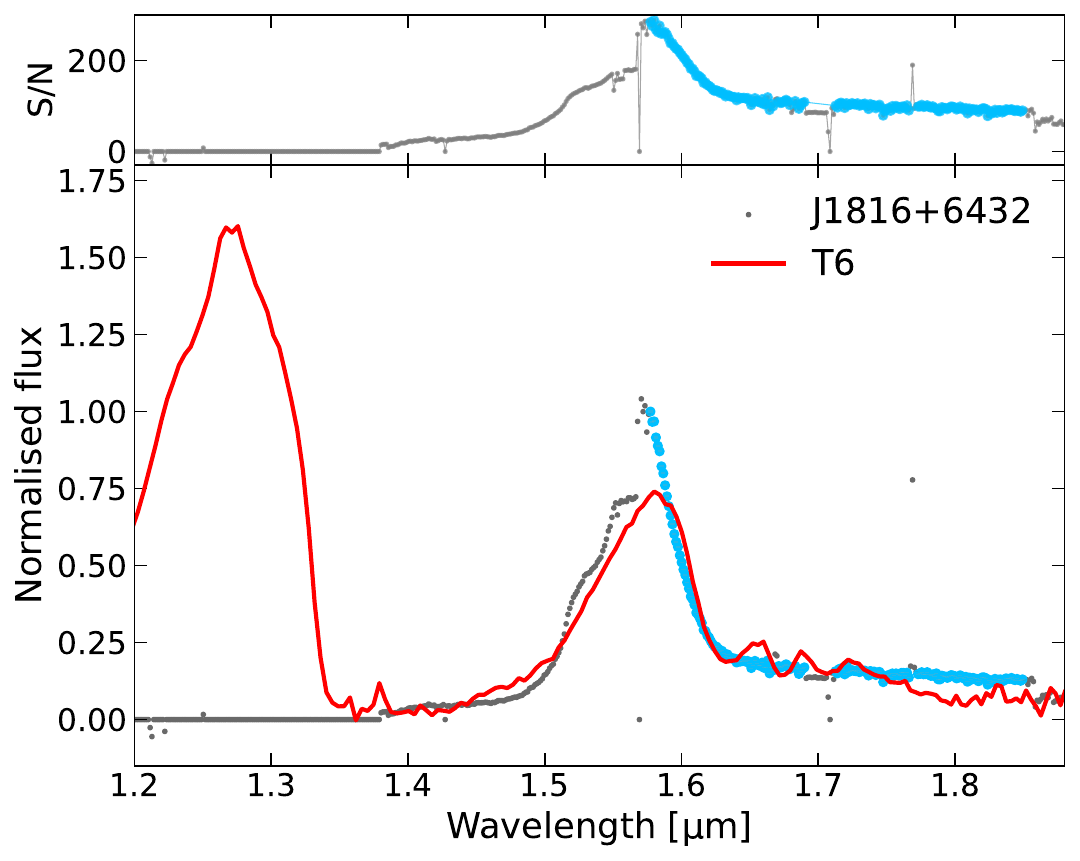}
\caption{High-S/N Q1 spectrum (\emph{bottom panel}) of J1816$+$6432 and its S/N distribution as a function of wavelength (\emph{top panel}). The colour scheme follows that used in Fig.~\ref{fig:maskeffect}.}
\label{fig:speactra_new_t6} 
\end{figure}

Figure~\ref{fig:speactra_new_t6} displays the Q1 spectrum of object J1816$+$6432 ($\texttt{OBJECT\_ID}=2741779784645475127$), which has a high mean S/N of 102 with 180 pixels meeting the filtering cuts. Its \HE-band peak resembles that of T dwarfs. However, half of the points are rejected during filtering, and there is no detectable signal in the \JE band. Nevertheless, it remains a promising late-T candidate for future  \Euclid observations.

% ===============================================
% ===============================================

% ===============================================
% ===============================================
\section{Photometric, astrometric, and spectroscopic analysis}
\label{sec:combinedtable} 

In total, we retrieved SpTs for \NSPECTRA{} UCDs from Q1 spectra, including \NNEWSPECTRA{} newly identified UCDs in \EDFN{}. Table~\ref{tab:fulldataset} lists the primary data used in this work. The corresponding spectra can be obtained directly from the \Euclid{} science archive using their \texttt{OBJECT\_ID}.

% ===============================================
\subsection{\Euclid colours and QSOs}
\label{sec:euclid_cl}

The Q1 MER catalogues provide fluxes derived through various methods for each \Euclid band \citep[][]{Q1-TP001}. These include fluxes from S\'ersic model-fitting photometry  \citep{2022arXiv221202428K} and aperture photometry \citep[\texttt{A-PHOT};][]{2019A&A...622A.169M}. \texttt{A-PHOT} provides four flux measurements using circular apertures with diameters of $1\times$, $2\times$, $3\times$, and $4\times$ the FWHM, where the FWHM is determined from the band with the lowest spatial resolution, typically one of the complementary ground-based bands used in conjunction with \Euclid observations \citep[Sect. 5;][]{Q1-TP004}. For the NIR bands, fluxes from template-fitting photometry \citep[\texttt{T-PHOT};][]{2015A&A...582A..15M} are also available, while for the VIS band, point-spread function (PSF) fitting photometry is provided.

To determine the optimal Q1 MER fluxes for point-like sources such as UCDs, we measured the deviation between \Euclid magnitudes and published DES and VHS magnitudes for the 326 DES objects, as reported in Appendix \ref{app:photometry}.  
Based on the standard deviation of the residuals, \texttt{A-PHOT} within a 2$\times$ FWHM aperture diameter provides the closest agreement for the DES and VHS magnitudes in \YE, \JE, and \HE filters and with a 4$\times$ FWHM aperture for \IE. We therefore adopted this flux measurement as our default for computing \Euclid magnitudes.

We also compared the \textit{Gaia} magnitudes with \Euclid magnitudes by deriving the transformation between the \textit{G} band and the \IE band (reported in Appendix \ref{app:photometry}). The residual scatter was found to be the lowest for the aperture photometric extraction methods.

\Euclid colour-colour diagrams for various filter combinations are presented in Fig.~\ref{fig:Q1x:YJvsJH}. On average, these diagrams reveal a correlation between SpT and colour, which is in agreement with the synthetic \Euclid UCD colour predictions of \cite{2024RNAAS...8..137S}.  
Each panel of Fig.~\ref{fig:Q1x:YJvsJH} also includes the predicted QSO tracks. These tracks overlap significantly with M- and L-type dwarfs in colour space, indicating that photometry alone may not be sufficient to distinguish QSOs from UCDs. However, at higher redshifts, QSOs can appear redder and may be separable from UCDs. 
\cite{2025MNRAS.542.1088B} demonstrated that a QSO with \JE = 20.10\,mag at $z=5.4$ can be identified using the \Euclid \RGE spectrum, whereas colour-based selection alone would not distinguish it from the more numerous M-type dwarfs. Furthermore, \cite{Q1-SP068} empirically established that \Euclid \RGE spectra are effective for discovering QSOs up to $z \sim 5$ for sources brighter than $\JE =21.5$\,mag. Consequently, for $\JE < 21.5$\,mag, \Euclid spectra provide a complementary means to differentiate between UCDs and QSOs.

% ===============================================
\subsection{Proper motions}
\label{sub:propermotions}

To derive the PMs of objects in the three Q1 fields, we needed an initial-epoch dataset that spans their full footprint and is deep enough to detect sources as faint as UCDs. The WISE survey is the only available option that meets these requirements. We used AllWISE\footnote{\url{https://wise2.ipac.caltech.edu/docs/release/allwise/expsup/sec2_5.html}} Data Release positions, as they are derived over a short time baseline and therefore represent a consistent single-epoch measurement (epoch = 2010.5589). Table~\ref{tab:EDFepochs} lists the mean epochs of the three Euclid Deep Fields (EDFs), giving a baseline of approximately 14 years when combined with AllWISE.

\Euclid entries were matched to AllWISE using a $10\arcsec$ search radius to account for sources with PMs up to $0.7\,\arcsec\,\mathrm{yr}^{-1}$. Because Q1 lacks a photometric band close to AllWISE wavelengths, magnitude criteria could not be used to refine the matches, leading to many spurious associations. All 33 new UCDs were therefore manually verified to ensure correct AllWISE matches.

AllWISE positions were determined using a combination of the Fourth United States Naval Observatory (USNO) CCD Astrograph Catalogue \citep[UCAC4;][]{2013AJ....145...44Z} and 2MASS, whereas \Euclid uses the \textit{Gaia} DR3 positions. This introduces potential systematic differences between the two reference frames. To mitigate this effect, we adjusted \textit{Gaia} DR3 matches to AllWISE objects to epoch 2010.5589 and applied the Infinity Overlapping Circles method \citep[][]{Bucciarelli01101993} to align them with the \textit{Gaia} reference frame. A comparison of the derived PMs to \textit{Gaia} PMs in the two coordinates has a reduced $\chi^2$ value of 0.7, indicating consistent values and uncertainties.

\begin{figure*}
\centering
\includegraphics[width=0.9\linewidth]{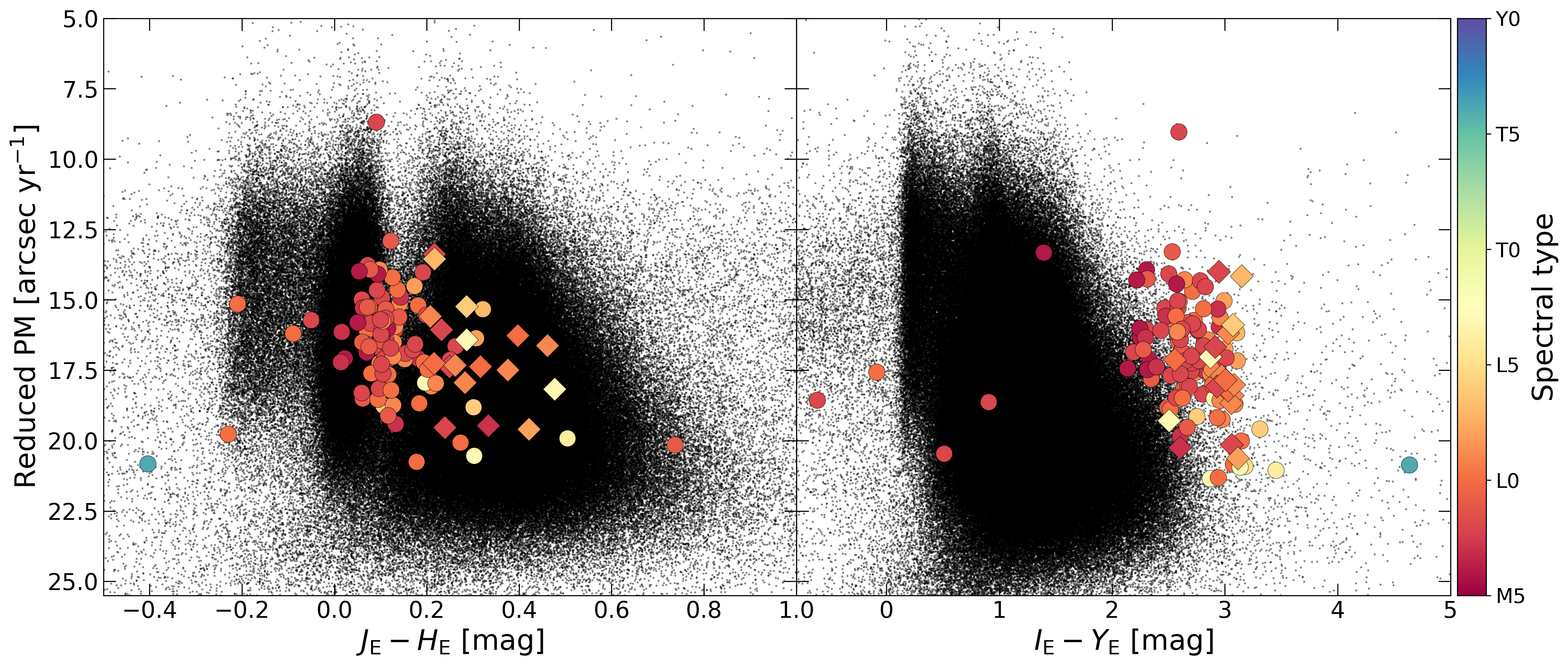}
\caption{Reduced proper motion diagrams for UCDs with reliable spectral typing. Diamonds represent newly identified UCDs, and dots show previously known UCDs, colour-coded by \Euclid SpT. Dark background points correspond to all sources matched between AllWISE and Q1. 
}
\label{fig:rpm}
\end{figure*}

Figure~\ref{fig:rpm} presents reduced proper motion diagrams for our comparison. Newly identified and previously known UCDs are represented as diamonds and circles, colour-coded by their respective SpT, while the black background points represent all sources in common between AllWISE and Q1. Different black cloud regions correspond to Galactic halo and disc populations. The left panel (\JE $-$ \HE diagram) shows a clear separation between these populations, with Galactic halo and disc populations centred around \JE $-$ \HE= 0.05\,mag and \JE $-$ \HE= 0.00\, mag, respectively. The right panel, which shows both visible and NIR bands (\IE $-$ \YE diagrams), demonstrates how these colours help isolate UCDs from halo and disc objects.

% ===============================================
\begin{table*}[]
\centering 
\caption{Euclid deep fields: coordinates, size, Q1 coverage, and mean epoch.}
\label{tab:EDFepochs}
\begin{tabular}{lccccc}
\toprule
\toprule
Field & RA & Dec & Size & Q1 cover & Mean epoch \\
      &  [h:m:s] & [d:m:s]  & [$\mathrm{deg}^2$] & [days] & [yr] \\
\midrule
\EDFN~(North)  & $17:58:55.9$   & $+66:01:03$ & 20 & 2.192 & 2024.5469 \\
\EDFF~(Fornax) & $03:31:43.6$   & $-28:05:18$ & 23 & 1.166 & 2024.5981 \\
\EDFS~(South)  & $04:04:57.8$   & $-48:25:22$ & 10 & 2.654 & 2024.6842 \\
\bottomrule
\end{tabular}
\end{table*}
% ===============================================

% ===============================================
% ===============================================

% ===============================================
\subsection{Spectroscopic distances}
\label{sub:distances}

For all UCDs in the combined catalogue, we calculated spectroscopic distances (\textit{d}) using the SpTs derived in this work and \Euclid magnitudes, following the empirical relations from \cite{2024RNAAS...8..137S}. These distances were compared to both trigonometric measurements from \textit{Gaia} and photometric estimates derived from transformed magnitudes using absolute magnitude calibrations in the literature. In the two cases, the spectroscopic distance estimates are consistent.

\begin{figure}
\centering
\includegraphics[width=1\linewidth]{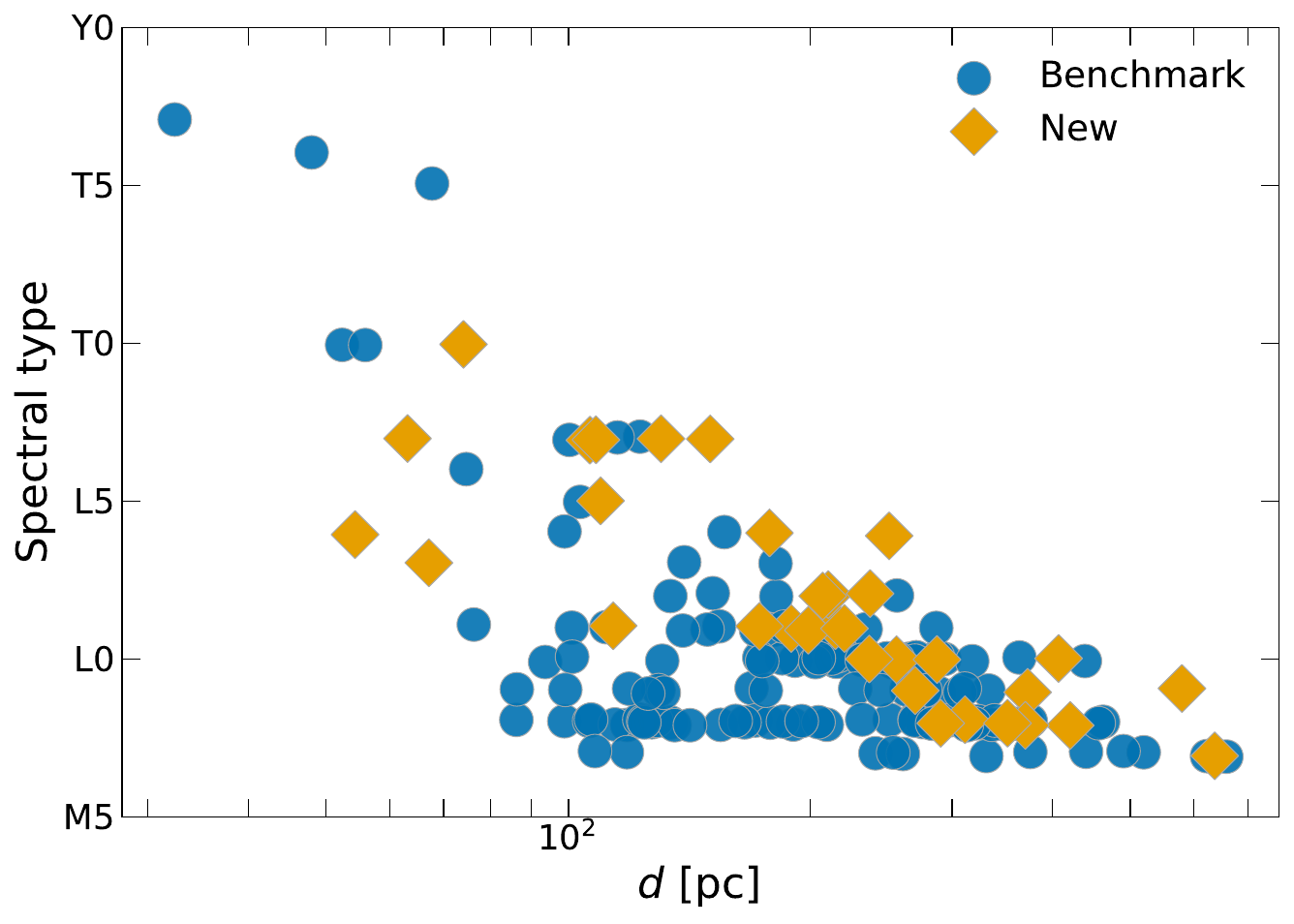}
\caption{SpT versus spectroscopic distance plot for UCDs. \NNEWSPECTRA new \EDFN UCDs and \NKNWSPECTRA known UCDs are represented as orange diamonds and blue dots, respectively. A uniformly distributed vertical jitter of $\pm 0.1$ is applied to the $y$-axis values for visual clarity.
}
\label{fig:dist}
\end{figure}

Figure~\ref{fig:dist} presents the SpT versus the \textit{d} distribution for both newly identified and previously known UCDs. As expected, late-L- and T-type dwarfs are located within 100\,pc, while late M dwarfs extend out to 500--600\,pc.

% ===============================================
% ===============================================

% ===============================================
\subsection{Model fitting and stellar parameter estimation}
\label{subsec:fitting}

We performed atmospheric model fitting on the \Euclid Q1 spectra to derive stellar parameters including $T_{\mathrm{eff}}$, [Fe/H], and $\log{g}$. The analysis is carried out using the ATMO atmospheric model \citep{2020A&A...637A..38P} with uniform priors set within the parameter ranges: $\mathrm{C/O}$ = 0.55, $\gamma$ = 1.03, $T_{\mathrm{eff}}$ = 800--3000\,K, $\log{g}$ = 4.5--5.5, and [Fe/H] = $-$0.6--0.6\,dex, 
where $\gamma$ is the adiabatic index. These ranges are consistent with the expected physical properties of field UCDs \citep{filippazzo2015, Ravinet2024}. The fitting is performed using the \texttt{stelparucds} package \citep{Ravinet2024}, which performs nested sampling implementation and log-likelihood estimation via the \texttt{ultranest} package \citep{2021JOSS....6.3001B} based forward modelling.

\begin{figure}
\centering
\includegraphics[width=1\linewidth]{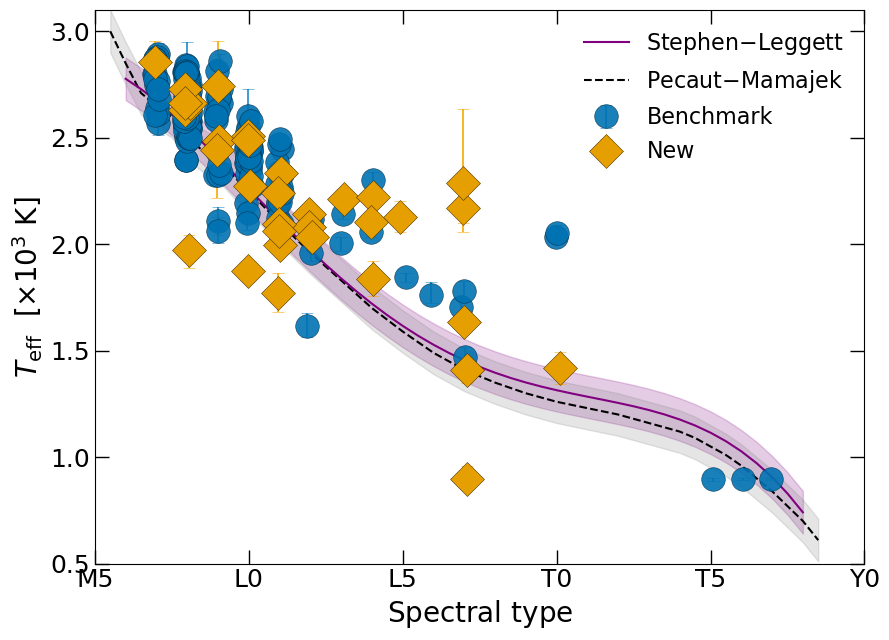}
\caption{$T_{\mathrm{eff}}$ versus SpT plot for UCDs. The blue dots denote the known UCDs, and the orange diamonds denote the new UCDs. A uniformly distributed horizontal jitter of $\pm 0.1$ is applied to the $x$-axis values for visual clarity. The purple and grey lines denote the empirical Stephens--Leggett and Pecaut--Mamajek $T_{\mathrm{eff}}$--SpT relations, respectively, with a $\pm$ 100\,K uncertainty range.
}
\label{fig:teff}
\end{figure}

Figure~\ref{fig:teff} displays the distribution of $T_{\mathrm{eff}}$ derived from Q1 spectra as a function of \Euclid SpTs for previously known and newly identified UCDs. The SpT--$T_{\mathrm{eff}}$ relation for UCDs from Stephens--Leggett \citep{2009ApJ...702..154S} and Pecaut--Mamajek \citep[][last updated on 22 March 2019]{pecaut2013}\footnote{\url{https://www.pas.rochester.edu/~emamajek/EEM_dwarf_UBVIJHK_colors_Teff.dat}} are overplotted, with a $\pm$100\,K uncertainty range. The obtained $T_{\mathrm{eff}}$ are in agreement with the literature predictions. Three L7 objects (J1746+6419, J1812+6616, and J1801+6623) show deviations from the literature relation, consistent with their larger deviations with respect to standard UCD templates.

The full set of stellar parameters derived from \Euclid NISP spectra is provided in the combined catalogue,  Table~\ref{tab:fulldataset}.
A comparison of stellar parameters obtained from \Euclid spectra with ATMO and BT-Settl \citep{2015A&A...577A..42B} models is presented in Appendix \ref{app:atmo_bt_comparison}.

% ===============================================
% ===============================================

% ===============================================
% ===============================================
\section{ \Euclid--\textit{Gaia} synergy}
\label{sec:gaia_euclid_spec_comb}

Among the \textit{Gaia} UCDs within Q1, 34 sources also have mean \textit{Gaia} XP continuous spectra. Of these, 26 have reliable spectral typing from both \Euclid and \textit{Gaia} spectra, and are analysed here to leverage the combined capabilities of the two spectral ranges.

\subsection{\Euclid--\textit{Gaia} combined spectra}
\textit{Gaia} DR3 mean XP spectra \citep{2023A&A...674A...2D}, constructed from multiple observations of the same object, are provided in the \textit{Gaia} archive as coefficients of base functions. We converted them into calibrated spectra using the \texttt{GaiaXPy} package \cite[v2.1.2, ][]{gaiaxpy2024}. Each XP spectrum comprises a blue photometer part (BP; 0.33--0.68\,\micron) and a red photometer part (RP; 0.63--1.05\,\micron), with the RP part generally achieving higher S/N.

We fitted the RP spectra with SPLAT’s library of UCD spectral templates to assign \textit{Gaia}-based SpTs. These classifications agree with the \Euclid Q1 SpTs with a standard deviation of less than one SpT.

\begin{figure}
    \centering
    \includegraphics[width=1\linewidth]{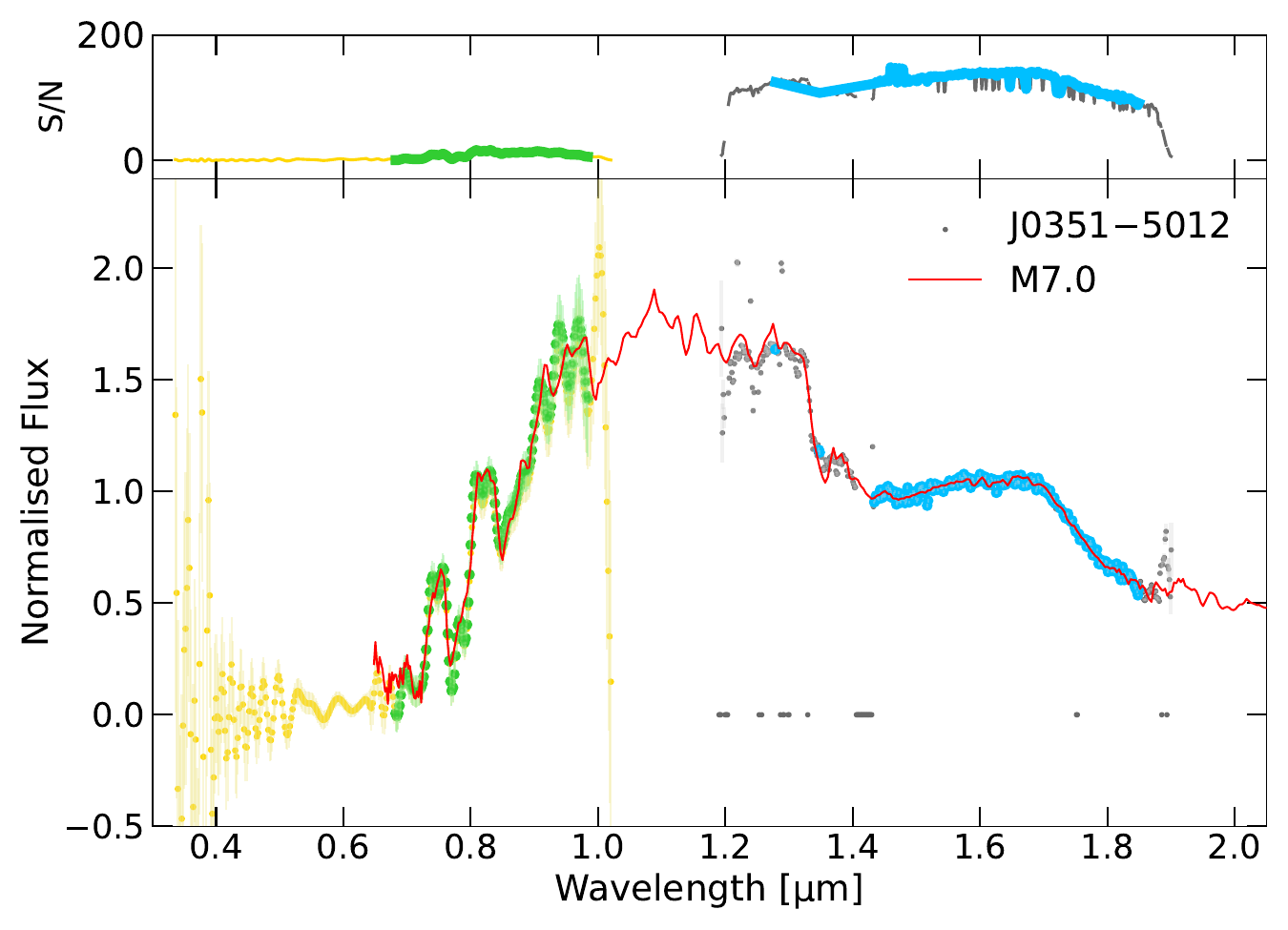}
    \caption{Normalised \Euclid and \textit{Gaia} spectra of object J0351$-$5012. \Euclid and \textit{Gaia} spectra are shown in grey and yellow, respectively. The selected \Euclid points are shown in sky blue and selected \textit{Gaia} points are shown in green. The red curve represents the SPLAT standard spectrum corresponding to the adopted \Euclid SpT.}
    \label{fig:gaia-scaling}
\end{figure}

Combining \textit{Gaia} and \Euclid{} spectra extends the wavelength coverage available for each object.
As an example, Fig.~\ref{fig:gaia-scaling} illustrates the normalised \Euclid and $\textit{Gaia}$ spectra of object J0351$-$5012 ($\textit{Gaia}$ DR3 4829833420345494016). The \Euclid{} spectrum is shown in grey, while the selected points are highlighted in sky blue. The \textit{Gaia} spectrum is shown in yellow, with the selected RP points highlighted in green. The BP component and RP data beyond 1~\micron{} are excluded due to their poor quality and the frequent presence of spectral artefacts. The red curve represents the SPLAT standard spectrum corresponding to the adopted SpT, which is consistent between \textit{Gaia} and \Euclid. 
The \textit{Gaia} and \Euclid spectra of all the 26 objects, in their original flux units, are displayed in Fig.~\ref{fig:26gaiaUCDs} in Appendix \ref{app:gaia_euclid_spectra}, showcasing the compatibility of the two missions.

% ===============================================
% ===============================================

% ===============================================
\subsection{Stellar parameter estimation}

The \textit{Gaia} and \Euclid spectra together are fitted with ATMO atmospheric models using \texttt{stelparucds} to derive stellar parameters.  Uniform priors identical to those described in Sect. \ref{subsec:fitting} were adopted. Radii were computed from the \textit{Gaia} parallax, $\varpi$, in mas, and scaling factor of the respective spectra, $\alpha$, using the relation, $R = \sqrt{\alpha} \times 1000/\varpi$, where $R$ is in pc.

\begin{figure}
    \centering
    \includegraphics[width=1\linewidth]{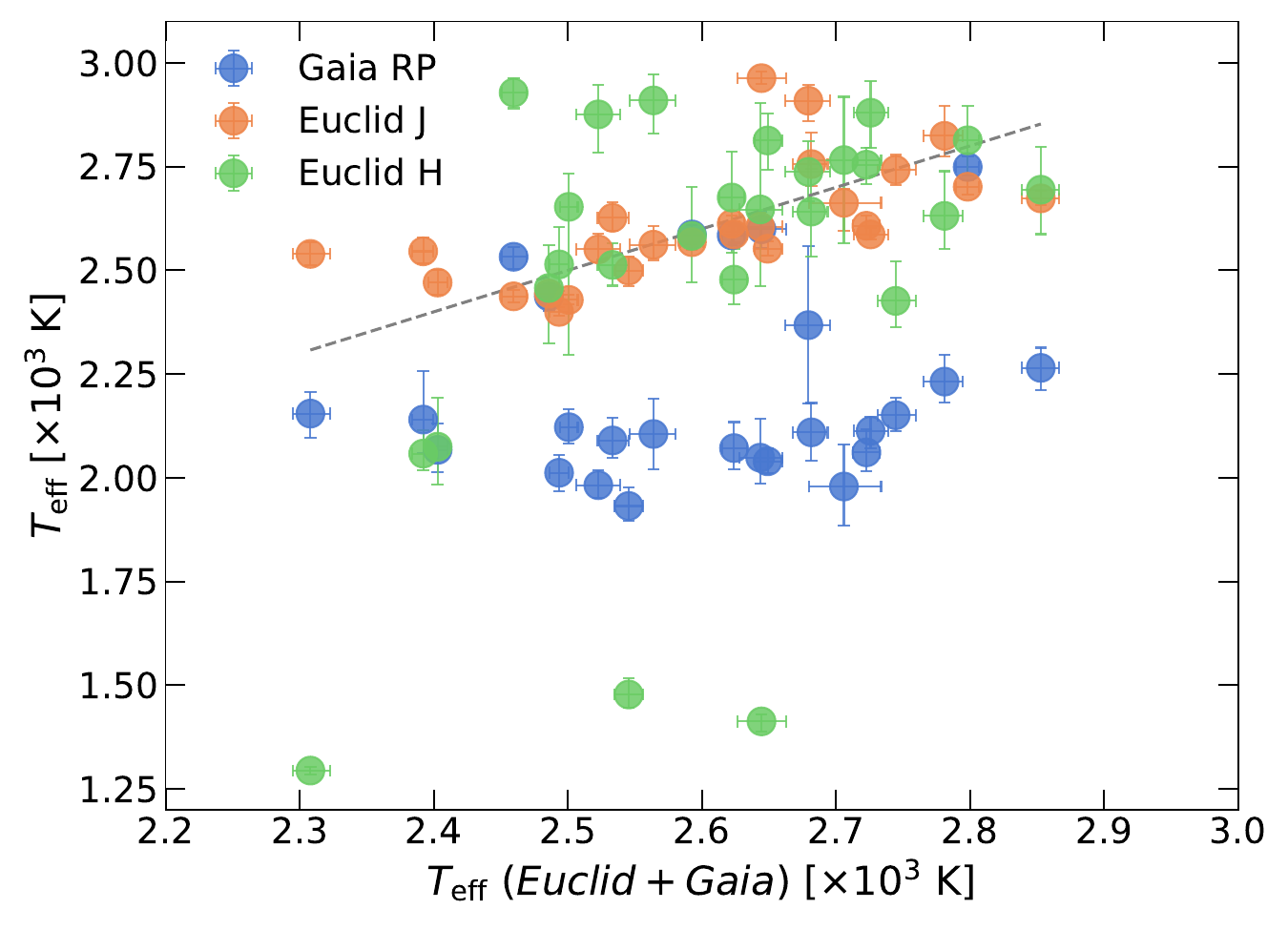}
    \caption{Comparison of $T_{\mathrm{eff}}$ from spectra. The $T_{\mathrm{eff}}$ derived from combined \Euclid $+$ \textit{Gaia} spectral fitting is compared with the $T_{\mathrm{eff}}$ obtained from individual bands. The dashed grey line indicates the 1:1 relation.
    }
    \label{fig:Teff_comparison plot}
\end{figure}

Figure~\ref{fig:Teff_comparison plot} compares $T_{\mathrm{eff}}$ derived from the \Euclid{} $+$ \textit{Gaia} spectra fitting with those obtained independently from RP part (blue), \JE band (orange), and \HE band (green). As shown, the estimates from the RP band are consistently lower than the results based on \Euclid{} $+$ \textit{Gaia}, while the results from \HE tend to show larger deviations. \cite{Ravinet2024} has also reported the reduced robustness of \Euclid \HE-band fitting using \texttt{stelparucds}.
This highlights the increased robustness of $T_{\mathrm{eff}}$ determinations when combining \textit{Gaia} and \Euclid{} spectra, which mitigates band-dependent systematics.

The derived $T_{\mathrm{eff}}$ and $R$ values are tabulated in Table \ref{tab:nophot_jrp}. The results are based on \JE $+$ \textit{Gaia} fitting, to avoid the larger deviations introduced by the \HE-band. The reported asymmetric uncertainties correspond to the 16th and 84th percentiles of the posterior distribution.

The scaling factors for \textit{Gaia} and \Euclid spectra were similar, and the corresponding $R$ values were comparable. The reported $R$ in Table \ref{tab:nophot_jrp} is the mean of the $R$ calculated from the \Euclid scaling factor and \textit{Gaia} scaling factor. Five objects show $R$ values greater than 1.5\,$R_{\rm{Jup}}$. Values of $R$ can vary depending on the bands and scaling used. More robust results are expected from future data releases.

The uncertainties in $R$ include only the statistical uncertainties from $\alpha$ and $\varpi$. They do not account for possible systematic uncertainties related to the atmospheric models, flux calibration, or assumptions in the spectral fitting procedure. Therefore, the true uncertainties are larger than the values reported here.

% ===============================================
% ===============================================

% ===============================================
% \input{tex_files/nophot_jrp}
\begin{table*}[htbp!]
\caption{Stellar parameters obtained from combined $\textit{Gaia}$ $+$ \Euclid model fitting. }
\label{tab:nophot_jrp}
\centering
\begin{tabular}{ccccccrc}
\toprule
\toprule
Short name & \Euclid name   & SpT & $\textit{Gaia}$ ID & \gmag & $T_{\mathrm{eff}}$ &  $\varpi$~~~~~~~~ & $R$ \\
& & & & [mag] & [K] & [mas]~~~~~ & [$R_{\mathrm{Jup}}$] \\
\midrule

J0327$-$2828 & J032711.95$-$282822.5 & M8 & 5059999313284104320 & $19.956$ & $2451^{+44}_{-40}$ & $14.91 \pm 0.49$ & $0.98 \pm 0.04$ \\
J0327$-$2906 & J032746.30$-$290657.3 & M8 & 5056950401899789824 & $18.910$ & $2431^{+10}_{-9}$ & $21.26 \pm 0.22$ & $1.09 \pm 0.01$ \\
J0328$-$2818 & J032838.79$-$281845.0 & M8 & 5060370948216377600 & $20.563$ & $2653^{+44}_{-37}$ & $5.93 \pm 0.73$ & $1.34 \pm 0.12$ \\
J0329$-$2900 & J032952.93$-$290022.3 & M8 & 5057283309110507904 & $20.040$ & $2548^{+22}_{-23}$ & $6.75 \pm 0.52$ & $1.65 \pm 0.09$ \\
J0329$-$2626 & J032958.34$-$262603.2 & L0 & 5060844047454500096 & $20.533$ & $2416^{+27}_{-19}$ & $8.91 \pm 0.68$ & $1.26 \pm 0.07$ \\
J0330$-$2912 & J033033.46$-$291209.2 & M8 & 5057265373327081728 & $20.499$ & $2583^{+32}_{-27}$ & $5.71 \pm 0.79$ & $1.53 \pm 0.15$ \\
J0332$-$2627 & J033204.00$-$262734.3 & M8 & 5084665963299825536 & $19.873$ & $2700^{+22}_{-21}$ & $8.63 \pm 0.45$ & $1.12 \pm 0.04$ \\
J0332$-$2616 & J033226.23$-$261639.3 & L0 & 5084692935694431488 & $18.868$ & $2410^{+12}_{-15}$ & $15.90 \pm 0.23$ & $1.45 \pm 0.02$ \\
J0332$-$2702 & J033243.64$-$270244.2 & M9 & 5060563740708156416 & $20.378$ & $2938^{+12}_{-13}$ & $7.93 \pm 0.71$ & $0.87 \pm 0.06$ \\
J0349$-$4958 & J034920.95$-$495838.0 & M9 & 4829854860822491904 & $20.476$ & $2890^{+44}_{-91}$ & $7.03 \pm 0.65$ & $0.88 \pm 0.07$ \\
J0351$-$5012 & J035103.70$-$501247.8 & M7 & 4829833420345494016 & $18.871$ & $2637^{+41}_{-28}$ & $14.84 \pm 0.20$ & $1.12 \pm 0.03$ \\
J0351$-$4854 & J035111.15$-$485455.4 & M8 & 4830054319104251264 & $20.150$ & $2495^{+20}_{-24}$ & $11.00 \pm 0.47$ & $1.12 \pm 0.04$ \\
J0357$-$5037 & J035748.02$-$503748.4 & M8 & 4828842107534832256 & $18.560$ & $2420^{+12}_{-11}$ & $24.58 \pm 0.18$ & $1.09 \pm 0.01$ \\
J0358$-$5026 & J035815.81$-$502631.0 & M8 & 4828938108643706880 & $18.643$ & $2577^{+13}_{-12}$ & $15.14 \pm 0.15$ & $1.29 \pm 0.02$ \\
J0403$-$4715 & J040348.51$-$471535.2 & M8 & 4831859339239484416 & $19.563$ & $2563^{+18}_{-16}$ & $8.10 \pm 0.30$ & $1.66 \pm 0.05$ \\
J0423$-$4703 & J042325.04$-$470353.4 & M8 & 4788943441902663808 & $18.311$ & $2565^{+16}_{-14}$ & $18.43 \pm 0.13$ & $1.29 \pm 0.02$ \\
J0424$-$4725 & J042432.13$-$472525.9 & M7 & 4788920214719251200 & $19.588$ & $2653^{+18}_{-18}$ & $9.98 \pm 0.30$ & $1.17 \pm 0.03$ \\
J1743$+$6404 & J174355.09$+$640428.5 & M8 & 1440622156672820864 & $19.940$ & $2495^{+29}_{-32}$ & $8.12 \pm 0.34$ & $1.45 \pm 0.06$ \\
J1745$+$6710 & J174501.84$+$671044.5 & M8 & 1634299343198545664 & $20.291$ & $2397^{+30}_{-27}$ & $10.80 \pm 0.57$ & $1.18 \pm 0.05$ \\
J1753$+$6622 & J175348.21$+$662204.6 & M8 & 1633293186981081984 & $20.461$ & $2641^{+35}_{-68}$ & $9.15 \pm 0.61$ & $0.99 \pm 0.06$ \\
J1755$+$6654 & J175515.41$+$665425.9 & M8 & 1633576075003342592 & $19.307$ & $2519^{+16}_{-18}$ & $13.05 \pm 0.27$ & $1.22 \pm 0.03$ \\
J1757$+$6741 & J175730.71$+$674136.5 & M8 & 1633676508518808576 & $17.515$ & $2674^{+17}_{-13}$ & $22.43 \pm 0.08$ & $1.29 \pm 0.02$ \\
J1800$+$6552 & J180056.17$+$655224.1 & M8 & 2161689063353927296 & $19.671$ & $2388^{+11}_{-12}$ & $10.83 \pm 0.32$ & $1.61 \pm 0.04$ \\
J1801$+$6559 & J180124.82$+$655957.2 & M7 & 2161711813796634880 & $19.693$ & $2617^{+20}_{-23}$ & $10.48 \pm 0.25$ & $1.12 \pm 0.03$ \\
J1809$+$6621 & J180934.55$+$662139.6 & M8 & 2257753364211398912 & $19.587$ & $2536^{+17}_{-15}$ & $14.12 \pm 0.34$ & $1.03 \pm 0.02$ \\
J1815$+$6447 & J181521.42$+$644721.6 & M9 & 2161379172873739264 & $19.804$ & $2494^{+19}_{-22}$ & $7.43 \pm 0.28$ & $1.76 \pm 0.06$ \\
\bottomrule
\end{tabular}
\end{table*}
% ===============================================

% ===============================================
% ===============================================
\section{GTC follow-up of new \EDFN UCDs}
\label{sec:GTC_spectra}

Among the 33 newly discovered UCDs reported in this work, two sources, J1744$+$6406 and J1807$+$6401 (\Euclid SpTs L4 and L7, respectively), were followed up with ground-based spectroscopic observation using the 10.4-m GTC at the Roque de los Muchachos Observatory, La Palma. These targets were prioritised based on their brightness suitability for ground-based spectroscopy and to provide an independent validation of the \Euclid-based UCD detection method. Observations were conducted with the Optical System for Imaging and low-Intermediate-Resolution Integrated Spectroscopy Plus \citep[OSIRIS$+$;][]{2013ApJ...762...46P} instrument in R300R grism mode and the Espectr\'ografo Multiobjeto Infra-Rojo \citep[EMIR;][]{2022A&A...667A.107G} instrument in low-resolution (LR) grism mode, as part of the GTC observing programme GTCMULTIPLE2D-25A (PI: E. L. Martín).

OSIRIS$+$ observations were carried out on 20 June 2025, with exposure times of 2$\times$1600\,s and 2$\times$1800\,s for objects J1744$+$6406 and J1807$+$6401, respectively. EMIR observations were conducted on 7 June 2025, with exposure times of 1287\,s in the \textit{YJ} band and 1930\,s in the \textit{HK} band, for both J1744$+$6406 and J1807$+$6401.

OSIRIS$+$ spectra cover the wavelength range from 0.37--1.05 \,\micron\ with a resolution of $\sim$350 while EMIR spectra cover the range 0.97--2.40\,\micron\ (\textit{Y}, \textit{J}, \textit{H}, and \textit{K} bands) with a resolution of $\sim$ 1000. OSIRIS$+$ optical spectra were reduced and calibrated using the \texttt{PyLongslit} package \citep{valeckas2025}, while EMIR \textit{YJ} and \textit{HK} spectra were reduced independently using \texttt{PyEmir} \citep{2019ASPC..523..317C} and \texttt{IRAF} \citep{fitzpatrick2025} packages. Telluric correction was performed using a standard star, 55~Dra, which was observed and reduced in the same manner.

\begin{figure*}
    \centering
    \includegraphics[width=0.7\linewidth]{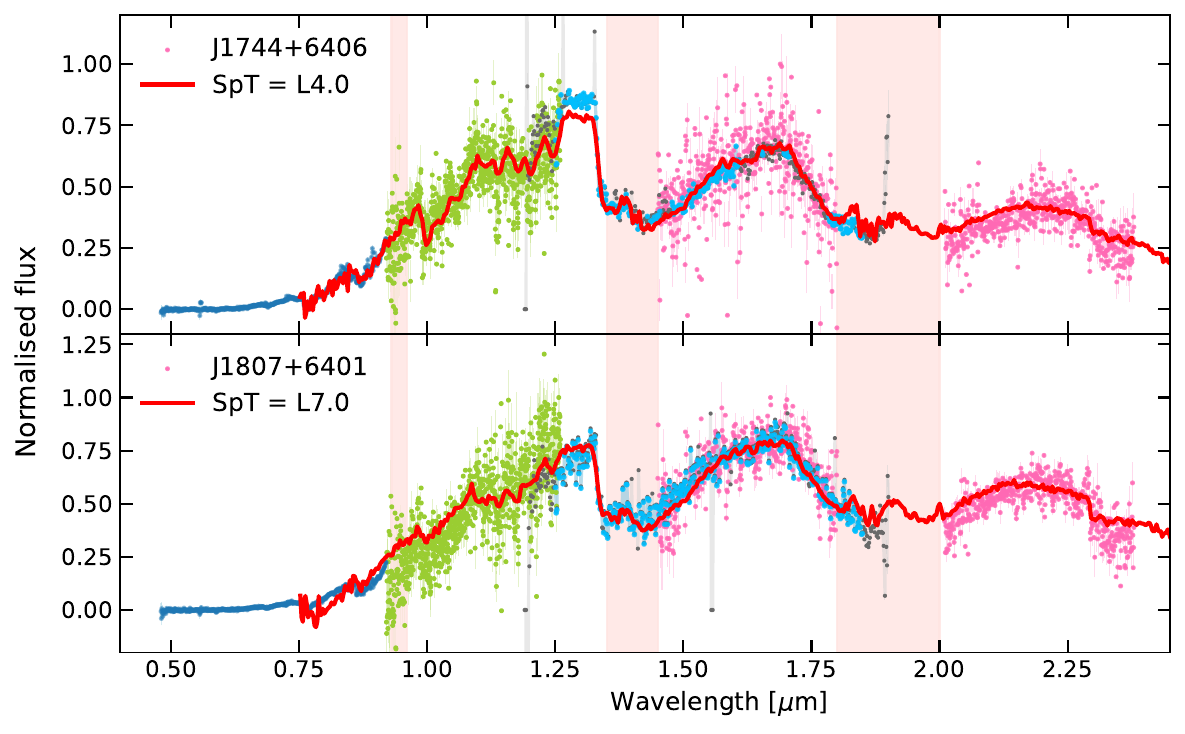}
    \caption{GTC spectra of two newly discovered UCDs in the \EDFN; J1744$+$6406 (\emph{top panel}) and J1807$+$6401 (\emph{bottom panel}). OSIRIS$+$, EMIR/\textit{YJ}, and EMIR/\textit{HK} observations are shown in dark blue, green, and rose, respectively.  \Euclid spectra are plotted following the colour scheme in Fig.~\ref{fig:maskeffect}. Each spectrum is normalised and scaled to the SPLAT spectral standard (red), corresponding to the \Euclid SpT determined in this work.}
    \label{fig:gtc}
\end{figure*}

The resulting GTC spectra for  J1744$+$6406 and J1807$+$6401 are shown in Fig.~\ref{fig:gtc}. Spectra from the different instruments were normalised and scaled to the SPLAT standard spectrum corresponding to their \Euclid-based SpTs for visualisation purposes. \Euclid spectra, OSIRIS$+$, EMIR/\textit{YJ}, and EMIR/\textit{HK} components, and the SPLAT spectral standard are displayed. Regions strongly affected by atmospheric absorption were removed from the GTC spectra.

The \Euclid spectra are consistent with the GTC data in both spectral shape and flux level, while exhibiting significantly reduced scatter due to the absence of telluric contamination. Notably, the spectral region from 1.35--1.45\,\micron{}, corresponding to the peak of the \ce{H2O} absorption band and \JE to \HE transition, is inaccessible to ground-based observations due to atmospheric opacity. \Euclid provides uninterrupted coverage in this region, revealing water absorption features that are critical for characterising UCD atmospheres \citep{2010JQSRT.111.2139R, 2022JQSRT.27707949G}, offering constraints on $T_{\mathrm{eff}}$, $\log{g}$, and SpT, as well as validating atmospheric models and refining molecular line lists.

The combined GTC spectra, spanning 0.48\,\micron{} to 2.4\,\micron{}, were assigned SpTs using SPLAT's classify-by-standard method, and they yield L5 and L7 SpTs for J1744$+$6406 and J1807$+$6401, respectively. These spectra were subsequently fitted with ATMO atmospheric models using the \texttt{stelparucds} package, yielding stellar parameters summarised in Table~\ref{tab:gtc}. The parameters were derived from \Euclid spectra, GTC spectra, and \Euclid{}$+$GTC combined spectra. 
The quoted uncertainties correspond to the formal Bayesian posterior uncertainties from the model fitting procedure and do not include possible systematic uncertainties associated with the atmospheric models, flux calibration, or the finite resolution of the model grid.

% ===============================================
\begin{table*}[]
     \centering 
     \caption{Stellar parameters obtained from model fitting of \Euclid, GTC, and \Euclid $+$ GTC  spectra.}
     \begin{tabular}{ccccccc}
 \toprule
 \toprule
     Parameter & & J1744$+$6406 & & & J1807$+$6401 & \\
      & \Euclid & GTC & \Euclid $+$ GTC &\Euclid & GTC & \Euclid $+$ GTC \\
\midrule

$T_{\mathrm{eff}}$  [K]    &  $2103^{+9}_{-8}$  &  $1829^{+8}_{-8}$   &   $1901^{+14}_{-13}$  & $1634^{+18}_{-16}$  &    $1550^{+10}_{-11}$  &    $1554^{+10}_{-10}$ \\

$\log{g}$ [dex]   & $4.51^{+0.01}_{-0.01}$ &  $4.54^{+0.03}_{-0.04}$   &   $4.52^{+0.01}_{-0.02}$   & $4.64^{+0.06}_{-0.06}$  &    $4.74^{+0.03}_{-0.03}$   &    $4.76^{+0.04}_{-0.04}$   \\

[Fe/H] [dex] & $0.18^{+0.01}_{-0.02}$ &   $0.19^{+0.03}_{-0.03}$  &   $0.15^{+0.02}_{-0.02}$  & $-0.36^{+0.04}_{-0.04}$ &   $-0.60^{+0.00}_{-0.01}$    &   $-0.58^{+0.02}_{-0.02}$   \\

\bottomrule
     \end{tabular}
     \label{tab:gtc}
 \end{table*}
% ===============================================

% ===============================================
% ===============================================

% ===============================================
% ===============================================
\section{Conclusions}
\label{sec:conclusions}

This work demonstrates the potential of \Euclid{} Q1 spectroscopy to identify and characterise UCDs based solely on their spectral features, without prior morphological filtering. We validated a template-matching spectral classification method against a comparison sample of known UCDs, confirmed the robustness of this approach, and applied it to newly identified candidates. We confirm \NNEWSPECTRA{} new UCDs in the \EDFN, six of which were also found in Q1 studies by \cite{dominguez2025} and \cite{zerjal2026}, with SpTs ranging from M7 to T1 ($\JE = 17$--21\,mag and $d = 50$--600\,pc). Two of these objects are independently validated through ground-based optical and NIR follow-up observations using the GTC OSIRIS$+$ and EMIR instruments.
Q1 spectra of new UCDs, along with \NKNWSPECTRA previously known UCDs, were analysed to obtain stellar parameters and SpTs. The known UCD sample includes 103 DES objects (M7--T0, $\JE = 17$--22\,mag, and $d = 50$--600\,pc), 6 LIT objects (M8--T7, $\JE = 18$--20\,mag, and $d = 30$--300\,pc), and 58 \textit{Gaia} sources (M7--L1, $\JE = 17$--24 \,mag, and $d = 70$--300\,pc).

Stellar parameters of the UCD sample were derived from NISP spectra using ATMO atmospheric models. The derived $T_{\rm eff}$ values are consistent with the Stephens–Leggett relation. The combination of \Euclid and \textit{Gaia} spectra enables the robust estimation of fundamental parameters, including stellar radii, which were computed with the help of \textit{Gaia} parallaxes. Estimates based on a single band exhibit systematic biases: using only \textit{Gaia} RP leads to underestimates, while relying solely on \HE\ results in overestimates. In contrast, the combined \Euclid{}$+$\textit{Gaia} approach yields more reliable and stable parameter estimates. GTC spectra extend the wavelength coverage and further support parameter validation. Among the various flux options in the \Euclid MER catalogue, aperture photometry is found to be the optimal choice for point-like sources. The photometry based on an aperture diameter of $2 \times$ FWHM for NISP filters and $4 \times$ FWHM for VIS shows the least deviation from published photometry for previously known DES UCDs in Q1. Astrometric and photometric analysis yields distance estimates and reduced proper motion diagrams, indicating that \Euclid UCDs are distributed up to $\sim$600\,pc. The late-type SpTs are closer, typically within 100\,pc for late-L- and T-type dwarfs.

Near-infrared colours provide an effective means of identifying UCDs, with high-redshift QSOs constituting the primary point-source contaminants. UCDs exhibit significantly redder VIS$-$NIR colours than low-redshift QSOs \citep[Fig.~1 of][]{Barnett-EP5, Q1-SP068} and may be undetectable in the optical wavelengths. For brighter objects ($\JE < 21.5$\,mag), \Euclid spectra provide sufficient diagnostic power to distinguish UCDs from high-redshift QSOs. However, the distinction becomes more challenging at fainter magnitudes or when relying solely on Q1 photometry, highlighting the need for continued methodological development.

Spectroscopic searches for UCDs using \Euclid\ Q1 data, sometimes combined with photometry, have proven highly effective. \cite{zhang2025} used a machine-learning-based approach that combines Vera C.~Rubin Observatory Legacy Survey of Space and Time and \Euclid Q1 data to identify UCDs. Their search has resulted in 15 new detections, with SpTs ranging from M6 to L2, in the \EDFS, which is non-overlapping and complementary to the search in \EDFN presented in this work. \cite{frank2026} conducted a study focused on T dwarfs in Q1, using a combined photometric and spectroscopic approach; they identified and characterised 15 T dwarfs with SpTs from T2 to T7. The work presented here extends the previous efforts by deriving stellar parameters for the \Euclid UCDs and explores their complementarity with \textit{Gaia}, which has not been addressed in previous studies. \textit{James Webb} Space Telescope spectra have revealed distant UCDs in the Galactic thick disc \citep{morrissey2025}. The upcoming \Euclid data releases in the EDFs may similarly provide the opportunity to observe more distant UCDs.

As discussed in Sect.~\ref{sec:qualitycheck}, identifying faint, fast-moving sources presents significant challenges. \Euclid{}’s unprecedented sensitivity results in a confusion-limited sky, where cross-matching is non-trivial and spectral blending is frequent. 
Despite these challenges, Q1 already delivered over 4 million spectra and 200 million photometric observations. \cite{zerjal2026} photometrically identified over 5000 UCDs, and \Euclid is expected to eventually observe more than 1.5 million UCDs. The next major data release (DR1), expected in late 2026, will expand the sky coverage by a factor of ~30, enabling a more complete substellar census. Future larger data releases will require an automated quality assessment framework based on parameters such as S/N, magnitude, and additional quality metrics, calibrated on spectra from forthcoming releases. The UCD sample and analysis framework developed in this work will provide a valuable foundation and serve as a training set for this effort.

The methods and results presented in this study serve as an important first step towards realising the full potential of \Euclid{} in the study of UCDs.
The resulting catalogue, containing both newly discovered and previously known UCDs with spectral classifications and derived parameters, provides a valuable resource for statistical studies of low-mass stellar populations. This work establishes reliable tools for identifying and characterising faint UCDs and demonstrates the power of combining \Euclid data with other space- or ground-based observations for detailed physical characterisations. These efforts lay the foundation for future large-scale studies of substellar objects and contribute to building a more complete understanding of the low-mass stellar population of the Galaxy.

% ===============================================
% ===============================================
% ===============================================
\begin{acknowledgements}

We would like to thank Dr. Matthew J. Temple for his help in constructing the QSO tracks and Davy Kirkpatrick and Federico Marocco for providing Keck MOSFIRE images.
\AckEC  

\AckQone

\cite{Q1cite}

Based on data from UNIONS, a scientific collaboration using three Hawaii-based telescopes: CFHT, Pan-STARRS, and Subaru \url{www.skysurvey.cc}\,. Based on data from the Dark Energy Camera (DECam) on the Blanco 4-m Telescope at CTIO in Chile \url{https://www.darkenergysurvey.org}\,. 
This work uses results from the ESA mission \textit{Gaia}, whose data are being processed by the Gaia Data Processing and Analysis Consortium \url{https://www.cosmos.esa.int/gaia}\,. 
RS acknowledges support in part from the Italian Space Agency (ASI) through contracts 2018-24-HH.0 and 2018-24-HH.1-2022 awarded to the National Institute for Astrophysics (INAF).
Funding for CDT, ELM, M{\v Z}, and JYZ was provided by the European Union (ERC Advanced Grant, SUBSTELLAR, project number 101054354).
NH, NL, MRZO acknowledge support from MCIN/AEI/10.13039/501100011033 through grants PID2023-150468NB-I00, PID2022-137241NB-C41, and PID2022-137241NB-C42, respectively.
TD acknowledges support from UKRI STFC AGP grant ST/W001209/1. For the purpose of open access, the author has applied a Creative Commons Attribution (CC-BY) licence to any Author Accepted Manuscript version arising from this submission.
\end{acknowledgements}
% ===============================================

% ===============================================
% ===============================================
% ===============================================
\bibliography{combined}
\onecolumn
% ===============================================

% ===============================================
% ===============================================
% ===============================================
\begin{appendix}

% ===============================================
% ===============================================
\section{UCD catalogue}
\label{app:table}

This appendix provides the details of the electronic table containing information on all the UCDs analysed in this study, including both the \NKNWSPECTRA previously known and \NNEWSPECTRA{} newly discovered UCDs in Q1.

% ===============================================
% \input{tex_files/fulldataset_combined.table}
\begin{table*}[htbp!]
\caption{Contents of the UCD catalogue. The first selected object is provided as an example. The complete catalogue, containing all UCDs used in this work, is made available online.
}
\label{tab:fulldataset}
\centering
\begin{tabular}{lllll}
\toprule
\toprule
Parameter & Format & Unit & Comment & Example  \\ 
\midrule
\texttt{SHORT\_NAME} &    a14 & ...    &                 Short name used in text            &    J0352-4910 \\
\texttt{SIMBAD\_NAME} &    a30 & ...    &        Common discovery name                       &     WISEA J035231.80-491059.4 \\
\texttt{EUCLID\_NAME} &    a27 & ...    &                  Positional IAU name in Q1               &   EUCL\,J035231.98-491058.8\\ 
\texttt{OBJECT\_ID} &    i20 & ...    &               Unique source identifier in Q1            &  $-$581332495491830038\\
\texttt{RIGHT\_ASCENSION} &  f11.7 & deg    &                   Right ascension (ep. $\sim$ 2024.6)    &  58.1332495\\
\texttt{DECLINATION} &  f11.7 & deg    &                   Declination (ep. $\sim$ 2024.6)        & $-$49.1830038\\
\texttt{PMRA} &   f9.3 & mas $\mathrm{yr}^{-1}$ &        Proper motion in RA                         & Null\\
\texttt{PMRA\_ERR} &   f9.3 & mas $\mathrm{yr}^{-1}$ &        RA proper motion error                      & Null\\
\texttt{PMDec} &   f9.3 & mas $\mathrm{yr}^{-1}$ &        Proper motion in  Dec                       & Null\\
\texttt{PMDec\_ERR} &   f9.3 & mas $\mathrm{yr}^{-1}$ &        Dec proper motion error                     & Null\\
\texttt{FLAG\_VIS} &     i3 & ...    &         VIS flux flag                              &   0\\
\texttt{FLUX\_VIS\_2FWHM\_APER} &  f11.3 & $\mu$Jy &              VIS flux at  2 FWHM aperture          &       0.691\\
\texttt{FLUXERR\_VIS\_2FWHM\_APER} &  f11.3 & $\mu$Jy &        VIS aperture photometry error               &       0.050\\
\texttt{FLAG\_Y} &     i3 & ...    &         \textit{Y} flux flag                                &   0\\
\texttt{FLUX\_Y\_2FWHM\_APER} &  f11.3 & $\mu$Jy &                  \textit{Y} flux at 2 FWHM aperture         &      37.511\\
\texttt{FLUXERR\_Y\_2FWHM\_APER} &  f11.3 & $\mu$Jy &        \textit{Y} aperture photometry error                 &       0.433\\
\texttt{FLAG\_J} &     i3 & ...    &         \textit{J} flux flag                                &   0\\
\texttt{FLUX\_J\_2FWHM\_APER} &  f11.3 & $\mu$Jy &                 \textit{J} flux at 2 FWHM aperture          &      56.488\\
\texttt{FLUXERR\_J\_2FWHM\_APER} &  f11.3 & $\mu$Jy &        \textit{J} aperture photometry error                 &       0.481\\
\texttt{FLAG\_H} &     i3 & ...    &         \textit{H} flux flag                                &   0\\
\texttt{FLUX\_H\_2FWHM\_APER} &  f11.3 & $\mu$Jy &                  \textit{H} flux at 2 FWHM aperture         &      33.353\\
\texttt{FLUXERR\_H\_2FWHM\_APER} &  f11.3 & $\mu$Jy &       \textit{H} aperture photometry error                 &       0.298\\
\texttt{FWHM} &   f4.1 &  arcsec &        Full width at half maximum                     &  1.2\\
\texttt{POINT\_LIKE\_PROB} &   f4.2 & ...    &                   Source point-like probability    & 0.18\\
\texttt{ELLIPTICITY} &   f4.2 & ...    &        SExtractor source ellipticity              & 0.19\\
\texttt{POSITION\_ANGLE} &   f5.1 & deg    &                       Position angle of the source &  11.0\\
\texttt{DATALABS\_PATH} &    a35 & ...    &                     ESA datalabs spectrum file path & /data/euclid\_q1/Q1\_R1/SIR/102020531\\
\texttt{DIST} &   f7.1 & pc     &        Spectroscopic distance                      &    32.4\\
\texttt{DIST\_ERR} &   f7.1 & pc     &        Spectroscopic distance error                &    11.4\\
\texttt{SPT} &    a10 & ...    &        Spectral type from Q1 NISP spectra                    &      T7.0 \\
\texttt{SPECTRA\_SNR}
&   f7.2 & ...    &        Spectra signal-to-noise                     &   15.32\\
\texttt{SPECTRA\_CHISQ} &     i4 & ...    &               Spectrum $\chi^{2}$ value        &  2.43 \\
\texttt{SPECTRA\_PIXELS} &     i4 & ...    &               Spectrum good pixels number           &  321\\
\texttt{T\_eff} &     i4 & K    &               $T_{\mathrm{eff}}$ from model fitting         &  2777 \\
\texttt{logg}  &     f2.2 & dex    &               $\log{g}$ from model fitting           &  5.35 \\
\texttt{[Fe/H]} &     f2.2 & dex    &               [Fe/H] from model fitting          &  0.08\\ 
\texttt{Gaia\_ID} & i20 & ... & $\textit{Gaia}$ source ID of the object &  2258140353645857536 \\
\texttt{G} & f2.3 & mag & $\textit{Gaia}$ \gmag\ magnitude & 20.313 \\              
\bottomrule
\end{tabular}
\end{table*}
% ===============================================

% ===============================================
% ===============================================
\section{Point-source photometry }
\label{app:photometry}
In this appendix, we compare the various flux measurements in the Q1 MER catalogue in order to select the most suitable magnitude for point-like sources. Thereafter, we analyse the UCDs in different colour-colour diagrams and compare them with the locus of QSOs.

\begin{figure*}
\centering
\includegraphics[angle=0,width=0.49\hsize]{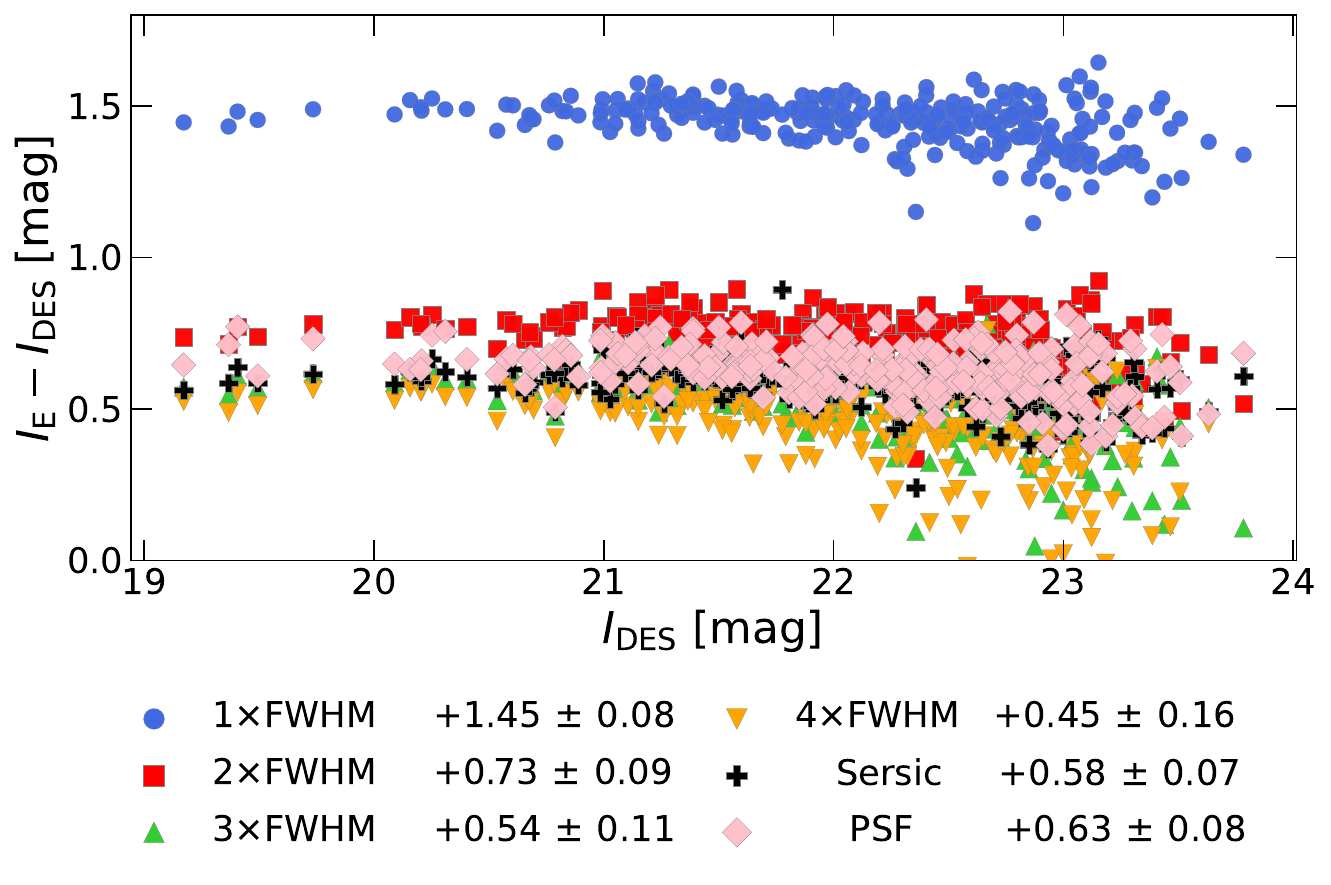}
\includegraphics[angle=0,width=0.49\hsize]{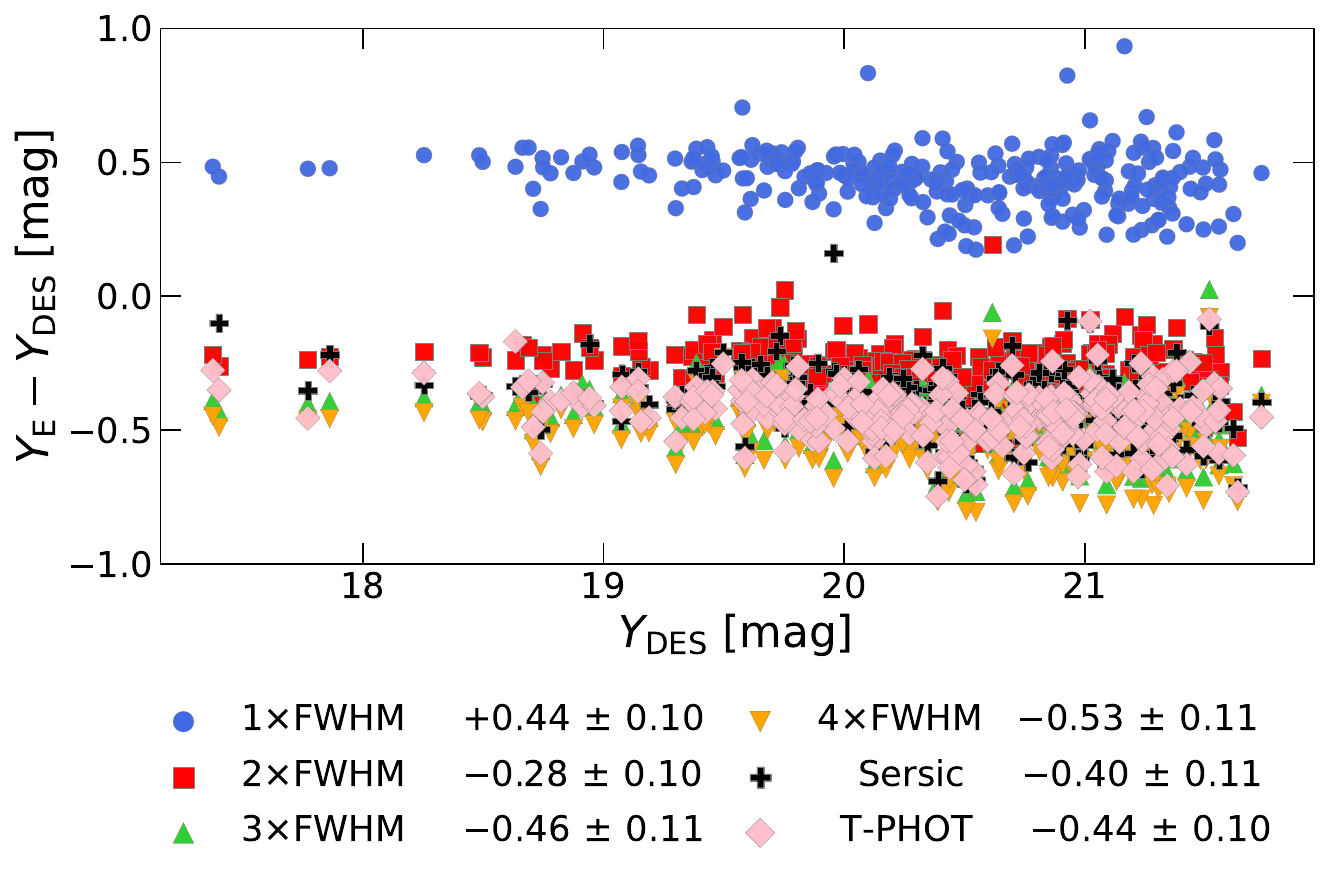}
\includegraphics[angle=0,width=0.49\hsize]{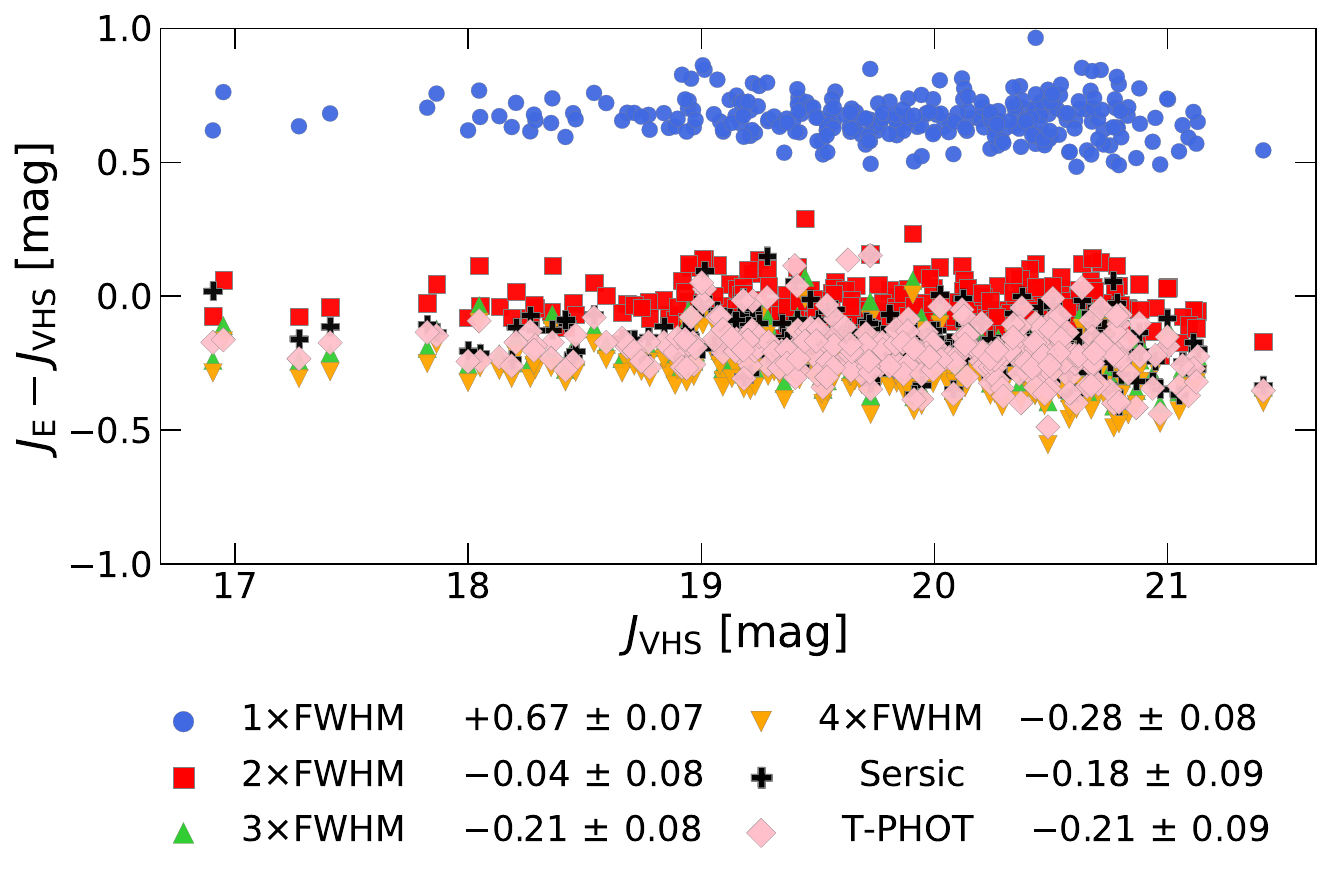}
\includegraphics[angle=0,width=0.49\hsize]{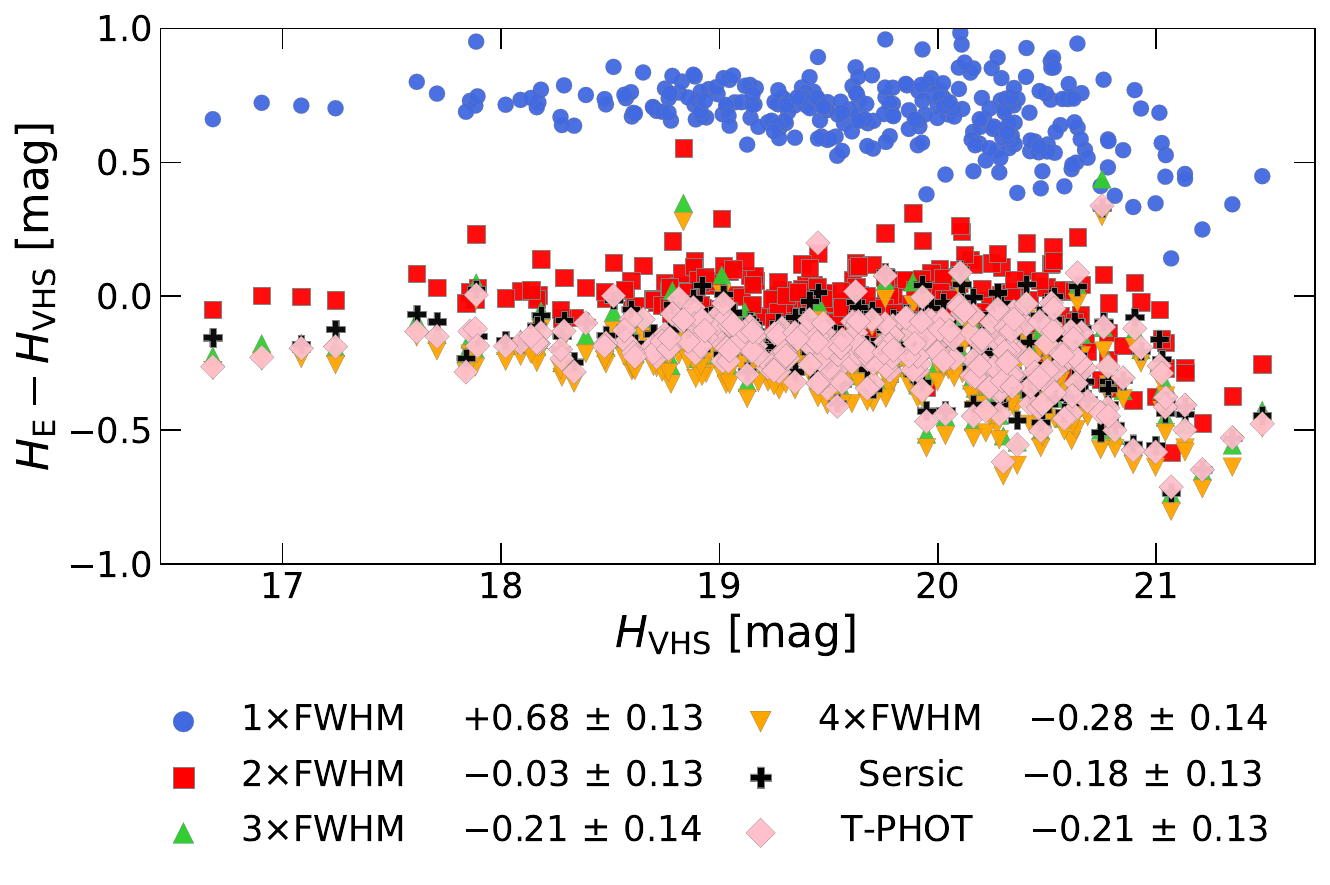}
\caption{Comparison of MER magnitudes and published magnitudes for the benchmark DES objects. \textit{Top left}: DES $i$ compared with \Euclid \IE. \textit{Top right}: DES $Y$ compared with \Euclid \YE. \textit{Bottom left}: VHS $J$, transformed to the AB system, compared with \Euclid \JE. \textit{Bottom right}: VHS $H$, transformed to the AB system, compared with \Euclid \HE. Points of different colours correspond to magnitudes measured with various \Euclid flux measurements: \texttt{A-PHOT} with aperture sizes 1$\times$, 2$\times$, 3$\times$, and 4$\times$ FWHM (blue, red, green, and yellow), S\'ersic magnitudes (black), and PSF (VIS) or \texttt{T-PHOT} (NISP) magnitudes (pink; see Sect.~\ref{sec:euclid_cl} for details). The legend reports the mean and standard deviation of each distribution after an iterative 5$\sigma$ clipping.}
\label{fig:merphoto}
\end{figure*}

Figure \ref{fig:merphoto} illustrates the difference between the various \Euclid magnitudes and the $i$, $Y$, $J$, and $H$ magnitudes measured in the DES and VHS surveys for the 326 DES objects in common with Q1 (Sect.~\ref{sec:des_sources}). Points of different colours denote the distinct Q1 MER flux options including \texttt{A-PHOT} apertures ($1\times$, $2\times$, $3\times$, $4\times\textrm{FWHM}$), S\'ersic, and PSF/\texttt{T-PHOT} measurements. Q1 fields have low extinction, and therefore, no reddening is applied for the transformations.
The top panel shows the DES $i$- and $Y$-band comparisons with \Euclid \IE and \YE, while the bottom panel shows the VHS $J$- and $H$-band comparisons with \Euclid \JE and \HE. An iterative 5$\sigma$ clipping is applied to the data, and the mean and standard deviation of each distribution are reported in the legend.

For NISP filters, \texttt{A-PHOT} flux measured within a $2\times$ FWHM aperture shows the least deviation, while for the \Euclid \IE filter, \texttt{A-PHOT} flux measured within a $4\times$ FWHM aperture provides the closest agreement with the literature values. We adopted these fluxes to calculate the respective \Euclid magnitudes.

\begin{figure*}
\centering
\includegraphics[angle=0,width=0.49\hsize]{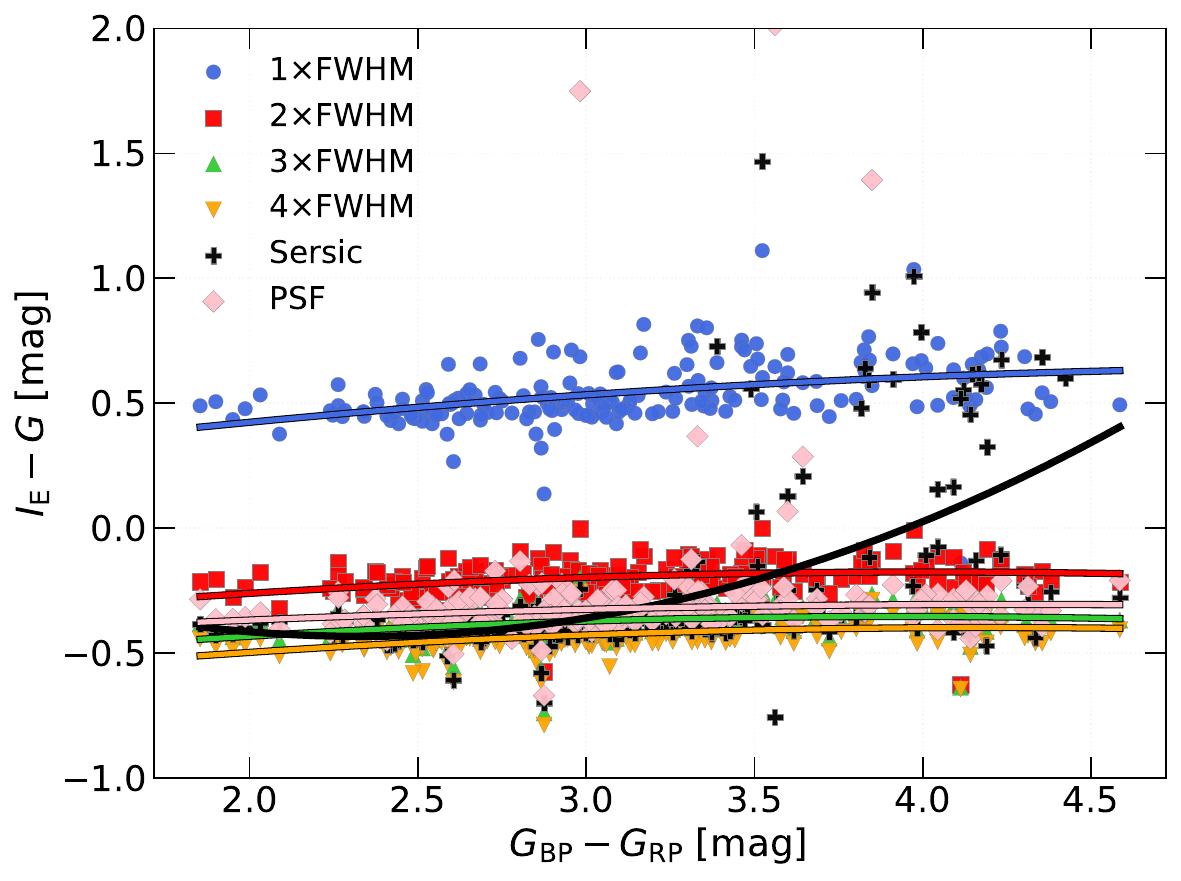}
\includegraphics[angle=0,width=0.49\hsize]{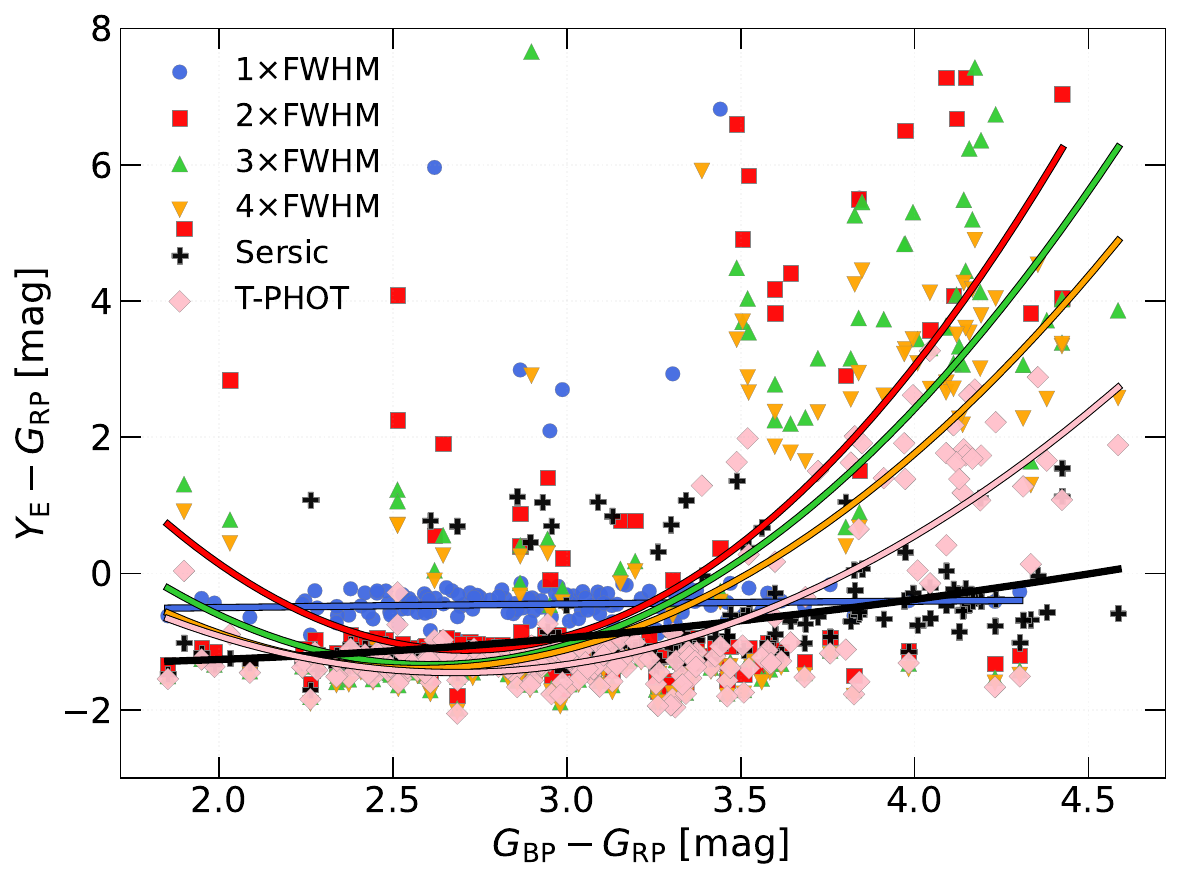}
\caption{
Comparison of \Euclid photometry (\IE and \YE bands) with \textit{Gaia} magnitudes. The magnitude differences are plotted as a function of $G_{\rm BP} - G_{\rm RP}$ colours. The different colours denote the various \Euclid photometric measurements. The scatter points correspond to the observations for the \textit{Gaia} UCD sample, while the line denotes the quadratic fit to the data. 
}
\label{fig:merphoto_gaia}
\end{figure*}

The \textit{Gaia} $G$ band overlaps with the \Euclid \IE band and extends into the blue end of the \Euclid \YE filter. The \textit{Gaia} $RP$ band partially overlaps with the \IE band and the blue edge of the \YE band, but cuts off before the \JE or \HE bands. The $BP$ band, being mostly optical, does not have a direct \Euclid counterpart. 
Figure~\ref{fig:merphoto_gaia} demonstrates the distribution of the $\IE - \gmag$ and $\YE - \grp$ colours as a function of \textit{Gaia} colour, $\gbp - \grp$, for the different photometric extraction methods. Using UCDs common with \Euclid and \textit{Gaia}, a quadratic polynomial fit is derived after applying iterative $5\sigma$ clipping to remove outliers. The residual scatter around each fit was used to assess the relative performance of the different extraction techniques.

The residual scatter is comparable in 2$\times$, 3$\times$, and 4$\times$ FWHM for the \IE band and smallest in 1$\times$ FWHM for the \YE band. The \YE-band transformations are inherently noisier due to limited colour overlap. We list the transformations between \IE-band and \textit{Gaia} magnitudes for different photometric extraction methods. The S\'ersic method shows a difference with the rest of the transformation and larger dispersion due to its incompatibility with stellar photometry.

% ===============================================
\begin{table}[ht]
\centering
\caption{Transformations between \IE\ and \textit{Gaia} photometry for different extraction methods. The relations are of the form $ \IE = \gmag + a(\gbp-\grp)^2 + b(\gbp-\grp) + c $.}
\label{tab:gaia_transform}
\begin{tabular}{lcccc}
\toprule
\toprule
Method & $a$ & $b$ & $c$ & $\sigma$ \\
 & [mag$^{-1}$] &  & [mag] & [mag] \\
\midrule
1$\times$FWHM & $-0.0190$ & $0.2050$ & $0.0896$  & $0.0985$ \\
2$\times$FWHM & $-0.0210$ & $0.1689$ & $-0.5151$ & $0.0585$ \\
3$\times$FWHM & $-0.0211$ & $0.1659$ & $-0.6805$ & $0.0538$ \\
4$\times$FWHM & $-0.0205$ & $0.1722$ & $-0.7601$ & $0.0560$ \\
S\'ersic      & $0.1654$  & $-0.7716$ & $0.4663$ & $0.2548$ \\
PSF           & $-0.0122$ & $0.1043$ & $-0.5279$ & $0.0666$ \\
\bottomrule
\end{tabular}
\end{table}
% ===============================================

\begin{figure*} 
\centering
\includegraphics[angle=0,width=0.39\hsize]{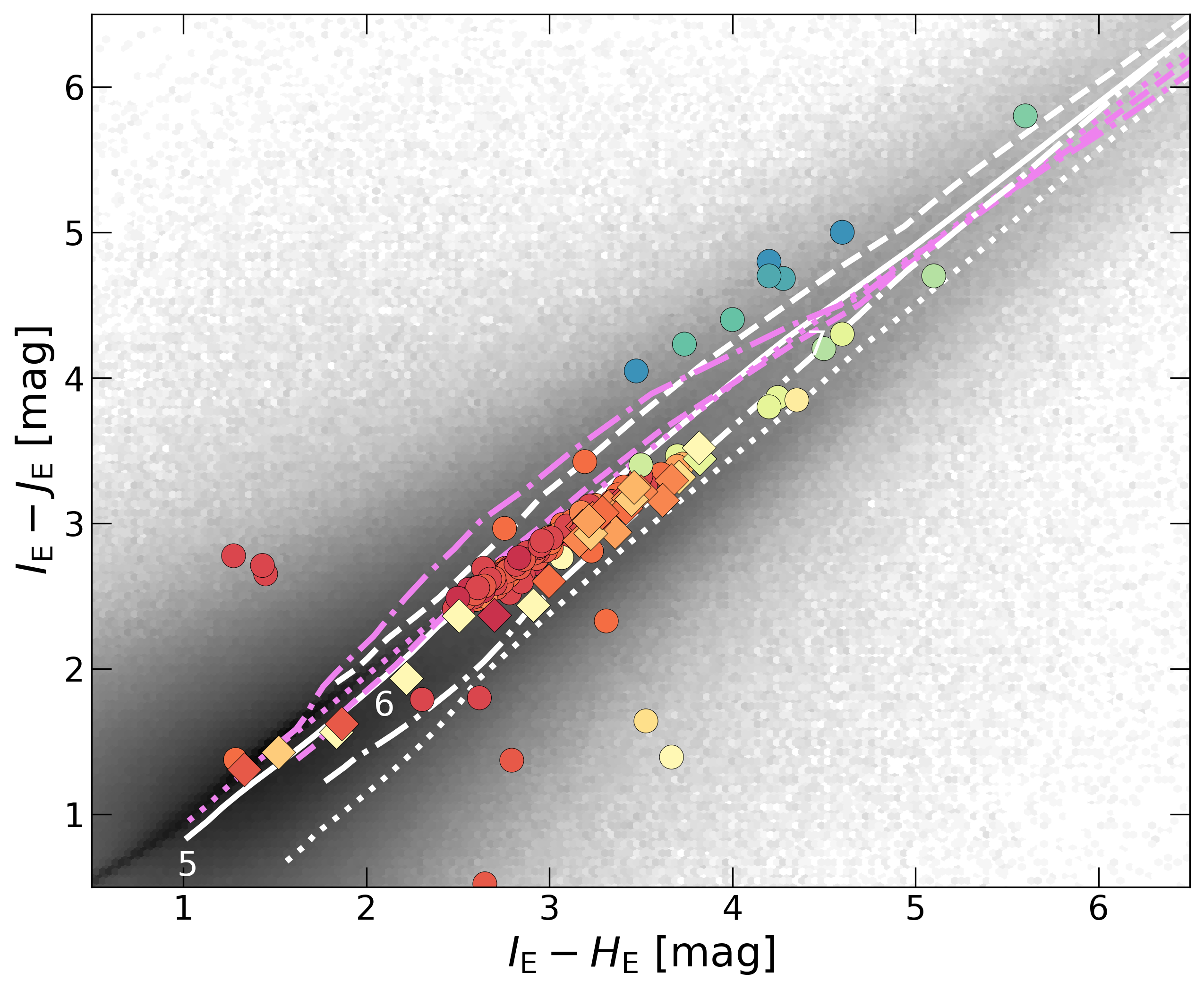}
\includegraphics[angle=0,width=0.45\hsize]{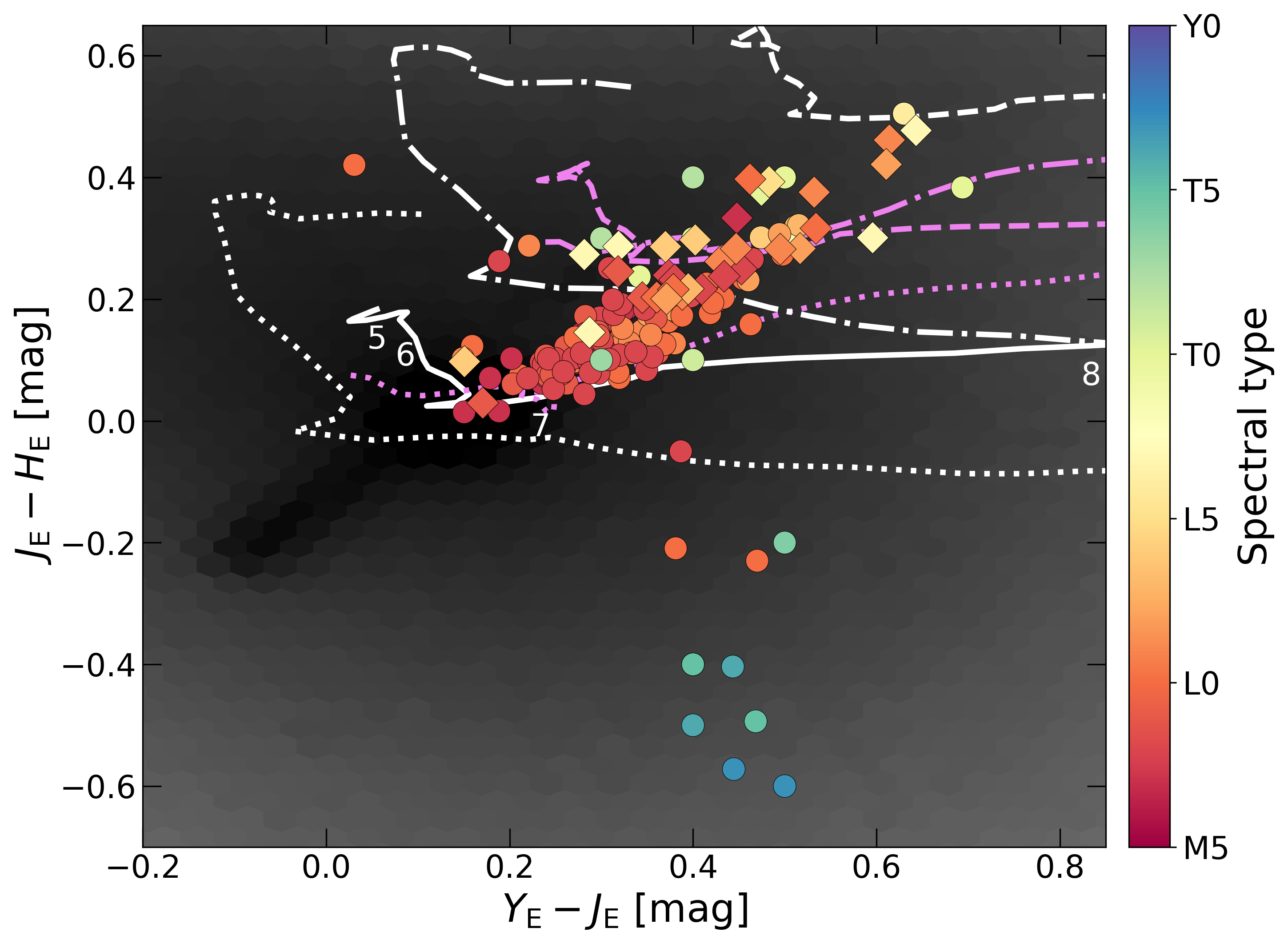}
\includegraphics[angle=0,width=0.41\hsize]{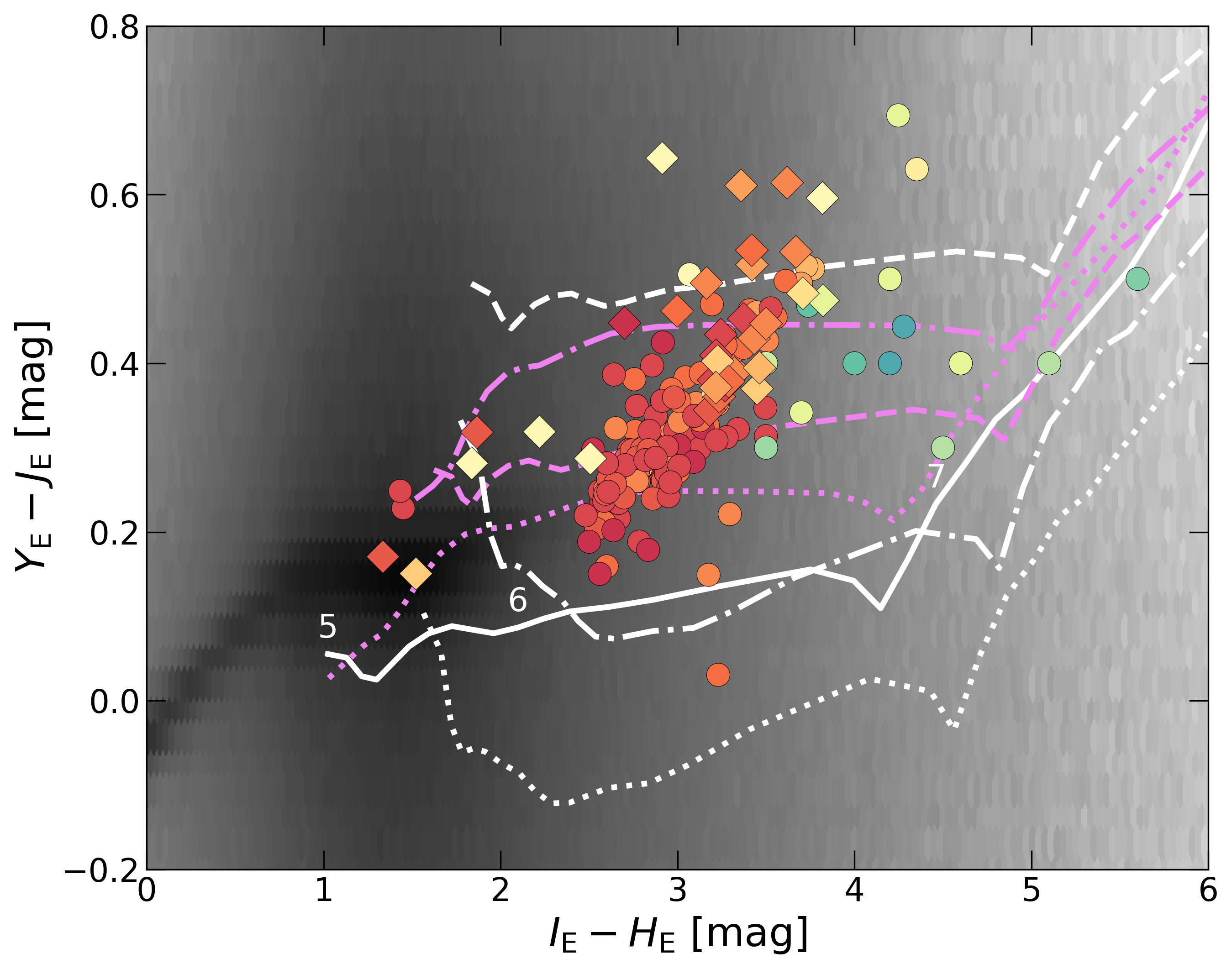}
\includegraphics[angle=0,width=0.45\hsize]{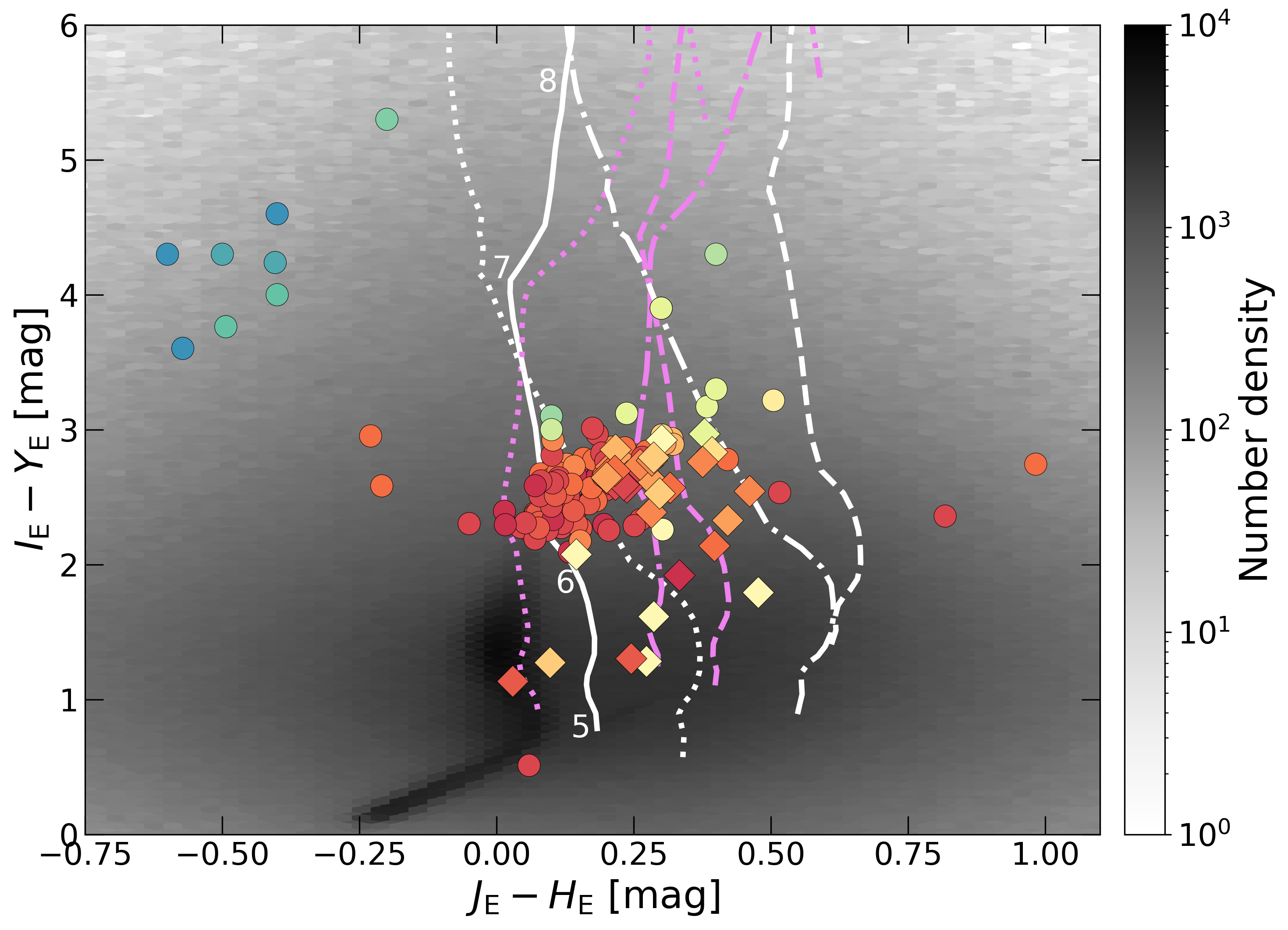}
\caption{Colour-colour diagrams of UCDs with reliable spectral typing. Dots represent \NKNWSPECTRA{} previously known UCDs, while diamonds denote the \NNEWSPECTRA{} new UCDs in \EDFN. They are colour-coded according to the \Euclid SpT assigned in this paper, as denoted in the scale. Grey points in the background represent the Q1 objects with  \texttt{point}\_\texttt{like}\_\texttt{prob} $\geq 0.8$, with shading indicating number density.
The lines indicate the expected locations of QSOs with redshifts from $z=5$ to $z=9$. The solid white line shows the default QSO model predictions. Solid white and purple lines denote extremely weak and extremely strong lined QSOs. Dashed white and purple lines denote \emph{E(B$-$V)} = 0.1\,mag and 0.2\,mag, in the QSO rest frame. Dash-dotted white and purple lines denote extremely weak and strong-lined QSOs with \emph{E(B$-$V)} = 0.1\,mag.}
\label{fig:Q1x:YJvsJH}
\end{figure*}

\Euclid colour-colour diagrams for various filter combinations are presented in Fig.~\ref{fig:Q1x:YJvsJH}. The sample includes \NSPECTRA{} UCDs with SpTs derived from their Q1 spectra. Owing to the low number of T dwarfs in our sample, we supplemented the data with ten T dwarfs from \cite{dominguez2025} to uniformly cover the spectral subtype range.

In each panel of Fig.~\ref{fig:Q1x:YJvsJH}, the predicted QSO tracks with redshifts from $z=5$ to $z=9$ are included. The tracks, constructed using \texttt{qsogen} QSO modelling codes, are taken from \cite{2021MNRAS.508..737T}. The default model predictions are also shown. 
Variants of the model explore different physical assumptions, such as line strength variations (extremely weak- and strong-lined QSOs), dust extinction values of \emph{E(B$-$V)} = 0.1\,mag and 0.2\,mag in the QSO rest frame, and combinations of the two effects. 
These tracks are consistent with the predicted loci in previous studies \citep{Barnett-EP5, 2023ApJ...956...52T} and show significant overlap with M- and L-type dwarfs in the colour space.

% ===============================================
% ===============================================

% ===============================================
% ===============================================

% ===============================================
% ===============================================

% ===============================================
% ===============================================
\section{Model fitting on \Euclid spectra }
\label{app:atmo_bt_comparison}

We performed atmospheric model fitting on the \Euclid spectra of UCDs discussed in this work using the ATMO and BT-Settl models. 
For the ATMO model, we used fixed parameters $\mathrm{C/O}$ = 0.55 and $\gamma$ = 1.03, with uniform priors on $T_{\mathrm{eff}} = 800$--3000\,K, $\log{g} = 4.5$--5.5\,dex, [Fe/H] $ = -0.6$ to 0.6\,dex, and $\mathrm{RV} = -$150 to $+$150\,km\,s$^{-1}$.
For the BT-Settl CIFIST model \citep{2011SoPh..268..255C}, we adopted fixed enrichment, $enrich$ = 0.0, and solar metallicity, [Fe/H] $ = 0.0$\,dex, with uniform priors on $T_{\mathrm{eff}} = 1500$--3000\,K, $\log{g} = 4.5$--5.5\,dex, and $\mathrm{RV} = -$150 to $+$150\,km\,s$^{-1}$.

\begin{figure*}
\centering
\includegraphics[angle=0,width=0.45\hsize]{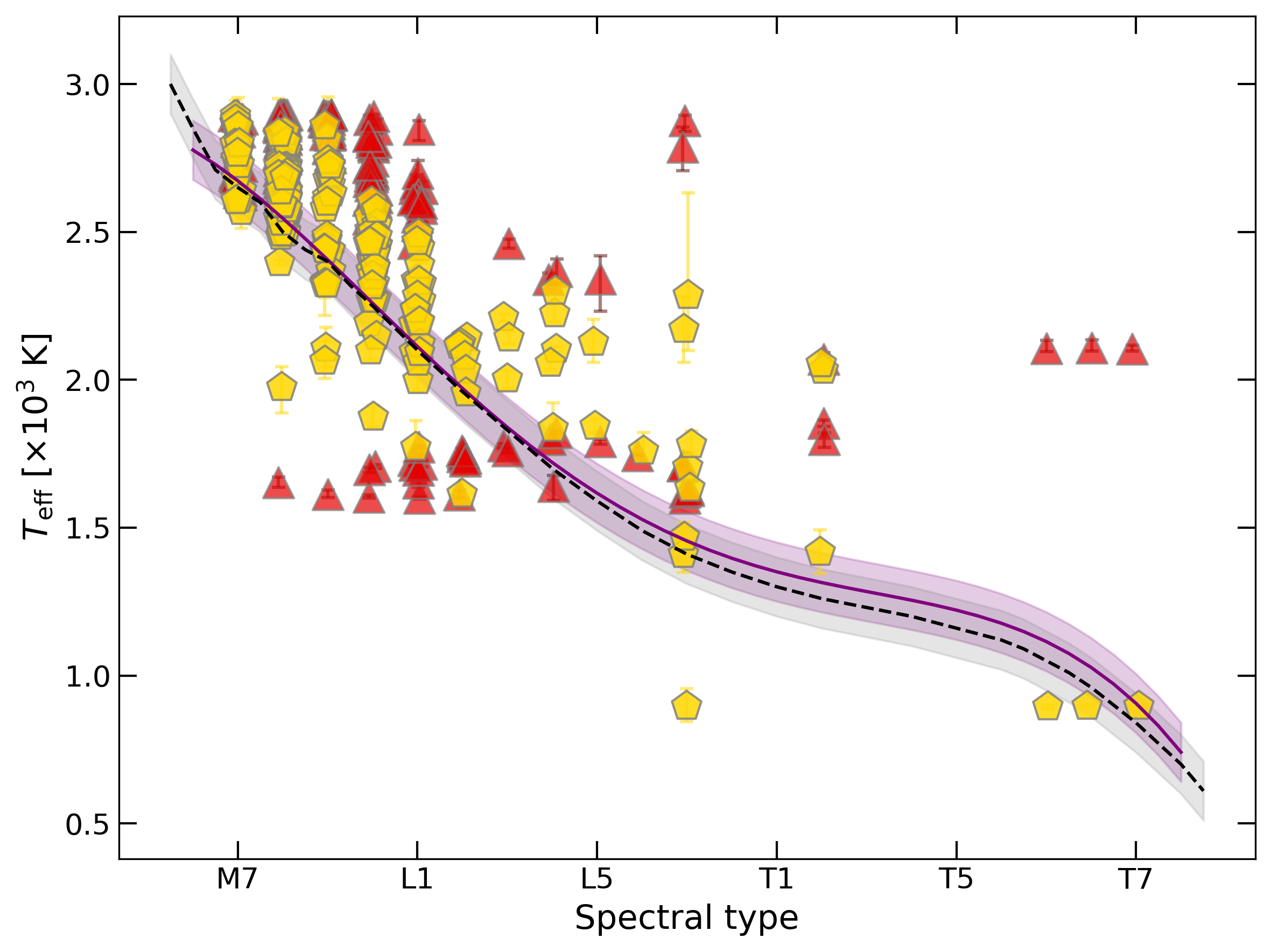}
\caption{Comparison of $T_{\mathrm{eff}}$ derived from \Euclid spectra using ATMO and BT-Settl models. $T_{\mathrm{eff}}$ is plotted as a function of \Euclid SpT. Results from the ATMO model are shown as yellow pentagons, while those from the BT-Settl model are red triangles. The purple and grey lines denote the empirical Stephens--Leggett and Pecaut--Mamajek $T_{\mathrm{eff}}$--SpT relations, respectively, with a $\pm$ 100\,K uncertainty range.}
\label{fig:model_compare}
\end{figure*}

Figure~\ref{fig:model_compare} demonstrates the distribution of stellar parameters derived from both models, $T_{\mathrm{eff}}$, as a function of the NISP spectra-based SpT. 
For $T_{\mathrm{eff}}$, ATMO shows significant agreement with the Stephens--Leggett and Pecaut--Mamajek relations, while BT-Settl shows a larger deviation, especially for later SpTs. The deviation shown by the BT-Settl model fitting with \texttt{stelparucds} is detailed in \cite{Ravinet2024}.

 % ===============================================
% ===============================================

% ===============================================
% ===============================================
\section{\textit{Gaia}--\Euclid spectra }
\label{app:gaia_euclid_spectra}

% ===============================================
\begin{figure*}
\centering
\includegraphics[angle=0,width=1\hsize]{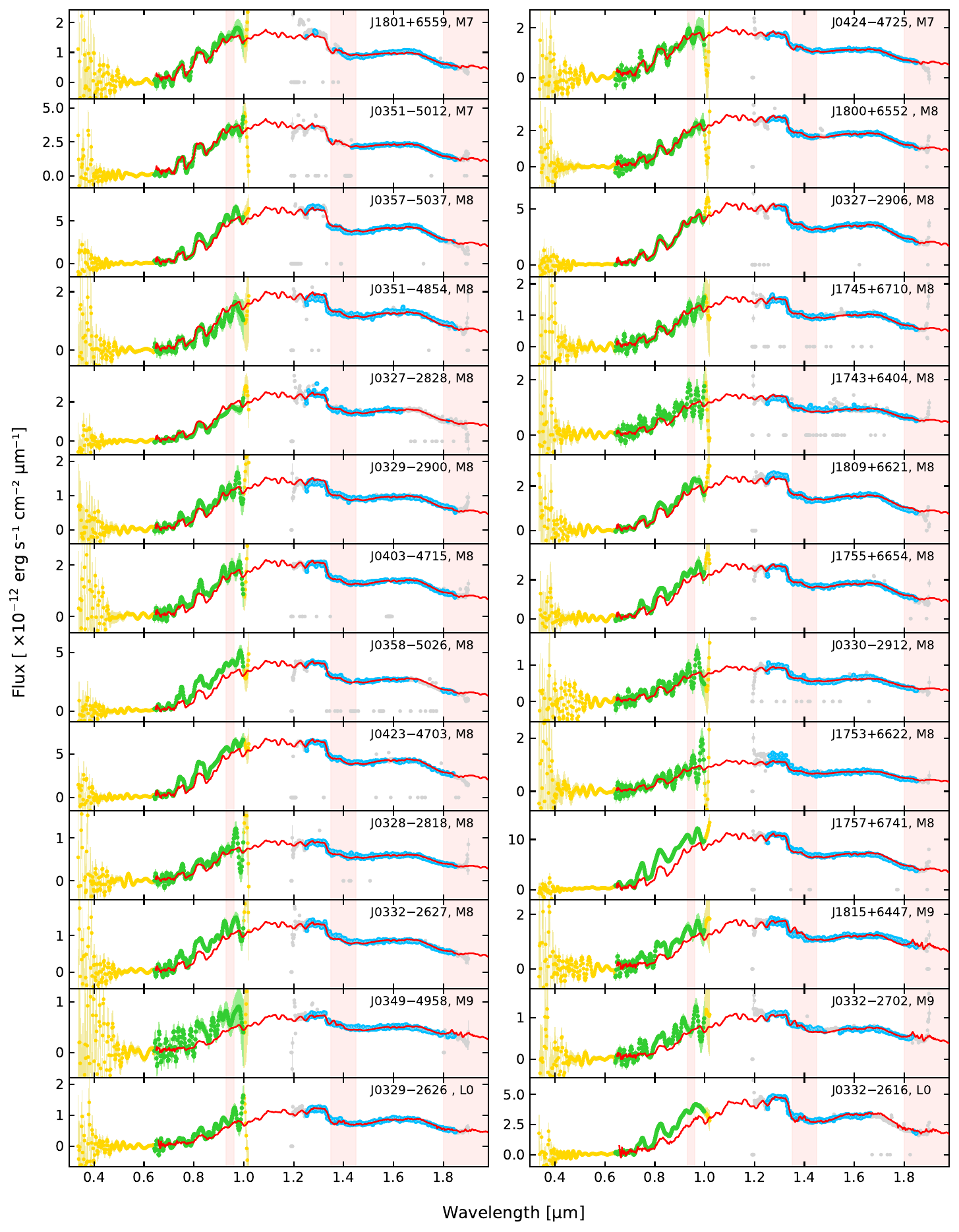}
\caption{\textit{Gaia} and \Euclid spectra of the 26 \textit{Gaia} UCDs in Q1. The full and the selected \Euclid spectral points are denoted in grey and sky blue colours. The full \textit{Gaia} XP spectrum and the selected RP spectral points are denoted in yellow and green colours. The SPLAT standard spectra corresponding to their SpT, scaled to the \Euclid spectra, are plotted in red. The dashed beige region denotes atmospheric absorption regions.}
\label{fig:26gaiaUCDs}
\end{figure*}
% ===============================================

Figure~\ref{fig:26gaiaUCDs} shows the \textit{Gaia} optical and \Euclid \RGE NIR spectra of the 26  \textit{Gaia} UCDs for which both \Euclid and \textit{Gaia} spectra are available and reliable. 
The \textit{Gaia} XP and the RP spectra are denoted in yellow and green colours. The spectra are plotted in original flux units, as obtained from the respective ESA archive. The SPLAT standard spectrum corresponding to their NIR SpT is scaled to the \Euclid flux levels. \textit{Gaia} spectra for most objects overlap exactly with this scaled spectral template.

The consistent flux calibration between \Euclid\ and \textit{Gaia} produces a continuous optical-to-NIR spectrum for these sources. The close agreement between the independently calibrated datasets highlights the high quality of the flux calibration and the strong cross-mission compatibility between the two ESA facilities.

\end{appendix}
%\label{LastPage}
\end{document}